\documentclass[%
aip,
 amsmath,amssymb,
preprint,%
]{revtex4-1}

\usepackage{graphicx}
\usepackage{dcolumn}
\usepackage{bm}

\usepackage[utf8]{inputenc}
\usepackage[T1]{fontenc}
\usepackage{mathptmx}
\usepackage{algorithm}
\usepackage{algorithmic}
\usepackage{multirow}
\usepackage{subfigure}
\usepackage{siunitx}
\usepackage{mathrsfs}
\usepackage{eucal}
\usepackage[dvipsnames]{xcolor}
\usepackage{dsfont}
\usepackage{bbold}
\usepackage{nomencl}
\usepackage{bm}
\usepackage{comment} 
\usepackage{xcolor}
\usepackage{tikz}
\usetikzlibrary{bayesnet}
\usetikzlibrary{arrows,graphs,matrix,positioning}

\usepackage[toc,page]{appendix}
\makenomenclature
\DeclareMathOperator*{\argmax}{arg\,max}
\def\drawline#1#2{\raise 2.5pt\vbox{\hrule width #1pt height #2pt}}
\def\spacce#1{\hskip #1pt}
\def\dashdotted{\hbox {\drawline{9.5}{.5}\spacce{3}\drawline{1}{.5}\spacce{3}\drawline{9.5}{.5}}\nobreak}

\begin{document}

\preprint{AIP/123-QED}

\title[Physics-Constrained Predictive Molecular Latent Space Discovery with Graph Scattering VAE]{Physics-Constrained Predictive Molecular Latent Space Discovery with Graph Scattering Variational Autoencoder}

\author{Navid Shervani-Tabar}
\email{nshervan@nd.edu}
\homepage{http://nsherv.weebly.com}

\author{Nicholas Zabaras}%
\email{nzabaras@gmail.com}
\homepage{https://www.zabaras.com/}

\affiliation{ 
Scientific Computing and Artificial Intelligence (SCAI) Laboratory, University of Notre Dame, Notre Dame, IN 46556, USA
}%

\date{\today}%

\begin{abstract}
Recent advances in artificial intelligence have propelled the development of innovative computational materials modeling and design techniques. Generative deep learning models have been used for molecular representation, discovery and design.
In this work, we assess the predictive capabilities of a molecular generative model developed based on variational inference and graph theory in the small data regime. Physical constraints that encourage energetically stable molecules are proposed. The encoding network is based on the scattering transform with adaptive spectral filters to allow for better generalization of the model. The decoding network is a one-shot graph generative model that conditions atom types on molecular topology. A Bayesian formalism is considered to capture uncertainties in the predictive estimates of molecular properties. The model's performance is evaluated by generating molecules with desired target properties.

\end{abstract}

\maketitle

\section{\label{sec:Intro}Introduction}

Designing molecules with desired properties is a challenging task. Molecular design~\cite{sanchez2018inverse} has applications in a wide variety of fields ranging from drug discovery~\cite{drews2000drug} to organic solar cells~\cite{hoppe2004organic}. Computer simulations of molecular systems remain the most prominent method for guiding molecular design~\cite{jorgensen2009efficient}. In drug design, molecular dynamics (MD) methods are used to predict how strongly a given molecule binds to a biological target, such as a protein or a nucleic acid. These computationally expensive methods usually rely on density functional theory (DFT).

With the recent advances in Artificial Intelligence (AI), deep learning techniques have become the lead contender for future direction in molecular design~\cite{elton2019deep}. AI approaches give accurate predictions of molecular properties faster than computer simulations~\cite{faber2017prediction}. 
Moreover, deep learning methods allow for an efficient exploration of the chemical space for molecular design through a low-dimensional latent representation of the molecular space and generative modeling~\cite{gomez2018automatic}. These have significantly reduced the required resources for synthesizing new drugs and molecules~\cite{stokes2020deep}.

The first attempts of such methods~\cite{gomez2018automatic} were based on Simplified Molecular-Input Line-Entry System (SMILES)~\cite{weininger1988smiles} and used variational autoencoders to learn a latent representation of molecules. One major problem with such methods was the validity of the generated molecules. To overcome this problem, Kusner et al.~\cite{kusner2017grammar} proposed a grammar Variational Autoencoder (VAE) which took advantage of parse trees to represent SMILES string structures and subsequently, learned these parse trees instead of SMILES. While this helped with syntactic validity, that is, the set of rules that define a correct combination of symbols for a particular representation, it did not guarantee semantic validity, which is the set of validity constraints that define a chemically viable molecule. With the recent advances in geometric deep learning~\cite{bronstein2017geometric}, researchers have shifted their focus to models utilizing graph-based molecular representations~\cite{gilmer2017neural}. A molecular graph consists of vertices that represent atoms and pair-wise connections that constitute covalent bonds. A weight matrix defines these bonds' order, and a signal vector captures the features for each atom.  Graphs allow the representation of topological molecular features and adding desired chemical parameters at each node. 
Moreover, 
molecules can be uniquely represented by a graph, while they may have several SMILES representations. Despite that, these methods may not be able to distinguish between two conformers, such as boat and chair conformations of cyclohexane.

Various deep learning frameworks have been used for generation of molecular graphs. These include Generative Adversarial Networks (GAN)~\cite{de2018molgan} and VAEs~\cite{simonovsky2018graphvae, liu2018constrained}. More recently, a flow-based generative model~\cite{dinh2014nice} was used~\cite{madhawa2019graphnvp}. Depending on the molecular representation used,  many encoding networks have been proposed in recent years. These include SMILES~\cite{gomez2018automatic, jastrzkebski2016learning, segler2018generating}, non-Euclidean (geometric deep learning)~\cite{hy2019covariant, chen2019graph, coley2019graph}, and Cartesian-based~\cite{thomas2018tensor, weiler20183d, kondor2018n} models. Similarly, there is an array of decoding networks for SMILES\cite{gomez2018automatic} and graph\cite{jin2018junction, you2018graph, simonovsky2018graphvae, de2018molgan, kipf2016variational} representations. More recently, methods have also been presented for the generation of 3D molecular structures~\cite{mansimov2019molecular, gebauer2019symmetry}.

The wavelet scattering transform~\cite{mallat2012group} is a deep convolutional neural network that uses a cascade of wavelet filters followed by a non-linearity operator to generate invariant features from an input signal. These features are constructed by linearizing the variations within predefined local symmetry groups~\cite{mallat2016understanding}. A map of these local invariants to a vector space can handle learning tasks such as regression and classification. In these networks, the number of nodes, layers, filters, and non-linearity layers are predefined. This eliminates the need for training the network and hence, limits the uncertainty introduced by the encoder. Moreover, scattering improves the generalization of a network in the presence of limited data~\cite{oyallon2018scattering}. In recent years, the scattering transform has been extended to non-Euclidean domains, including graphs~\cite{chen2014unsupervised, zou2019graph, gama2018diffusion}. A graph scattering network takes in the signal and its underlying graph and transforms it into features that are invariant to permutation and stable to graph and signal manipulation~\cite{zou2019graph}. This transform typically uses spectral-design wavelets to perform convolutions; that is, wavelets designed in the eigenspace defined by the graph's matrix representation. Gama et al.~\cite{gama2018diffusion} used diffusion wavelets~\cite{coifman2006diffusion}, which are defined using diffusion processes on graphs, as filters. On the other hand, Zou and Lerman~\cite{zou2019graph} used Hammond et al.'s~\cite{hammond2011wavelets} spectral wavelets for feature extraction. These wavelets, however, are only adapted to the maximum eigenvalue of the matrix representation. That is, they are defined for an interval from zero to the maximum eigenvalue, but they do not take the density of the eigenvalues in this interval into account. Therefore, in the cases where the graph eigenvalues are irregularly spaced, this may result in highly-correlated wavelets.

Graph VAEs can be divided into two categories based on their decoding network. The first group consists of the autoregressive models~\cite{jin2018junction, li2018learning}, which generate a molecule by sequentially adding components. The second group includes one-shot models~\cite{simonovsky2018graphvae, bresson2019two} that simultaneously evaluate all nodes and edges in the form of a probabilistic graph. In the autoregressive models, each iteration is conditioned on the previous iterations. These models become harder to train as the length of the sequence increases. On the other hand, one-shot models are not suitable for generating graphs with more than approximately $50$ nodes even though they are computationally more efficient. One-shot models need the maximum graph size to be predetermined~\cite{simonovsky2018graphvae} and generalize to smaller size molecules by introducing empty nodes. Many recent works on one-shot graph VAEs assume independence of nodes and edges in their graph model. Despite their success in many applications, these implementations may not learn validity constraints that dictate certain combinations of nodes and edges, i. e. a single connected graph that complies with the nodes' valency. Simonovsky and Komodakis~\cite{simonovsky2018graphvae} investigated a remedy by making the probability that a node contains an atom a function of the probability that at least a covalent bond exists that connects the node to the rest of the molecule. While this helps with such problems as isolated nodes, it does not consider the atoms' valence capacity. Furthermore, methods have been proposed to impose chemical constraints on the training process. Ma et al.~\cite{ma2018constrained} have outlined a regularization framework for one-shot models to improve the generated molecules' validity. They implemented constraints as penalty terms on the VAE's objective function to discourage the generation of invalid molecules.

Apart from the validity of the molecules based on valency and connectivity, one major challenge with molecular generative models is constructing  molecular geometries while accounting for their formation energy. In many cases, sampled molecules contain strongly strained rings requiring an excessive amount of energy to close. This includes small-sized rings, such as cyclopropane, and rings containing triple bonds that need strained angles to accommodate the prescribed geometry, such as cycloalkynes, in which the molecules would have to be far from their optimal molecular geometry as suggested by Valence-Shell Electron-Pair Repulsion Theory (VSEPR). These molecules have high formation energy and, thus, are energetically very unstable. This raises the need for shifting molecular generation toward more stable molecules.

Using limited training data induces uncertainties in the parameters of the model. Bayesian approaches have been used to model predictive uncertainty in neural networks~\cite{schoberl2019predictive}. It is common to approximate the full posterior over model parameters using the Laplace approximation~\cite{mackay2003information}. However, in practice, approximating the distribution of model parameters of a graph generative model with a Gaussian distribution may not be feasible as the sampled model could suffer from low validity of its predictions. To develop a predictive method for cases in which an accurate approximation to the posterior distribution over parameters is not available, Harris~\cite{harris1989predictive} proposed a bootstrap predictive distribution as an approximation to the predictive distribution. Fushiki et al.~\cite{fushiki2005nonparametric} further extended this to non-parametric bootstrap distribution. Fushiki recently proposed Bayesian bootstrap prediction~\cite{fushiki2010bayesian}, which takes advantage of Rubin's~\cite{rubin1981bayesian} Bayesian bootstrap by imposing a non-informative prior over bootstrap sampling weights. 

In this work, we are interested in constructing new molecular structures through a deep generative model using a limited-sized training dataset of molecular graphs. To this end, we develop a VAE that uses a hybrid graph scattering encoder and a one-shot decoder that formulates the probabilities of the nodes as conditional distributions given the edges. We further introduce constraints that increase the number of energetically stable molecules in the generated samples. Our small-sized training dataset is not sufficient to directly compute the statistics of molecular properties. We use the predicted molecular structures to perform accurate Monte Carlo estimation of these properties. To account for the model uncertainties, we consider Bayesian bootstrap resampling to yield a predictive distribution over the estimated properties and show our probabilistic confidence on them. We further inspect the generated structures by comparing their chemical space with that of the full database. This chemical space is constructed by two physicochemical properties of the molecules. Then, we look into the frequency of different functional groups in the generated samples and compare them with their frequency in the full database. We next modify the model to perform the conditional molecular generation task. In this task, we train the model on a dataset of molecular graphs and their corresponding molecular property values. We use this trained model to design molecular structures with desired target property values. Throughout this work, we chose molecular properties that can be readily estimated from the generated molecular structures. These properties include Polar Surface Area, octanol-water partition coefficient, and molecular weight. We use the RDKit chemoinformatic package to evaluate the frequency of the functional groups in the molecular structures and compute property values.

The rest of this work is structured as follows. In Section~\ref{sec:Inference}, we describe the inference problem and the general setup. Section~\ref{sec:model} discusses the encoding and decoding network architectures used. Physical constraints are introduced in Section~\ref{sec:constraint}. Section~\ref{sec:UQ} considers a Bayesian formalism to quantify uncertainties in the model. In Section~\ref{sec:results}, we train the network with limited data to discover a latent representation for the QM9 molecular database and to generate new molecules from samples in the latent space. We further assess the uncertainty of the network, providing a probabilistic evaluation of a number of properties. Finally, we report results for the generative model conditioned on a particular property.

\section{\label{sec:Inference}Problem definition}

Graphs~\cite{professor1874lvii, sylvester1878application}, along with SMILES~\cite{weininger1988smiles}, are two of the most common ways to represent a molecule in computational chemistry. A graph $G = (V, E)$ consists of a set of vertices $V=\{v_1, \dots, v_N\}$ with $N = |V|$ and a set of edges $E\subseteq V\times V$, defined by distinctive pairs of vertices $\varepsilon_{m,n}:=(v_m, v_n) \in V\times V$ with $1\leq m, n \leq N$ and $m\neq n$. A weighted graph $G = (E, V, W)$ is made of weights assigned to each edge $W=\{W_{m,n}|\varepsilon_{m,n}\in E\}$. The assigned value $W_{m,n}$ can represent proximity or strength of the relationship between the pair of vertices $(v_m, v_n)$. In a molecular graph, atoms are represented by graph vertices while edges represent the atomic bonds. In this setting, the order of the bond is shown by the weights assigned to the edges. In a graph, we can represent a signal $f:V\rightarrow \mathbb{R}^N$. Here, signals are the type of atoms sitting on the corresponding vertices. In some applications, these signals are extended to include extra atomic features, including hybridization, hydrogen neighbors, etc.

For a molecular graph $\mathcal{G}$, we define the structure of the graph using a weight matrix $\boldsymbol{W}$ and the data on the graph using a signal vector $\boldsymbol{f}$. Each element $f_i$ of the signal vector represents the atom label for the corresponding vertex $v_i$. A categorical distribution $p(f_i)$ for the type of the atom on node $v_i$ consists of probability values for each atom type. Similarly, each element of the weight matrix $W_{i,j}$ gives the order of the covalent bond. $p(W_{i,j})$ denotes the categorical probability distribution for $W_{i,j}$, which is composed of probability values for each bond order $\{\varnothing, I, II, III\}$, where label $\varnothing$ represents a disconnected pair of atoms and labels $I$, $II$, and $III$ represent single, double, and triple bonds, respectively. With these two components in hand, we can write the probability distribution $p(\mathcal{G})$ of a molecular graph as a joint distribution of atom types and bond orders for all the nodes and edges in the molecular graph
\begin{equation}
	p(\mathcal{G})=p(\boldsymbol{f}, \boldsymbol{W}).
\end{equation}

Given a finite-sized dataset of $K$ molecular graphs $\mathscr{G}=\{\mathcal{G}^{(i)}\}^{K}_{i=1}$, the objective is to reveal an underlying latent representation for this dataset, which later is used for predictive purposes, i. e. generating new molecules and estimating properties. Given the latent space variable $\boldsymbol{z}$ with prior distribution $p(\boldsymbol{z})$, the joint distribution over the observed data $\mathcal{G}$ and latent variable $\boldsymbol{z}$ can be written as
\begin{equation}
	\label{eq:joint}
	p(\mathcal{G}, \boldsymbol{z}) = p(\mathcal{G}|\boldsymbol{z})p(\boldsymbol{z}).
\end{equation}
In Eq.~(\ref{eq:joint}), $p(\mathcal{G}|\boldsymbol{z})$ represents the probability of the molecular graph $\mathcal{G}$ conditioned on its $J$-dimensional latent representation $\boldsymbol{z}$. A generative model $p(\mathcal{G})$ for the molecular graph is determined by marginalizing the joint distribution $p(\mathcal{G}, \boldsymbol{z})$ over the latent representation
\begin{equation}
    \label{eq:joint_marg}
	p(\mathcal{G})=\int p(\mathcal{G}, \boldsymbol{z}) d \boldsymbol{z}=\int p(\mathcal{G} | \boldsymbol{z}) p(\boldsymbol{z}) d \boldsymbol{z}.
\end{equation}

With the proper choice of $p(\boldsymbol{z})$ and $p(\mathcal{G}|\boldsymbol{z})$, the resulting distribution $p(\mathcal{G})$ from Eq.~(\ref{eq:joint_marg}) should mimic the true distribution of the data $p_{target}(\mathcal{G})$. Finding these distributions can be formulated as minimizing the KL distance between $p(\mathcal{G})$ and $p_{target}(\mathcal{G})$. We introduce a model $p(\mathcal{G}|{\boldsymbol{\theta}})$, parameterized by $\boldsymbol{\theta}$, to approximate the distribution $p_{target}(\mathcal{G})$, such that $p(\mathcal{G}|{\boldsymbol{\theta}}) = \int p_{\boldsymbol{\theta}}(\mathcal{G}|\boldsymbol{z})p_{\boldsymbol{\theta}}(\boldsymbol{z})d\boldsymbol{z}$. It can be shown that using a data-driven approach, this optimization problem boils down to maximizing the marginal log-likelihood $\log p(\mathcal{G}|\boldsymbol{\theta})$ with respect to the parameters $\boldsymbol{\theta}$.

Given the observed data $\mathscr{G}$, the parameters $\boldsymbol{\theta}$ of the approximate distribution $p(\mathcal{G}|\boldsymbol{\theta})$ can be found through the Maximum Likelihood Estimate (MLE)
\begin{equation}
    \label{eq:MLE}
    \boldsymbol{\theta}_{\text{MLE}}=\argmax_{\boldsymbol{\theta}}\log p(\mathscr{G} \mid \boldsymbol{\theta})=\argmax_{\boldsymbol{\theta}}\sum_{k=1}^{K} \log p\left(\mathcal{G}^{(k)} \mid \boldsymbol{\theta}\right),
\end{equation}
which involves an intractable integration. To address the intractable calculation of the marginal likelihood $p_{\boldsymbol{\theta}}(\mathscr{G})$, we introduce a standard variational inference and maximize the lower-bound on the marginal log-likelihood. This is also known as evidence lower-bound (ELBO) and takes the form
\begin{align}
    \label{eq:ELBO}
    \begin{split}
        \mathscr{L}_{ELBO}(\boldsymbol{\theta}, \boldsymbol{\phi} ; \mathscr{G})=&\sum_{i=1}^{K} \mathscr{L}_{ELBO}\left(\boldsymbol{\theta}, \boldsymbol{\phi} ; \mathcal{G}^{(i)}\right)\\
        =&\sum_{i=1}^{K}\mathbb{E}_{q_{\boldsymbol{\phi}}(\boldsymbol{z}^{(i)} | \mathcal{G}^{(i)})}\left[\log p_{\boldsymbol{\theta}}(\mathcal{G}^{(i)} | \boldsymbol{z}^{(i)})\right]\\
        &-\sum_{i=1}^{K}\mathrm{D_{KL}}\left[q_{\boldsymbol{\phi}}(\boldsymbol{z}^{(i)} | \mathcal{G}^{(i)}) \| p_{\boldsymbol{\theta}}(\boldsymbol{z}^{(i)})\right].
    \end{split}
\end{align}
Here, $q_{\boldsymbol{\phi}}(\boldsymbol{z} | \mathcal{G})$ represents an auxiliary density, parameterized by $\boldsymbol{\phi}$, and $\mathrm{D_{KL}}$ refers to the Kullback-Leibler distance between two distributions. The first term in the lower-bound represents the negative expected reconstruction error, while the second term introduces regularization by forcing the posterior of the latent variables to be close to the prior $p_{\boldsymbol{\theta}}(\boldsymbol{z})$, here taken as the standard Gaussian. For a Gaussian $q_{\boldsymbol{\phi}}$, it can be shown~\cite{kingma2013auto}
\begin{equation}
    -\mathrm{D_{KL}}\left[q_{\boldsymbol{\phi}}(\boldsymbol{z} | \mathscr{G}) \| p_{\boldsymbol{\theta}}(\boldsymbol{z})\right]=\frac{1}{2} \sum_{j=1}^{J}\left(1+\log \left(\left(\sigma_{j}\right)^{2}\right)-\left(\mu_{j}\right)^{2}-\left(\sigma_{j}\right)^{2}\right), 
    \label{eq:KL}
\end{equation}
where $q_{\boldsymbol{\phi}}\left(\boldsymbol{z} | \mathscr{G}\right)=\mathcal{N}\left(\boldsymbol{z}; \boldsymbol{\mu}, \operatorname{diag}\left(\boldsymbol{\sigma}^{2}\right)\right)$ and the prior model is a standard Gaussian distribution $p_{\boldsymbol{\theta}}(\boldsymbol{z})=\mathcal{N}(\boldsymbol{z};\boldsymbol{0},\boldsymbol{I})$. As the prior distribution's parameters are fixed values, for convenience, we drop $\boldsymbol{\theta}$ and represent the prior with $p(\boldsymbol{z})$ in the remaining of the paper. The reconstruction loss term in $\mathscr{L}_{ELBO}$ is discussed in Section~\ref{subsec:decode}.

\begin{figure}
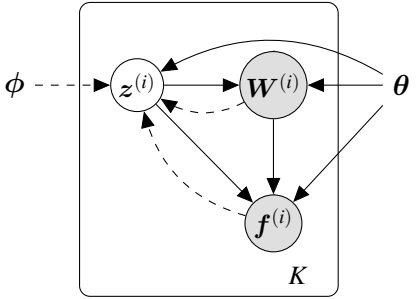

  \centering
    \tikz{ %
    \node (phi) {$\boldsymbol{\phi}$} ; %
    \node[latent, right=of phi] (z) {$\boldsymbol{z}^{(i)}$} ; %
    \node[obs, right=of z] (w) {$\boldsymbol{W}^{(i)}$} ; %
    \node[obs, below=of w] (f) {$\boldsymbol{f}^{(i)}$} ; %
    \node[right=of w] (theta) {$\boldsymbol{\theta}$} ; %
    \plate[inner sep=0.4cm, yshift=0.24cm] {plate} {(f) (z) (w)} {$K$}; %
    \edge[dashed] {phi} {z} ; %
    \edge {z,theta} {w} ; %
    \edge {w,z,theta} {f} ; %
    \graph[use existing nodes] {f--f->[dashed, bend left]z;};
    \graph[use existing nodes] {w--w->[dashed, bend left]z;};
    \graph[use existing nodes] {theta--theta->[bend right]z;};
  }
  \caption{Probabilistic graphical representation of the model. The latent representation (encoding) $\boldsymbol{z}^{(i)}$ of the molecular graph $\mathcal{G}^{(i)}=\{\boldsymbol{W}^{(i)}, \boldsymbol{f}^{(i)}\}$ is computed by the variational distribution $q_{\boldsymbol{\phi}}$, with parameterization $\boldsymbol{\phi}$. The topology of the molecular structure is reconstructed from this latent variable using $p_{\boldsymbol{\theta}}(\boldsymbol{W}^{(i)}|\boldsymbol{z}^{(i)})$ and the atom types are reconstructed using  $p_{\boldsymbol{\theta}}(\boldsymbol{f}^{(i)}|\boldsymbol{z}^{(i)},\boldsymbol{W}^{(i)})$. Solid arrows represent the generative model $p_{\theta}(\boldsymbol{f}|\boldsymbol{z}, \boldsymbol{W})p_{\theta}(\boldsymbol{W}|\boldsymbol{z})p_{\boldsymbol{\theta}}(\boldsymbol{z})$, parameterized by $\boldsymbol{\theta}$, and dashed arrows indicate the encoder. \label{fig:pgm}}
\end{figure}

Independent sampling of atoms and bonds makes it harder for a generative model to satisfy validity constraints in the predicted molecular graphs. In the present problem, we formulate the probabilistic mapping from the latent space to the molecular graph domain as
\begin{equation}
    \label{eq:graph_model}
	p_{\theta}(\mathcal{G}|\boldsymbol{z})=p_{\theta}(\boldsymbol{W}, \boldsymbol{f}|\boldsymbol{z})=p_{\theta}(\boldsymbol{f}|\boldsymbol{z}, \boldsymbol{W})p_{\theta}(\boldsymbol{W}|\boldsymbol{z}).
\end{equation}
In other words, sampling of the atoms is conditioned on the topology of the molecular graph. The model's graphical representation is shown in Fig.~\ref{fig:pgm}.

Limited size of the training dataset $\mathscr{G}$ gives rise to epistemic uncertainties in the parameters of the approximated model $p_{\boldsymbol{\theta}}(\mathcal{G})$, which subsequently propagate to the predictive samples. To study the uncertainty in the trained model, we can take advantage of the predictive distribution
\begin{equation}
    \label{eq:BP}
    p(\bar{\mathcal{G}} \mid \mathscr{G})=\int p(\bar{\mathcal{G}} \mid \boldsymbol{\theta}) p(\boldsymbol{\theta} \mid \mathscr{G}) d \boldsymbol{\theta},
\end{equation}
which requires the posterior distribution $p(\boldsymbol{\theta} \mid \mathscr{G})$ over model parameters. In Eq.~(\ref{eq:BP}), quantifying uncertainties in $\boldsymbol{\theta}$ enables us to capture the uncertainties in the generated molecules and their estimated properties. This is further investigated in Section~\ref{sec:UQ}.

\section{\label{sec:model}Model}

In this work, we take advantage of the Variational Autoencoder (VAE) framework to formulate the problem of approximating the distribution $p(\mathcal{G})$ outlined in Section~\ref{sec:Inference}. In a VAE setting, $q_{\boldsymbol{\phi}}$ provides the embedding of molecular graphs, whereas $p_{\boldsymbol{\theta}}$ allows the generation of molecular graphs using different samples of $\boldsymbol{z}$ in the latent space. Maximizing the lower-bound (Eq.~(\ref{eq:ELBO})) will lead to the desired maximum likelihood estimate for the parameters $\boldsymbol{\phi}$ and $\boldsymbol{\theta}$ of the variational posterior $q_{\boldsymbol{\phi}}$ and decoding distribution $p_{\boldsymbol{\theta}}$, respectively.

With the rise of signal processing methods on graphs~\cite{shuman2013emerging, bronstein2017geometric, ortega2018graph}, various graph-based networks have been introduced for encoding in VAEs. Similarly, there are various decoding networks used for graph generation. In this section, we provide details of the generative model and the approximate variational posterior inference network. The generative model is constructed of a prior on the latent space variable $p(\boldsymbol{z})$ and a mapping from the latent space to the graph domain $p_{\boldsymbol{\theta}}(\mathcal{G}|\boldsymbol{z})$. The variational posterior $q_{\boldsymbol{\phi}}(\boldsymbol{z}|\mathcal{G})$ approximates the posterior on $\boldsymbol{z}$. In this model, we assume a standard normal prior distribution over $\boldsymbol{z}$.

\begin{figure}[h] 
    \setlength{\unitlength}{0.01\textwidth} 
    \begin{picture}(100,83.5)
    
    \put(0,0){\includegraphics[width=1.\textwidth]{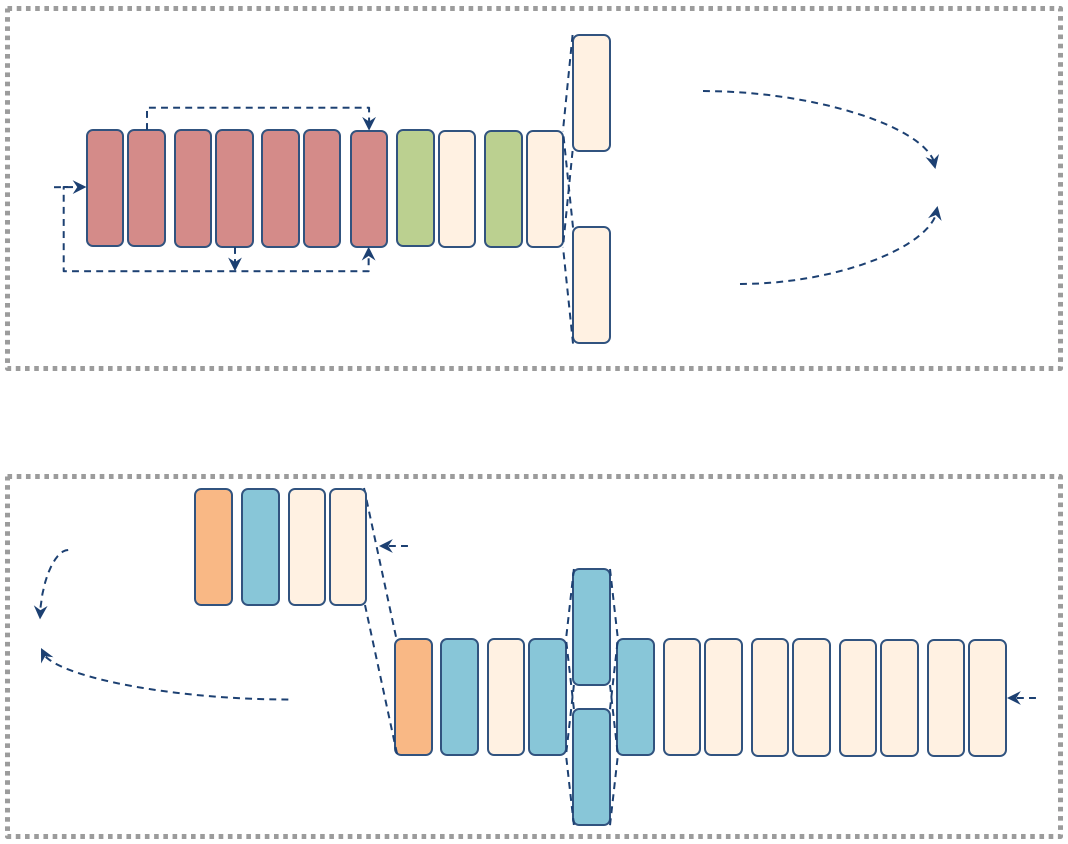}}
    \put(1,79.5){ Encoder}
    \put(1,35.5){ Decoder}
    \put(8,39){ $\mathscr{L}_{ELBO}(\boldsymbol{\theta}, \boldsymbol{\phi} ; \mathscr{G})=\sum_{i=1}^{K}\mathbb{E}_{q_{\boldsymbol{\phi}}(\boldsymbol{z}^{(i)} | \mathcal{G}^{(i)})}\left[\log p_{\boldsymbol{\theta}}(\mathcal{G}^{(i)} | \boldsymbol{z}^{(i)})\right]-\sum_{i=1}^{K}\mathrm{D_{KL}}\left[q_{\boldsymbol{\phi}}(\boldsymbol{z}^{(i)} | \mathcal{G}^{(i)}) \| p_{\boldsymbol{\theta}}(\boldsymbol{z}^{(i)})\right]$}
    
    \put(1.5,61){$\mathcal{G}^{(i)}$}
    \put(9.05,61.2){$\hat{\textbf{\textsf{G}}}$}
    \put(13.2,61.5){$\rho$}
    \put(17.25,61.2){$\hat{\textbf{\textsf{G}}}$}
    \put(21.4,61.5){$\rho$}
    \put(25.45,61.2){$\hat{\textbf{\textsf{G}}}$}
    \put(29.6,61.5){$\rho$}
    \put(33.85,61.2){$\eta$}
    \put(37.5,61.2){\footnotesize $n_{\boldsymbol{\phi}}^{(1)}$}
    \put(41.6,61.2){\footnotesize $l_{\boldsymbol{\phi}}^{(1)}$}
    \put(45.7,61.2){\footnotesize $n^{(2)}_{\boldsymbol{\phi}}$}
    \put(49.7,61.2){\footnotesize $\tilde{a}^{(1)}$}
    \put(54.2,70.1){\footnotesize $l^{(3)}_{\boldsymbol{\phi}}$}
    \put(54.2,51.7){\footnotesize $l^{(2)}_{\boldsymbol{\phi}}$}
    \put(57.7,60.8){$q_{\boldsymbol{\phi}}(\boldsymbol{z}^{(i)}|\mathcal{G}^{(i)})=\mathcal{N}(\boldsymbol{z}^{(i)}|\boldsymbol{\mu}_{\boldsymbol{\phi}}(\mathcal{G}^{(i)}), \boldsymbol{S}_{\boldsymbol{\phi}}(\mathcal{G}^{(i)}))$}
    \put(57.6,70.1){$\boldsymbol{\mu}_{\boldsymbol{\phi}}(\mathcal{G}^{(i)})$}
    \put(57.6,51.9){$\log\boldsymbol{\sigma}^2_{\boldsymbol{\phi}}(\mathcal{G}^{(i)})$}
    
    \put(97.2,13.3){$\boldsymbol{z}$}
    \put(91.2,13){\footnotesize $l_{\boldsymbol{\theta}}^{(1)}$}
    \put(87.2,13){\footnotesize $a^{(1)}$}
    \put(83.0,13){\footnotesize $l_{\boldsymbol{\theta}}^{(2)}$}
    \put(78.9,13){\footnotesize $a^{(2)}$}
    \put(74.8,13){\footnotesize $l_{\boldsymbol{\theta}}^{(3)}$}
    \put(70.7,13){\footnotesize $a^{(3)}$}
    \put(66.6,13){\footnotesize $l_{\boldsymbol{\theta}}^{(4)}$}
    \put(62.5,13){\footnotesize $a^{(4)}$}
    \put(46.,13){\footnotesize $a^{(5)}$}
    \put(37.8,10){\rotatebox{90}{\footnotesize $softmax$}}
    \put(27.3,13.5){\footnotesize $p_{\boldsymbol{\theta}}(\boldsymbol{W}|\boldsymbol{z})$}

    \put(38.2,27.5){$\boldsymbol{z}$}
    \put(31.4,27.5){\footnotesize $l_{\boldsymbol{\theta}}^{(6)}$}
    \put(27.3,27.5){\footnotesize $a_{f}^{(6)}$}
    \put(19.1,24.2){\rotatebox{90}{\footnotesize $softmax$}}
    \put(7,27.5){\footnotesize $p_{\boldsymbol{\theta}}(\boldsymbol{f}|\boldsymbol{W}, \boldsymbol{z})$}

    \put(2,19.5){\footnotesize
    $p_{\boldsymbol{\theta}}(\boldsymbol{W},\boldsymbol{f}| \boldsymbol{z})$}
    \end{picture}
    \caption{Schematic representation of the network architecture. We show the scattering layers in red (Alg.~\ref{alg:graphScat}). Shown in light beige are the fully-connected layers   $l$ with $a$ and $\tilde{a}$ being the activation functions for the decoding and encoding networks, respectively. Batch normalization layers are displayed in green. Permutation, transpose, and multiplication layers are colored blue, and softmax layer is depicted by orange blocks. The model is trained by maximizing the lower-bound $\mathscr{L}_{ELBO}(\boldsymbol{\theta}, \boldsymbol{\phi} ; \mathscr{G})$ for the training dataset $\mathscr{G}=\{\mathcal{G}^{(i)}\}_{i=1}^K$.}
    \label{fig:ModelArch} 
\end{figure}
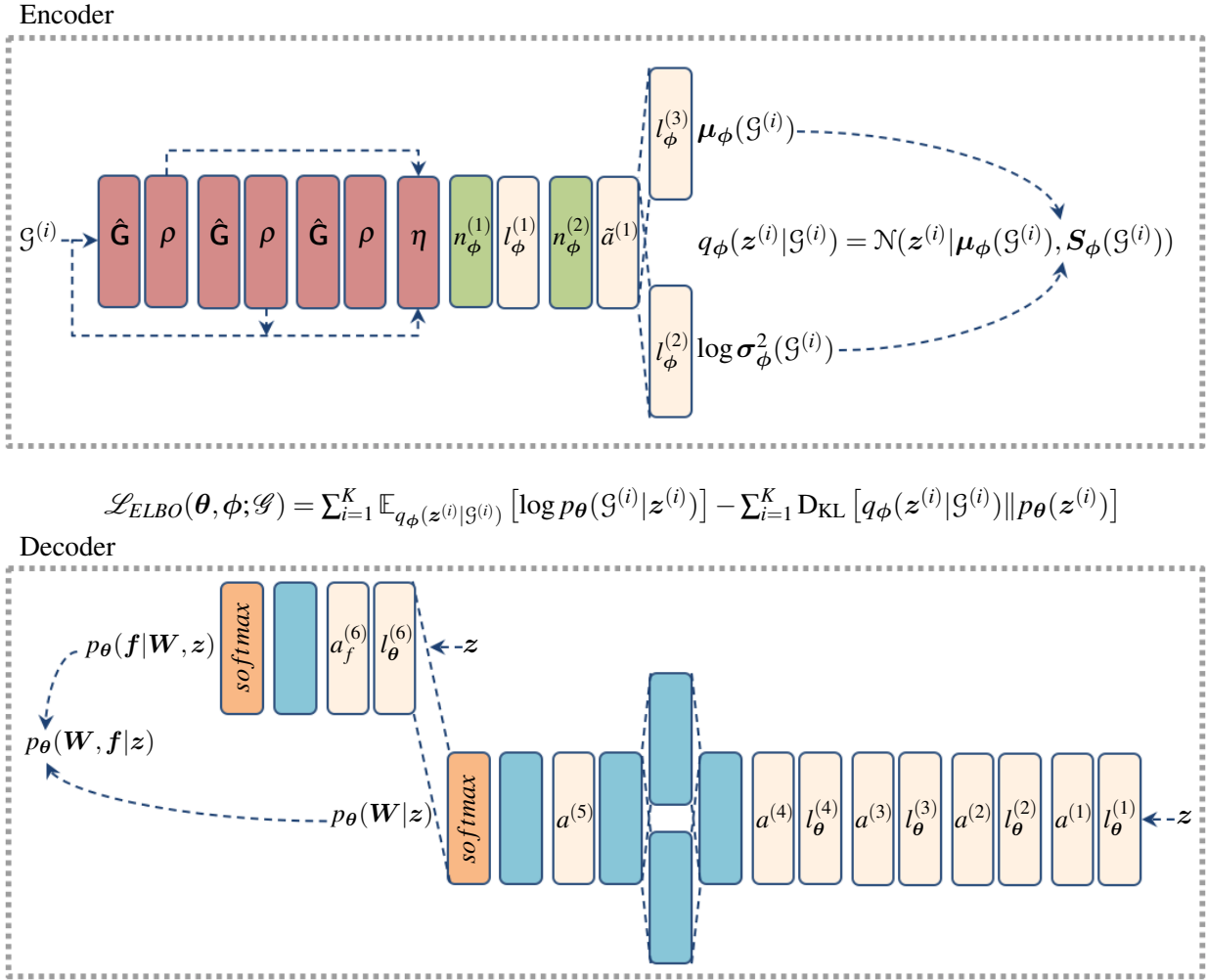

Figure~\ref{fig:ModelArch} depicts the encoding and decoding network architectures that we implemented as well as the training procedure. In this diagram, a linear layer $l$ indicates the operation $l(\boldsymbol{x})=\mathscr{W}\boldsymbol{x}+\boldsymbol{b}$ and a batch normalization layer $n$ represents $n(\boldsymbol{x})=\boldsymbol{\tau}\odot\hat{\boldsymbol{x}}+\boldsymbol{\beta}$, where $\mathscr{W}$, $\boldsymbol{b}$, $\boldsymbol{\tau}$, and $\boldsymbol{\beta}$ are learnable parameters, $\hat{\boldsymbol{x}}$ is normalized input, and $\odot$ is the element-wise product. The indices $\boldsymbol{\phi}$ and $\boldsymbol{\theta}$ distinguish between the learnable parameters of the encoding layers $\boldsymbol{\phi}=\{\mathscr{W}^{(1)}_{\phi},$ $\mathscr{W}^{(2)}_{\phi},$ $\mathscr{W}^{(3)}_{\phi},$ $\boldsymbol{b}^{(1)}_{\phi},$ $\boldsymbol{b}^{(2)}_{\phi},$ $\boldsymbol{b}^{(3)}_{\phi},$ $\boldsymbol{\tau}^{(1)}_{\phi},$ $\boldsymbol{\tau}^{(2)}_{\phi},$ $\boldsymbol{\beta}^{(1)}_{\phi},$ $\boldsymbol{\beta}^{(2)}_{\phi} \}$ and the decoding layers $\boldsymbol{\theta}=\{\mathscr{W}^{(1)}_{\theta},$ $\mathscr{W}^{(2)}_{\theta},$ $\mathscr{W}^{(3)}_{\theta},$ $\mathscr{W}^{(4)}_{\theta},$ $\mathscr{W}^{(6)}_{\theta},$ $\boldsymbol{b}^{(1)}_{\theta},$ $\boldsymbol{b}^{(2)}_{\theta},$ $\boldsymbol{b}^{(3)}_{\theta},$ $\boldsymbol{b}^{(4)}_{\theta},$ $\boldsymbol{b}^{(6)}_{\theta}\}$. Lastly, $a$ and $\tilde{a}$ denote the non-linear activation functions for the decoding and encoding networks, respectively.

\subsection{\label{subsec:encode}Encoding}

We are interested in learning a latent distribution over molecular data using an unsupervised VAE framework. In this framework, we jointly train an encoder and a decoder to map the molecular data onto a latent space and vice versa. The encoder represents a variational approximation to the posterior $p(\boldsymbol{z}|\mathcal{G})$, which is modeled by a Gaussian distribution
\begin{equation}
    \label{eq:var_approx}
	q_{\boldsymbol{\phi}}(\boldsymbol{z}|\mathcal{G})=\mathcal{N}(\boldsymbol{z};\boldsymbol{\mu}_{\boldsymbol{\phi}}(\mathcal{G}), \boldsymbol{S}_{\boldsymbol{\phi}}(\mathcal{G})),
\end{equation}
with $\boldsymbol{S}_{\boldsymbol{\phi}}$ representing a diagonal covariance matrix. The hyperparameters of this distribution are represented by
\begin{equation}
    \label{eq:var_approx_param}
    \boldsymbol{\mu}_{\boldsymbol{\phi}}(\mathcal{G})=h_{\boldsymbol{\phi}}^{\mu}(\mathcal{G})
    \qquad\text{ and }\qquad
    \log\boldsymbol{\sigma}^2_{\boldsymbol{\phi}}(\mathcal{G})=h_{\boldsymbol{\phi}}^{\sigma}(\mathcal{G}),
\end{equation}
where $h_{\boldsymbol{\phi}}$ is a non-linear mapping $\mathcal{G} \mapsto h_{\boldsymbol{\phi}}(\mathcal{G})$, represented by a neural network and parameterized by $\boldsymbol{\phi}$, and superscripts $\boldsymbol{\sigma}$ and $\boldsymbol{\mu}$ denote mappings to mean and log variance, respectively. To make the training easier, we train the model to find $\log\boldsymbol{\sigma}^2_{\boldsymbol{\phi}}$ instead and use its exponential to define the covariance matrix as $\boldsymbol{S}_{\boldsymbol{\phi}}=diag(\boldsymbol{\sigma}^2_{\boldsymbol{\phi}}(\mathcal{G}))$.

In this work, we are interested to learn the generative model using a limited training dataset. 
One way to alleviate this is to use prior information to construct feature maps of the input data. 
Moreover, this can assist in reducing the number of parameters that need to be learned. 
One tool that provides this is the graph scattering network~\cite{chen2014unsupervised, zou2019graph, gama2018diffusion}, a feature extraction algorithm that mimics deep neural networks. We use this tool to construct a hybrid network by following the graph scattering network with fully-connected layers. As opposed to the fully-connected layers, which need training data to learn the parameters, layers of the scattering network are constructed by predefined filters. Hence, they do not need data or training to learn them during optimization. They provide feature maps that are invariant to graph permutation and stable to graph and signal manipulation~\cite{zou2019graph}. 

Scattering layers have previously been used in supervised learning as a way to initialize the network~\cite{oyallon2018scattering}. 
Here, we use the scattering layers in an unsupervised fashion, replacing the first layers of the encoding network. The graph scattering network is constructed of few scattering layers, each of which performs a scattering transform. Each layer comprises three components: an average pooling layer $\eta$ that constructs feature maps, a convolution layer that extracts high-frequency information from the data using multiresolution filters $\textbf{\textsf{G}}$, and a non-linear layer $\rho$ that follows the convolution and prepares its outputs for the next scattering layer.

Graph filters can roughly be divided into two categories: vertex-design filters and spectral-design filters. We compose and employ the latter type of filters, which are defined in the eigenspace of the matrix representation of the graph~\cite{shuman2013emerging, shuman2015spectrum}. The most commonly used matrix for this purpose is the Laplacian matrix, defined as
\begin{equation}
    \label{eq:Laplacian}
    \mathcal{L} := \boldsymbol{\Delta} - \boldsymbol{W},
\end{equation}
where $\boldsymbol{W}$ is the weight matrix, and the diagonal matrix $\boldsymbol{\Delta}$ is the degree matrix, whose elements are $\boldsymbol{\Delta}_{n,n}=\sum_m W_{m,n}$. The Laplacian matrix is analogous to the Laplace operator defined in the Euclidean space. As the Fourier basis in Euclidean space is defined using the Laplace operator's eigen-decomposition, the Laplacian matrix, is used to define the Fourier basis on graphs. This notion forms the foundation of spectral analysis in graphs. 

A Laplacian matrix has an eigen-decomposition of the form
\begin{equation}
    \label{eq:eigL}
    \mathcal{L}=\boldsymbol{\chi}\boldsymbol{\Lambda}\boldsymbol{\chi}^*.
\end{equation}
Here, the matrix $\boldsymbol{\chi}$ is the set of eigenvectors $\{\chi_0, \dots, \chi_{N-1}\}$ of the Laplacian matrix, and the diagonal matrix $\boldsymbol{\Lambda}$ contains the set of eigenvalues $\{\lambda_0, \dots, \lambda_{N-1}\}$ of $\mathcal{L}$, where $N$ represents the number of nodes. The set of eigenvalues $\{\lambda_{\ell}|0\leq\ell\leq N-1\}$ constitutes the spectrum of the graph.

Based on the convolution theorem, the Fourier transform of the convolution of signal $\boldsymbol{f}$ with filter $\textsl{g}$ is defined as the point-wise multiplication of their Fourier transforms. This definition is used to reformulate convolution on graphs as
\begin{equation}
    \label{eq:convMat}
    \boldsymbol{f}*\textsl{g}=\hat{\textsl{g}}(\mathcal{L})\boldsymbol{f},
\end{equation}
where $\hat{\textsl{g}}$ is the filter defined in the spectral domain, and $\hat{\textsl{g}}(\mathcal{L})=\boldsymbol{\chi}\hat{\textsl{g}}(\boldsymbol{\Lambda})\boldsymbol{\chi}^*$ is called a frame, with filtering matrix $\hat{\textsl{g}}(\boldsymbol{\Lambda})$ denoting a diagonal matrix of filter responses in the spectral domain $\{\hat{\textsl{g}}(\lambda_0),\dots,\hat{\textsl{g}}(\lambda_{N-1})\}$.

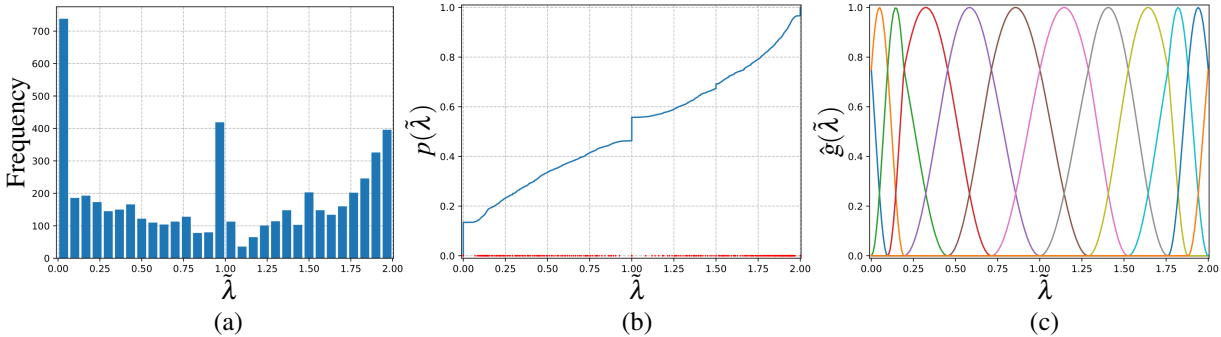
\begin{figure}[H] 
    \setlength{\unitlength}{0.01\textwidth} 
    \begin{picture}(100,25)
    \put(2,1.5){\includegraphics[width=0.31\textwidth]{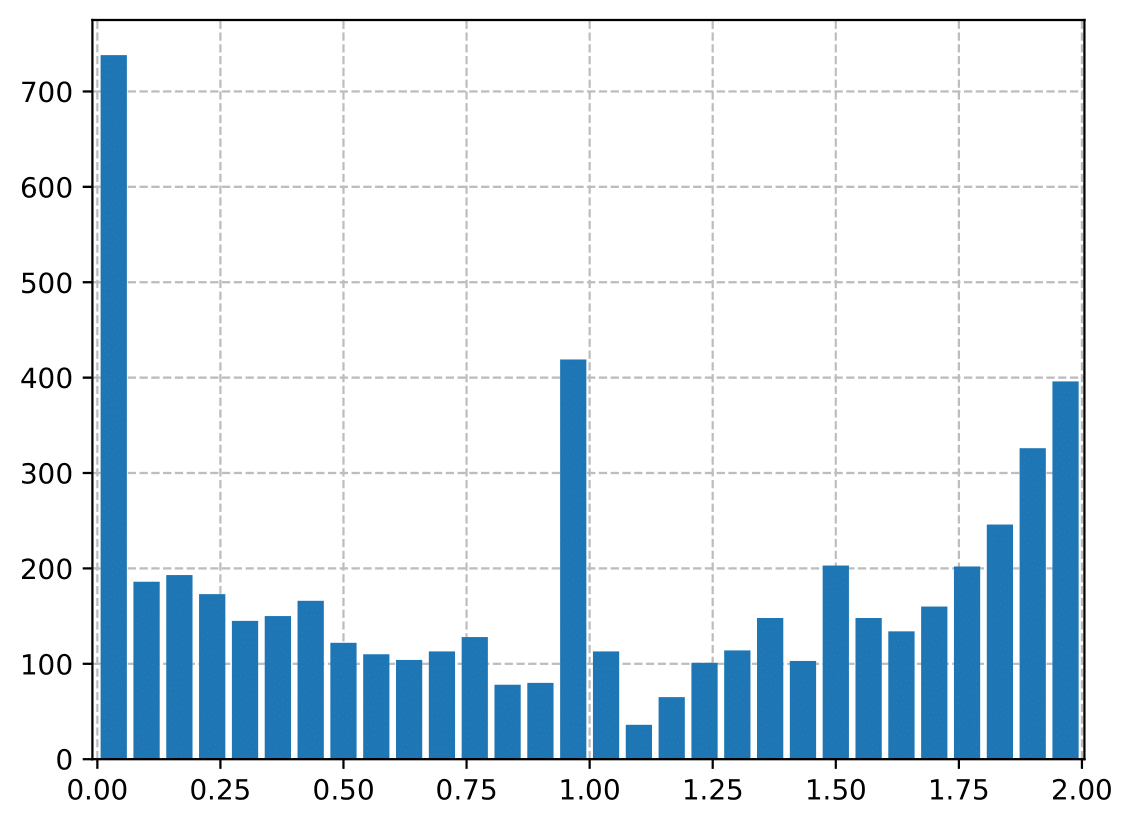}}
    \put(35,1.5){\includegraphics[width=0.31\textwidth]{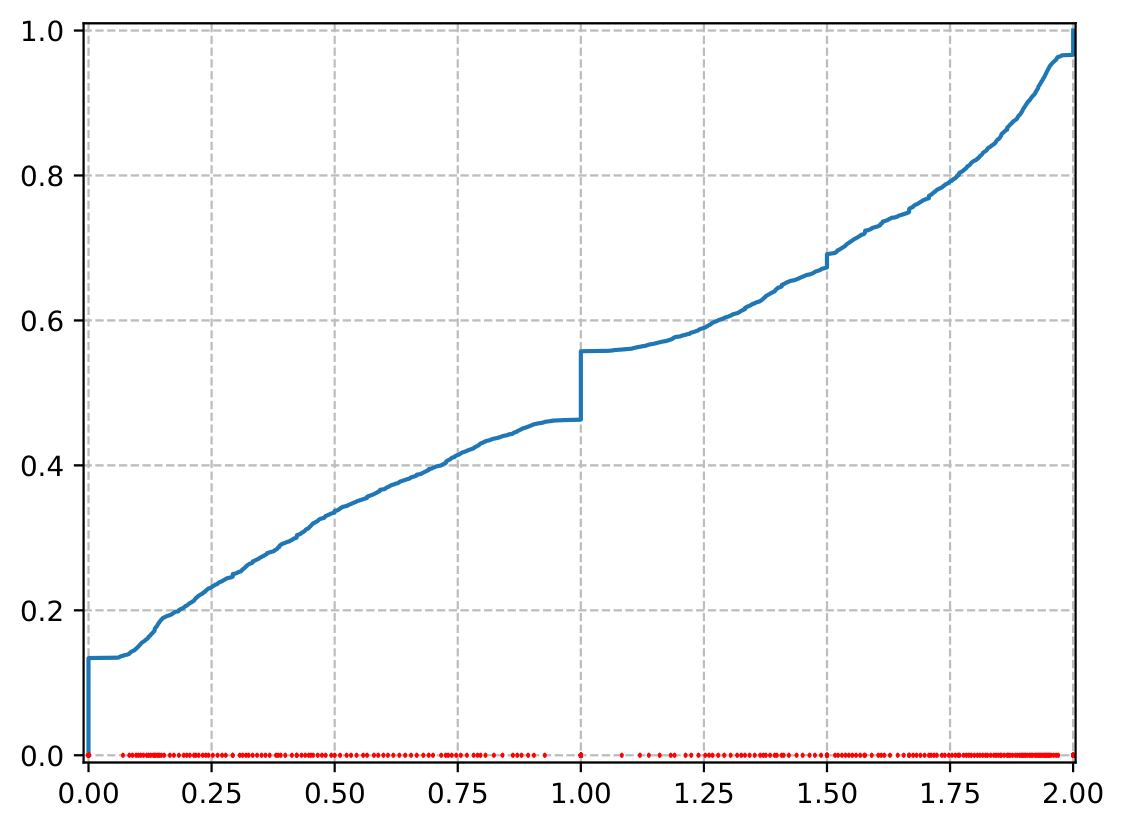}}
    \put(68,1.5){\includegraphics[width=0.31\textwidth]{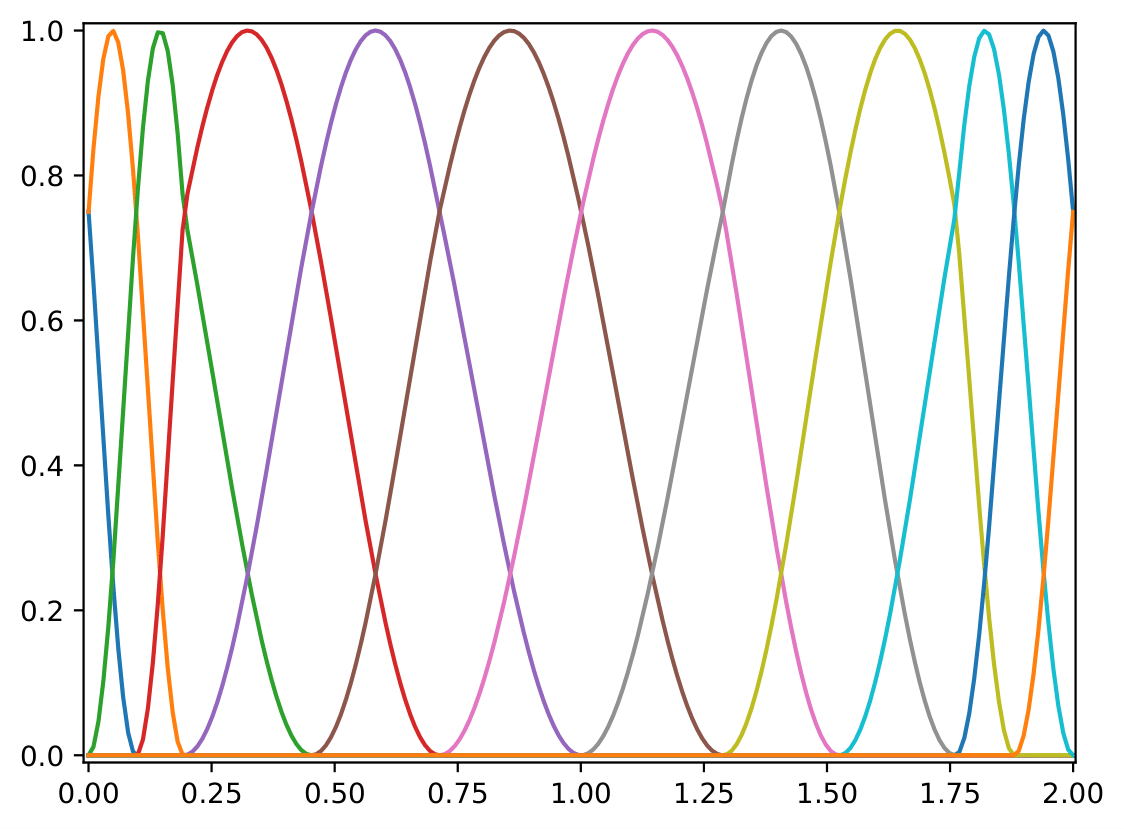}}
    \put(0.5,9){\rotatebox{90}{\footnotesize Frequency}}
    \put(32.7,11.5){\rotatebox{90}{\footnotesize $p(\tilde{\lambda})$}}
    \put(65.7,11.5){\rotatebox{90}{\footnotesize $\hat{\textsl{g}}(\tilde{\lambda})$}}
    \put(17.8,0.0){\footnotesize  $\tilde{\lambda}$}
    \put(50.8,0.0){\footnotesize  $\tilde{\lambda}$}
    \put(83.8,0.0){\footnotesize  $\tilde{\lambda}$}
    \put(17.2,-2.5){\footnotesize  (a)}
    \put(50.2,-2.5){\footnotesize  (b)}
    \put(83.2,-2.5){\footnotesize  (c)}
    \end{picture}
    \caption{Spectral graph filters: (a) histogram of the eigenvalues $\{\tilde{\lambda}_{\ell}^{(i)}\}_{i=1,\dots,K,\ell=0,\dots,N-1}$ of the normalized graph Laplacians for a training set of $K=600$ graphs and (b) cumulative spectral density function of eigenvalues $\tilde{\lambda}$ in (a) (Eq.~\ref{eq:CSDF}). Normalized eigenvalues are shown with red dots on the $x-$axis. For ease of visualization, we display one point for each interval of 18 eigenvalues. We construct the warping function $\omega$ as a smooth approximation of the cumulative spectral density function. (c) Adaptive kernels $\{\hat{\textsl{g}}_j\}_{j=1}^{\mathcal{J}}$, which are tailored to the set of normalized Laplacian spectra shown in (a), with $\mathcal{J}=12$. We set the parameters $\mathcal{K}=1$, $d_0 = d_1 = 0.5$, and $R = 3$ to define the main window (Eq.~\ref{eq:HC_window}). \label{fig:filters}}
\end{figure}

In the scattering transform, we perform convolutions with a bank of filters. A filter-bank is a set of $\mathcal{J}$ filters $\hat{\textbf{\textsf{G}}}=\{\hat{\textsl{g}}_j\}_{j=1}^\mathcal{J}$. A convolution with $\hat{\textbf{\textsf{G}}}$ takes the form
\begin{equation}
    \label{eq:filtering}
    \hat{\textbf{\textsf{G}}}(\mathcal{L}) \boldsymbol{f}:= \{\hat{\textsl{g}}_j(\mathcal{L})\boldsymbol{f}\}_{j=1}^\mathcal{J},
\end{equation}
where $\hat{\textbf{\textsf{G}}}(\mathcal{L}) := \{\hat{\textsl{g}}_j(\mathcal{L})\}_{j=1}^\mathcal{J}$ is a frame for the filter-bank $\hat{\textbf{\textsf{G}}}$. This operation extracts $\mathcal{J}$ features from $\boldsymbol{f}$. Spectral-design filters $\hat{\textsl{g}}$ are defined in the interval $[0, \lambda_{\max}]$, where $\lambda_{\max}$ is the maximum possible eigenvalue for the graphs in the study. When eigenvalues of the molecules' matrix representation are unevenly spaced in this interval, some filters are highly correlated with the neighboring filters, while some are left unused, i.e. the molecular graph does not have any eigenvalues in the part of the spectral domain that the filter covers. Fig.~\ref{fig:filters}(a) shows an example of distribution of unevenly spaced eigenvalues of a matrix representation for a dataset of molecular graphs. We observe higher densities towards the limits of the interval and lower densities in the central region except for a spike in the middle.

In this work, we use adaptive kernels~\cite{shuman2015spectrum} for feature extraction. These kernels are adapted to the density of the set of eigenvalues of the training molecular graph Laplacians in the interval $[0, \lambda_{\max}]$. This way, we have more filters in the dense regions, reducing the correlations between the neighboring filters and better discriminating the data. For this purpose, we use the tight frame spectral filter design~\cite{shuman2015spectrum, leonardi2013tight}, which defines a filter-bank as translations of a main window. A warping function is then used to adapt these kernels to the areas with a higher density of eigenvalues. The resulting filter-bank is precomputed and adapted to the training data and does not change during the training process. 

We start by defining the main window using a half-cosine kernel

\begin{equation}
    \label{eq:HC_window}
    \hat{{\textsl{g}}}^{'}(\lambda):=\sum_{k=0}^{\mathcal{K}} d_{k}\left[\cos \left(2 \pi k\left(\frac{\lambda}{\textsl{a}}-\frac{1}{2}\right)\right) \cdot \mathbb{1}\{0\leq \lambda<\textsl{a}\}\right], 
\end{equation}
where
\begin{equation}
    \label{eq:dilfac}
    \textsl{a}=\frac{R\omega(\gamma)}{\mathcal{J}-R+1},
\end{equation}
is the dilation factor, $R$ is the kernel overlap, which tunes the kernel's width in the spectral domain, $\mathcal{J}$ is the number of filters in the filter-bank, $\gamma$ is the upper limit $\lambda_{\max}$ of the set of eigenvalues, which adapts these filters to the length of the spectrum, $\omega$ is the warping function, and $\mathcal{K}$ is the number of cosine terms, satisfying $\mathcal{K}<R/2$. 

We customize the filters for a given dataset $\mathscr{G}=\{\mathcal{G}^{(i)}\}^{K}_{i=1}$ by finding the set of eigenvalues for all the training data $\{\tilde{\lambda}_{\ell}^{(i)}\}_{i=1,\dots,K,\ell=0,\dots,N-1}$ and then computing the cumulative spectral density function for these eigenvalues
\begin{equation}
    \label{eq:CSDF}
    P_{\tilde{\lambda}}(s):=\frac{1}{NK} \sum_{i=1}^{K}\sum_{\ell=0}^{N-1}\mathbb{1}\{\tilde{\lambda}^{(i)}_{\ell} \leq s\}.
\end{equation}
We translate the main window (Eqs.~(\ref{eq:HC_window}) and~(\ref{eq:dilfac})) to uniform distances in the spectral domain and use a smooth approximation of the  cumulative spectral density function (Eq.~\ref{eq:CSDF}) as the warping function $\omega$ to define the filters 
\begin{equation}
    \label{eq:warp}
    \hat{\textsl{g}}_{j}(\lambda):=\hat{\textsl{g}}^{'}\left(\omega(\lambda)- \frac{\textsl{a}(j-R+1)}{R}\right), \quad\text{for } j=1,\dots,\mathcal{J}.
\end{equation}
The set of filters in Eq.~(\ref{eq:warp}) defines the filter-bank $\hat{\textbf{\textsf{G}}}$. Note that we design the filters with $\gamma=\lambda_{\max}$ from the training dataset. In the test set, however, the spectrum may contain eigenvalues larger than $\lambda_{\max}$. To avoid this issue, we instead use the normalized Laplacian~\cite{chung1997spectral}
\begin{equation}
    \label{eq:norm_Laplacian}
	\tilde{\mathcal{L}} = \boldsymbol{\Delta}^{-1/2}\mathcal{L}\boldsymbol{\Delta}^{-1/2},
\end{equation}
to define the spectral domain. Eigenvalues of the normalized Laplacian fall in the interval $[0,2]$. Hence, $\tilde{\lambda}_{\max}$ is limited by $2$ for both the training and test datasets.

In Fig.~\ref{fig:filters}, the set of graph spectra for a training dataset of $K=600$ molecules is computed (Fig.~\ref{fig:filters}(a)) and its cumulative spectral density function is determined (Fig.~\ref{fig:filters}(b)). In Fig.~\ref{fig:filters}(c), a smooth approximation of this function is used to adapt the kernels to regions of higher population in the spectral domain. This is the only data dependent part of the scattering layers and is computed prior to training.

Given a graph with signal $\boldsymbol{f}$ defined on its vertices, the scattering transform generates feature maps by applying an average pooling operator~\cite{gama2019stability} $\eta$ to the signal, where
\begin{equation}
    \label{eq:avg_pooling}
    \eta = \frac{\textbf{1}^T}{N}.
\end{equation}
This operator averages the signal over all $N$ nodes of a graph. For a molecular graph, this averages over $N$ one-hot vectors of atom types. These feature maps, however, may not be sufficient to discriminate between different graphs. They capture low-frequency information, but the higher frequency information is lost. Convolutions of the signal with filters in the filter-bank $\hat{\textbf{\textsf{G}}}$ restore this information. The restored information has the same topology as the original signal. We construct higher-order feature maps from the restored information by applying a non-linear layer and propagating information to the next scattering layer. The same procedure recurs in the next layer.

At layer $m$, the scattering feature maps are generated by 
\begin{equation}
    \mathcal{S}_{m}\boldsymbol{f} = \eta(\mathcal{U}_{m-1}\boldsymbol{f}),
    \label{eq:Sm}
\end{equation}
where $\eta$ is the average pooling operator (Eq.~(\ref{eq:avg_pooling})),  $\mathcal{U}_{0}\boldsymbol{f}=\boldsymbol{f}$, and for $m>0$, $\mathcal{U}_{m}\boldsymbol{f}$ represents the information that is extracted from layer $m-1$ by convolution with filter-bank $\hat{\textbf{\textsf{G}}}$ (Eq.~(\ref{eq:filtering})) and propagated to the layer $m$ after applying the non-linear layer $\rho$
\begin{equation}
    \mathcal{U}_m\boldsymbol{f} = \rho\left( \hat{\textbf{\textsf{G}}}(\tilde{\mathcal{L}})\mathcal{U}_{m-1}\boldsymbol{f}\right),
    \label{eq:Um}
\end{equation}
where $\rho$ represents the modulus (absolute value) non-linear layer, which is a point-wise operator. We observed that a four layer scattering network has an optimal discriminative power.

In Fig.~\ref{fig:ModelArch}, scattering layers are shown with red blocks. The input graph $\mathcal{G}$ contains signals $\boldsymbol{f}$ and weight matrix $\boldsymbol{W}$. The first set of features are generated by applying operator $\eta$ on the input signal $\boldsymbol{f}$ (Eq.~(\ref{eq:Sm})). Then, the signal is propagated to the next layer after convolution with spectral filters $\hat{\textbf{\textsf{G}}}$ and applying non-linear point-wise operator $\rho$ (Eq.~(\ref{eq:Um})). The second set of features are generated by applying operator $\eta$ on the inputs of the second layer. Since the scattering network's features are generated at each layer, we use arrows to connect inputs at the first, second, and third layers to layer $\eta$ to show the flow of the information in the network.

In summary, after going through a cascade of convolutions with spectral-design filters $\hat{\textbf{\textsf{G}}}$ and modulus non-linear activation function $\rho$, information from different scattering layers passes through the average pooling layer $\eta$, which constructs feature maps from the input graphs. Using a $4$ layer scattering network, the scattering feature maps of a molecular graph $\mathcal{G}$ are determined by
\begin{equation}
    \label{eq:scat_coef}
    \mathcal{A}(\mathcal{G})=[(\eta\circ\rho\circ\hat{\textbf{\textsf{G}}}\circ\rho\circ\hat{\textbf{\textsf{G}}}\circ\rho\circ\hat{\textbf{\textsf{G}}})(\mathcal{G}), (\eta\circ\rho\circ\hat{\textbf{\textsf{G}}}\circ\rho\circ\hat{\textbf{\textsf{G}}})(\mathcal{G}), (\eta\circ\rho\circ\hat{\textbf{\textsf{G}}})(\mathcal{G}), (\eta)(\mathcal{G})].
\end{equation}
This is summarized in Algorithm~\ref{alg:graphScat}.

\begin{algorithm}[H]
	\caption{Graph scattering transform  with adaptive kernels. \label{alg:graphScat}}
	\begin{algorithmic}
    \STATE Input molecular graph training dataset $\{\mathcal{G}^{(i)}\}^{K}_{i=1}=\{\boldsymbol{W}^{(i)}, \boldsymbol{f}^{(i)}\}^{K}_{i=1}$ and window $\hat{\textsl{g}}^{'}$ (Eq.~(\ref{eq:HC_window})).
    \FOR{$i=1,\dots,K$}
        \STATE Compute the normalized Laplacian matrix $\tilde{\mathcal{L}}^{(i)}$ from $\boldsymbol{W}^{(i)}$ (Eqs.~(\ref{eq:Laplacian}) and~(\ref{eq:norm_Laplacian})).
        \STATE Compute the eigenvalues $\tilde{\lambda}_{\ell}^{(i)}$ and eigenvectors $\tilde{\chi}_{\ell}^{(i)}$ of the normalized Laplacian $\tilde{\mathcal{L}}^{(i)}$.
    \ENDFOR
    \STATE Compute the warping function $\omega$ from the normalized Laplacian spectra $\{\tilde{\lambda}_{\ell}^{(i)}\}_{\begin{subarray}{l} i=1,\dots,K\\ \ell=0,\dots,N-1\end{subarray}}$ (Eq.~(\ref{eq:CSDF})).
    \STATE Form adaptive kernels $\hat{\textbf{\textsf{G}}}=\{\hat{\textsl{g}}_j\}_{j=1}^{\mathcal{J}}$ using window $\hat{\textsl{g}}^{'}$ and the warping function $\omega$ (Eq.~(\ref{eq:warp})).
    \FOR{$i=1,\dots,K$}
        \STATE Compute frame $\hat{\textbf{\textsf{G}}}(\tilde{\mathcal{L}}^{(i)})=\{\hat{\textsl{g}}_j(\tilde{\mathcal{L}}^{(i)})\}_{j=1}^{\mathcal{J}}$.
        \STATE Compute first layer scattering feature maps $\mathcal{S}_1\boldsymbol{f}^{(i)}$ (Eq.~(\ref{eq:Sm})).
        \FOR{$m=2,\dots,M$}
            \STATE Compute $\mathcal{U}_m\boldsymbol{f}^{(i)}$ (Eq.~(\ref{eq:Um})).
            \STATE Compute $m$-th layer scattering feature maps $\mathcal{S}_m\boldsymbol{f}^{(i)}$ (Eq.~(\ref{eq:Sm})).
        \ENDFOR
        \STATE Determine scattering feature maps $\mathcal{A}(\mathcal{G}^{(i)})=[\mathcal{S}_1\boldsymbol{f}^{(i)},\dots,\mathcal{S}_M\boldsymbol{f}^{(i)}]$ (Eq.~(\ref{eq:scat_coef})).
    \ENDFOR
	\end{algorithmic}
\end{algorithm}

After extracting feature maps from the training data in the scattering layers, we pass these features to conventional FCN layers. The scattering layers reduce the need for a deeper FCN. As a result, we merely incorporate a single linear layer combined with two batch normalization layers and a non-linear layer to learn features that were not captured in the scattering layers. Subsequently two linear layers $l_{\boldsymbol{\phi}}^{(2)}$ and $l_{\boldsymbol{\phi}}^{(3)}$ are used to compute the parameters $\boldsymbol{\mu}_{\boldsymbol{\phi}}$ and $\boldsymbol{\sigma}_{\boldsymbol{\phi}}$ of the variational posterior $q_{\boldsymbol{\phi}}$ of the variable $\boldsymbol{z}$. To sum it up,
\begin{equation}
    h_{\boldsymbol{\phi}}^{\mu}(\mathcal{G})=(l^{(3)}_{\boldsymbol{\phi}}\circ h^{'}_{\boldsymbol{\phi}})(\mathcal{G})\qquad \text{ and } \qquad h_{\boldsymbol{\phi}}^{\sigma}(\mathcal{G})=(l^{(2)}_{\boldsymbol{\phi}}\circ h^{'}_{\boldsymbol{\phi}})(\mathcal{G}), 
    \label{eq:enc_map}
\end{equation}
with the shared network structured as
\begin{equation}
    h^{'}_{\boldsymbol{\phi}}(\mathcal{G})=(\tilde{a}^{(1)}\circ n^{(2)}_{\boldsymbol{\phi}}\circ l_{\boldsymbol{\phi}}^{(1)}\circ n_{\boldsymbol{\phi}}^{(1)}\circ\mathcal{A})(\mathcal{G}).
    \label{eq:enc_map_shared}
\end{equation}
Here, $\mathcal{A}$ summarizes the operations within $4$ scattering layers (Eq.~\ref{eq:scat_coef}), which initialize the network by extracting scattering feature maps from the molecular graph $\mathcal{G}$.

In Eq.~(\ref{eq:enc_map}), the linear layers $l_{\boldsymbol{\phi}}^{(2)}$ and $l_{\boldsymbol{\phi}}^{(3)}$ take the high-dimensional data from the previous hidden layer and project it to a lower-dimensional latent space. In this sense, they learn a probabilistic projection of the extracted feature maps to the latent space. Together, Eqs.~(\ref{eq:var_approx}),~(\ref{eq:var_approx_param}), and~(\ref{eq:scat_coef})-(\ref{eq:enc_map_shared}) define the approximate posterior distribution.

While evaluating $\mathscr{L}_{ELBO}$ (Eq.~(\ref{eq:ELBO})), the reconstruction loss term is intractable and is evaluated through Monte Carlo approximation by sampling from the encoder $q_{\boldsymbol{\phi}}(\boldsymbol{z}|\mathcal{G})$
\begin{align}
    \label{eq:ELBO_MC}
    \begin{split}
        \mathscr{L}_{ELBO}(\boldsymbol{\theta}, \boldsymbol{\phi} ; \mathscr{G})=&\frac1L\sum_{i=1}^{K}\sum_{l=1}^{L}\log p_{\boldsymbol{\theta}}(\mathcal{G}^{(i)} | \boldsymbol{z}^{(i,l)})\\
        &-\sum_{i=1}^{K}\mathrm{D_{KL}}\left[q_{\boldsymbol{\phi}}(\boldsymbol{z}^{(i)} | \mathcal{G}^{(i)}) \| p(\boldsymbol{z}^{(i)})\right],
    \end{split}
\end{align}
where $L$ denotes the number of the Monte Carlo samples. Maximizing Eq.~(\ref{eq:ELBO_MC}) updates the decoder parameters $\boldsymbol{\theta}$ and the parameters $\boldsymbol{\phi}$ of the fully-connected layers of the encoder. Updating the model parameters needs evaluating the gradient of the terms in $\mathscr{L}_{ELBO}$. Evaluating the gradient estimator for the reconstruction loss term by directly sampling from $q_{\boldsymbol{\phi}}(\boldsymbol{z}|\mathcal{G})$ (Eqs.~(\ref{eq:var_approx})) shows very high variance. We sample from this distribution using the differentiable transform $\boldsymbol{z}^{(i,l)}=\boldsymbol{\mu}_{\boldsymbol{\phi}}(\mathcal{G}^{(i)})+ \boldsymbol{S}_{\boldsymbol{\phi}}(\mathcal{G}^{(i)})\odot\boldsymbol{\epsilon}^{(l)}$, where $\boldsymbol{\epsilon}^{(l)}$ is sampled from $ p(\boldsymbol{\epsilon})=\mathcal{N}(\boldsymbol{\epsilon};\boldsymbol{0},\boldsymbol{I})$~\cite{kingma2013auto}, and $\odot$ denotes element-wise product.

In practice, the model is not trained with the whole training data at once. Instead, the model is trained in few iterations with minibatches of size $K'$. In each iteration, $\mathscr{L}_{ELBO}$ is rescaled using
\begin{equation}
    \mathscr{L}_{ELBO}(\boldsymbol{\theta}, \boldsymbol{\phi} ; \mathscr{G})=\frac{K}{K'}\sum_{i=1}^{K'} \mathscr{L}_{ELBO}\left(\boldsymbol{\theta}, \boldsymbol{\phi} ; \mathcal{G}^{(i)}\right)
    \label{eq:ELBO_minibatch}
\end{equation}
Kingma and Welling~\cite{kingma2013auto} suggest that for minibatches of size larger $100$, the $\mathscr{L}_{ELBO}\left(\boldsymbol{\theta}, \boldsymbol{\phi} ; \mathcal{G}^{(i)}\right)$ term in Eq.~(\ref{eq:ELBO_minibatch}) can be evaluated using a single MC sample $L=1$. In this work, we are dealing with small training datasets and $K'$ may not be larger than $100$. Therefore, we select $L$ such that $K'\times L$ is larger than $100$.

\subsection{\label{subsec:decode}Decoding}

The decoding network represents a probabilistic mapping from latent space to molecular structures. In Eq.~(\ref{eq:graph_model}), the discrete probability distribution $p_{\boldsymbol{\theta}}(\boldsymbol{W}|\boldsymbol{z})$ is the joint distribution of the covalent bond orders for all the possible edges $\varepsilon_{i,j}$ between all pairs of atoms
\begin{equation}
	p_{\boldsymbol{\theta}}(\boldsymbol{W} | \boldsymbol{z})=\prod_{i=1}^{N} \prod\limits_{\substack{j=1 \\ j>i}}^{N} p_{\boldsymbol{\theta}}\left(W_{i,j} | \boldsymbol{z}\right),
\end{equation}
and $p_{\boldsymbol{\theta}}(\boldsymbol{f} | \boldsymbol{z}, \boldsymbol{W})$ is the joint categorical distribution that contains the probability values of different atom types for all possible atoms $v_i$ in the molecule
\begin{equation}
	p_{\theta}(\boldsymbol{f} | \boldsymbol{z}, \boldsymbol{W})=\prod_{i=1}^{N}p_{\theta}\left(f_{i} | \boldsymbol{z}, \boldsymbol{W}\right).
\end{equation}
These probabilities are computed using 
\begin{equation}
    p_{\boldsymbol{\theta}}(\boldsymbol{W}|\boldsymbol{z})=\text{softmax}(h_{\boldsymbol{\theta}}^{\boldsymbol{W}}(\boldsymbol{z}))\qquad\text{and}\qquad p_{\boldsymbol{\theta}}(\boldsymbol{f}|\boldsymbol{z}, \boldsymbol{W})=\text{softmax}(h_{\boldsymbol{\theta}}^{\boldsymbol{f}}(\boldsymbol{z}, \boldsymbol{W})),
    \label{eq:decoders}
\end{equation}
where $h_{\theta}$ denotes the non-linear mapping $\boldsymbol{z} \mapsto h_{\boldsymbol{\theta}}(\boldsymbol{z})$, which is parameterized by a deep neural network with learnable parameters $\boldsymbol{\theta}$, and superscripts $\boldsymbol{W}$ and $\boldsymbol{f}$ indicate the weight and signal generators, respectively. Thereby, the decoder is constructed of two non-linear maps that are jointly trained on a graph dataset. Given a sample from the latent space, Eq.~(\ref{eq:decoders}) presents a probabilistic graph in terms of the probabilities for all nodes and edges.  

In Eq.~(\ref{eq:decoders}), a non-linear map $h_{\theta}$ yields unnormalized scores, which subsequently are transformed into probability values for the categorical distribution by the softmax layer. Given the input latent space variable $\boldsymbol{z}$, $h_{\boldsymbol{\theta}}^{\boldsymbol{W}}(\boldsymbol{z})$ is trained to generate the weight matrix $\boldsymbol{W}$. To this end, we incorporate the following structure: First, we define a mapping 
\begin{equation}
    h''_{\boldsymbol{\theta}}(\boldsymbol{z})=\left(a^{(4)}\circ l_{\boldsymbol{\theta}}^{(4)}\circ a^{(3)}\circ l_{\boldsymbol{\theta}}^{(3)}\circ a^{(2)}\circ l_{\boldsymbol{\theta}}^{(2)}\circ a^{(1)}\circ l_{\boldsymbol{\theta}}^{(1)}\right)(\boldsymbol{z}),
    \label{eq:h_theta}
\end{equation}
which takes samples from the latent space. Then, unnormalized score values for each bond order are constructed by
\begin{equation}
	h_{\boldsymbol{\theta}}^{\boldsymbol{W}}(\boldsymbol{z})=\left(a^{(5)}\circ h^{''}_{\boldsymbol{\theta}} h^{''T}_{\boldsymbol{\theta}}\right)(\boldsymbol{z}).
	\label{eq:h_theta_W}
\end{equation}
In Eq.~(\ref{eq:h_theta_W}), the output from Eq.~(\ref{eq:h_theta}) is multiplied by its transpose to ensure symmetry of the final output probability tensor since molecules are represented by undirected graphs. Lastly, a softmax layer turns these scores into probability values $p_{\boldsymbol{\theta}}(\boldsymbol{W}|\boldsymbol{z})$. 

Based on Eq.~(\ref{eq:graph_model}), in addition to a probabilistic weight matrix, we need to obtain the joint distribution over all atom types for all the nodes in the graph. This is accomplished through Eq.~(\ref{eq:decoders}), where the non-linear map $h_{\boldsymbol{\theta}}^{\boldsymbol{f}}$ is specified by an FCN
\begin{equation}
	h_{\boldsymbol{\theta}}^{\boldsymbol{f}}(\boldsymbol{z})=\left(a^{(6)} \circ l_{\boldsymbol{\theta}}^{(6)}\right)(\boldsymbol{z}, \boldsymbol{W}).
\end{equation}
Fig.~\ref{fig:ModelArch} summarizes the architecture of the decoding network.

In essence, the decoding network tackles a classification problem for every node and edge in the graph. When dealing with molecular graphs, for the node signal $\boldsymbol{f}$, the possible classes include the heavy atom types in the dataset along with the case of an empty node. The empty node, or null vertex, means that no atoms reside on this node and the molecule has fewer atoms than the predefined maximum possible number of atoms. Similarly, the classes for each edge include the possible types of the covalent bond between the respective atoms plus null, which means that there are no covalent bonds between the corresponding pair of atoms.

Using the probabilistic graph model (Eq.~(\ref{eq:graph_model})), we can write the negative expected reconstruction loss term in Eq.~(\ref{eq:ELBO}) as
\begin{align}
    \begin{split}
     \mathbb{E}_{q_{\boldsymbol{\phi}}}\left[\log p_{\boldsymbol{\theta}}\left(\mathcal{G} | \boldsymbol{z}\right)\right] =&  \mathbb{E}_{q_{\boldsymbol{\phi}}}\left[\log p_{\boldsymbol{\theta}}\left(\boldsymbol{W}, \boldsymbol{f}  | \boldsymbol{z}\right)\right]\\
    =&  \mathbb{E}_{q_{\boldsymbol{\phi}}}\left[\log \left(p_{\boldsymbol{\theta}}\left(\boldsymbol{W} | \boldsymbol{z}\right)p_{\boldsymbol{\theta}}\left(\boldsymbol{f}  | \boldsymbol{z}, \boldsymbol{W}\right)\right)\right]\\
    =&  \mathbb{E}_{q_{\boldsymbol{\phi}}}\left[\log p_{\boldsymbol{\theta}}\left(\boldsymbol{W} | \boldsymbol{z}\right)+\log p_{\boldsymbol{\theta}}\left(\boldsymbol{f}  | \boldsymbol{z}, \boldsymbol{W}\right)\right]\\
    =&  \mathbb{E}_{q_{\boldsymbol{\phi}}}\left[\sum_{i=1}^{N}\sum_{\substack{j=1 \\ j>i}}^{N}\log p_{\boldsymbol{\theta}}\left(W_{i,j} | \boldsymbol{z}\right)+\sum_{i=1}^{N}\log p_{\boldsymbol{\theta}}\left(f_{i}  | \boldsymbol{z}, \boldsymbol{W}\right)\right]\\
    =&  \sum_{i=1}^{N}\sum_{\substack{j=1 \\ j>i}}^{N}\mathbb{E}_{q_{\boldsymbol{\phi}}}\left[\log p_{\boldsymbol{\theta}}\left(W_{i,j} | \boldsymbol{z}\right)\right]+\sum_{i=1}^{N}\mathbb{E}_{q_{\boldsymbol{\phi}}}\left[\log p_{\boldsymbol{\theta}}\left(f_{i}  | \boldsymbol{z}, \boldsymbol{W}\right)\right].
    \end{split}
\end{align}

Note that computing the loss for each node and each edge is a multi-class classification problem where we want to find the correct class for each node and edge in the graph. We take advantage of  the  generalized form of cross-entropy for multi-class problems to compute the reconstruction error. This computes the relative entropy between the predicted probability and the true probability over the node and edge classes. Given the probability $p_{\boldsymbol{\theta}}\left(W_{i, j} | \boldsymbol{z}\right)$ for edge $\varepsilon_{i, j}$, we can write the reconstruction error term for edge $\varepsilon_{i, j}$ as
\begin{equation}
    \mathcal{H}(\boldsymbol{t}^{\varepsilon_{i,j}},p_{\boldsymbol{\theta}}\left(W_{i,j} | \boldsymbol{z}\right))=-\sum_{c=1}^{C_{W}} t^{\varepsilon_{i,j}}_c\log p_{\boldsymbol{\theta}}\left(W_{i,j}=c | \boldsymbol{z}\right), 
\end{equation}
where $\mathcal{H}$ denotes the cross-entropy between the target distribution $\boldsymbol{t}^{\varepsilon_{i,j}}$ and the predicted distribution $p_{\boldsymbol{\theta}}\left(W_{i, j} | \boldsymbol{z}\right)$ for node $\varepsilon_{i,j}$, index $c$ denotes class label, and $C_W$ is the total number of the edge classes. Furthermore, given the probability $p_{\boldsymbol{\theta}}\left(f_{i} | \boldsymbol{z}, \boldsymbol{W}\right)$ for node $v_i$, we compute the reconstruction error term for node $v_i$ as
\begin{equation}
    \mathcal{H}(\boldsymbol{t}^{v_{i}},p_{\theta}\left(f_{i} | \boldsymbol{z}, \boldsymbol{W}\right))=-\sum_{c=1}^{C_{f}} t^{v_{i}}_c\log p_{\theta}\left(f_{i}=c | \boldsymbol{z}, \boldsymbol{W}\right),
\end{equation}
where $\boldsymbol{t}^{v_{i}}$ is the target distribution for node $v_{i}$, and $C_f$ is the total number of the node classes.

Hence, we can write the reconstruction loss of the decoding network by summing the loss over all possible nodes and edges in the graph. The reconstruction loss for data point $t$ has the form

\begin{equation}
	L^{(t)}=\sum_{i=1}^{N}\sum_{c=1}^{C_{f}} -t^{v_{i}}_c\log p_{\theta}\left(f_{i}=c | \boldsymbol{z}, \boldsymbol{W}\right)+\sum_{i=1}^{N}\sum_{\substack{j=1\\j> i}}^{N}\sum_{c=1}^{C_{W}} -t^{\varepsilon_{i,j}}_c\log p_{\boldsymbol{\theta}}\left(W_{i,j}=c | \boldsymbol{z}\right).
\end{equation}

\section{\label{sec:constraint}Physical Constraints}

In this section, the implementation of physical constraints on the generative model is discussed. We employ a regularization algorithm to impose constraints in the VAE framework~\cite{ma2018constrained}. The first groups of constraints target connectivity and valence capacities in the molecule. The former promotes a single connected output graph, while the latter deals with the combination of valence capacity of the atoms and the incident bonding electron pairs.

\begin{figure}[H]
    \centering
    \includegraphics[width=5cm]{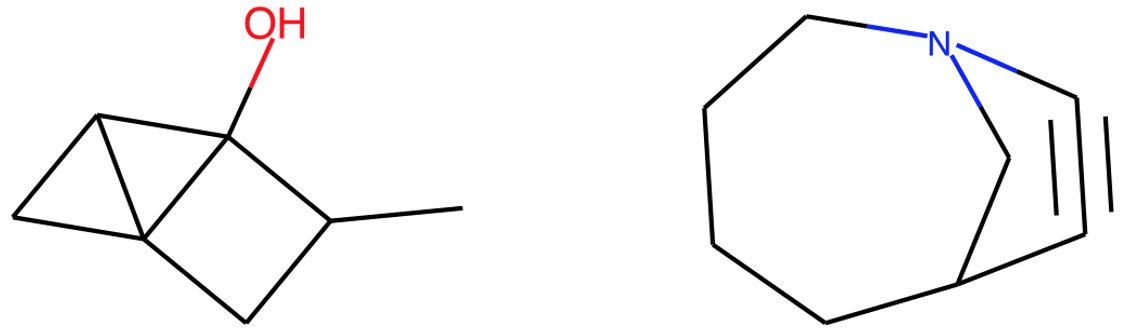}
    \caption{Some examples of energetically unstable molecules.
    \label{fig:unstable}}
\end{figure}

While the above constraints result in valid Lewis structures, not all such molecules are feasible. One aspect that is overlooked in the literature is that many of these valid combinations would be energetically unstable. Two examples of such molecules are the alkyne bridging over a 7-member cycle and the two $3-$member cycles in Fig.~\ref{fig:unstable}. These molecules would have to be very far from the optimal geometry as suggested by Valence Shell Electron Pair Repulsion (VSEPR) theory to have the represented atomic bonds. Here, we impose physical constraints to discourage the generation of such energetically unstable molecular structures.

In the VAE framework, the approximation of model parameters is accomplished by maximizing the lower-bound for log-evidence (Eq.~(\ref{eq:ELBO})), or equivalently
\begin{equation*}
    \min_{\boldsymbol{\theta}, \boldsymbol{\phi}}  \ -\mathscr{L}_{E L B O}(\boldsymbol{\theta},\boldsymbol{\phi};\mathscr{G}).
\end{equation*}
In a regularization framework, validity and physical constraints are implemented as regularizing terms. A constraint optimization problem is often transformed to an unconstrained optimization problem using Lagrange multipliers methods. The regularization method used here varies in that rather than solving for the Lagrange multipliers, constraint terms are accompanied by regularization parameters that are tuned until we observe the desired output in terms of the number of valid and energetically stable molecules. The regularized objective of VAE takes the form 
\begin{equation}
    \min_{\boldsymbol{\theta}, \boldsymbol{\phi}}  \ -\mathscr{L}_{E L B O}(\boldsymbol{\theta},\boldsymbol{\phi};\mathscr{G})+ \sum_{r=1}^\mathcal{R} \mu_r \left(\int \mathcal{C}_{+}^{(r)}(\boldsymbol{z}, \boldsymbol{\theta})^2p(\boldsymbol{z})d\boldsymbol{z}\right)^{\frac{1}{2}},
    \label{eq:reg_loss}
\end{equation}
where $\mu_r$ is the regularization parameter corresponding to the constraint $\mathcal{C}^{(r)}$, $\mathcal{R}$ is the total number of the constraints, and $\mathcal{C}_{+}$ represents the ramp function, defined as $\max (\mathcal{C}, 0)$. Essentially, the regularization term is added to the objective function of a classic VAE in order to penalize the violations of the physical constraints.

The constraint term in Eq.~(\ref{eq:reg_loss}) is marginalized to drop its dependence on the latent variable $\boldsymbol{z}$. As this integration is computationally not feasible, Monte Carlo approximation is used to evaluate the regularization term
\begin{equation}
    \int \mathcal{C}_{+}(\boldsymbol{z},\theta)^2 p(\boldsymbol{z}) \mathrm{d} \boldsymbol{z} \approx \frac{1}{L} \sum_{l=1}^{L} \mathcal{C}_{+}(\boldsymbol{z}^{(l)},\theta)^2, \quad \text{with } \boldsymbol{z}^{(l)} \sim p(\boldsymbol{z}),
    \label{eq:MC_int}
\end{equation}
where $L$ is the number of the Monte Carlo samples~\cite{ma2018constrained}. Therefore, we can evaluate the objective function (Eq.~(\ref{eq:reg_loss})) as
\begin{align}
    \label{eq:ELBO_regularized}
    \begin{split}
       \mathscr{L}(\boldsymbol{\theta},\boldsymbol{\phi}, \boldsymbol{\mu};\mathscr{G})=&-\sum_{i=1}^{K}\left(\frac1L\sum_{l=1}^{L}\log p_{\boldsymbol{\theta}}(\mathcal{G}^{(i)} | \boldsymbol{z}^{(i,l)})\right)\\
        &+\sum_{i=1}^{K}\mathrm{D_{KL}}\left[q_{\boldsymbol{\phi}}(\boldsymbol{z}^{(i)} | \mathcal{G}^{(i)}) \| p(\boldsymbol{z}^{(i)})\right]\\
        &+\sum_{i=1}^{K}\sum_{r=1}^{\mathcal{R}} \mu_r \left(\frac{1}{L} \sum_{l=1}^{L} \left(\mathcal{C}^{(r)}_{+}(\boldsymbol{z}^{\left((i-1)L+l\right)},\theta)\right)^2\right)^{\frac{1}{2}}.
    \end{split}
\end{align}
Note that $\boldsymbol{z}^{(i,l)}$ is sampled from $q_{\boldsymbol{\phi}}(\boldsymbol{z}| \mathcal{G})$ using the transformation $\boldsymbol{z}^{(i,l)}=\boldsymbol{\mu}_{\boldsymbol{\phi}}(\mathcal{G}^{(i)})+ \boldsymbol{S}_{\boldsymbol{\phi}}(\mathcal{G}^{(i)})\odot\boldsymbol{\epsilon}^{(l)}$, with $\boldsymbol{\epsilon}$ sampled from $ p(\boldsymbol{\epsilon})=\mathcal{N}(\boldsymbol{\epsilon};\boldsymbol{0},\boldsymbol{I})$, and $\odot$ denotes element-wise product, while $\boldsymbol{z}^{(l)}$ is sampled from the prior $p(\boldsymbol{z})$. In the training process, $\mathscr{L}_{ELBO}$ terms are computed using the training data, while the constraint terms are computed using synthetic data; that is, samples from $p(\boldsymbol{z})$, which then are mapped using $p_{\boldsymbol{\theta}}(\boldsymbol{W}|\boldsymbol{z})$. The resulting probabilistic representation of the weight matrix is used to compute the constraint terms, as discussed later in this section. We choose the number of the Monte Carlo samples $L$ to be the same as the number of the MC samples used to compute the expectation term in Eq.~(\ref{eq:ELBO_MC}).

In the rest of the section, we introduce physical constraints to encourage the generation of energetically more stable molecular graphs. For validity constraints on connectivity and valence capacity and detailed discussion on the regularization framework, we refer the reader to reference~\onlinecite{ma2018constrained}.

\subsection{\label{subsec:3_mem_cyc}$3-$member cycles}

Here, we introduce constraints to avoid the generation of molecules containing $3-$member cycles. To this end, we need to define a differentiable constraint $\mathcal{C}$ based on the probabilistic representation of the weight matrix $p(\boldsymbol{W})$.

Let $\boldsymbol{A}$ represent the adjacency matrix of graph $\mathcal{G}$. That is, element $A_{m,n}$ is $1$ if nodes $v_m$ and $v_n$ are connected and is zero otherwise. A closed walk is a sequence of connected edges and vertices that start and end at the same node and can visit an edge or node more than once. A closed walk is called a cycle when the sequence only consists of unique edges and vertices. $(\boldsymbol{A}^k)_{m,m}$ computes the number of closed walks on a graph $\mathcal{G}$ that start from node $v_m$ and end at the same node and have size $k$. Since a closed walk with size $3$ has unique edges and vertices, it also is a cycle. Therefore, we can find the number of the $3-$member cycles of graph $\mathcal{G}$ by summing the diagonal elements $\text{tr}(\boldsymbol{A}^3)=\sum_m (\boldsymbol{A}^3)_{m,m}$. So the graph should satisfy
\begin{equation}
    \sum_m (\boldsymbol{A}^3)_{m,m}\leq 0.
    \label{eq:3_mem_cyc_non_diff}
\end{equation}
Note that each cycle is counted once per node and for both clockwise and counterclockwise directions. Hence, to count the unique cycles, the sum above needs to be divided by $6$.

In this work, instead of an adjacency matrix, we have a probabilistic representation of the weight matrix $p_{\boldsymbol{\theta}}(\boldsymbol{W}^{(l)}|\boldsymbol{z}^{(l)})$, where $\boldsymbol{z}^{(l)}$ is sampled from the prior distribution $p(\boldsymbol{z})$. This probability distribution includes the probability of each class of the bonds $\{\varnothing, I, II, III\}$. A differentiable version of the adjacency matrix, denoted by $\bar{\boldsymbol{A}}$, can be obtained by summing the probabilities of the covalent bonds of different orders. Equivalently, we can subtract the probability of disconnected nodes from $1$. In other words,
\begin{equation}
    \label{eq:prob_adj}
     \bar{A}_{m,n}=1-p_{\boldsymbol{\theta}}(W_{m,n}=\varnothing|\boldsymbol{z}).
\end{equation}
Using the relation for the number of the $3-$member cycles, we can formulate the constraint term (Eq.~(\ref{eq:3_mem_cyc_non_diff})) as
\begin{equation}
	\mathcal{C}^{(1)}(\boldsymbol{\theta}, \boldsymbol{z})= \frac{1}{6}\text{tr}(\bar{\boldsymbol{A}}^3).
\end{equation}

\subsection{\label{subsec:triple_cyc}Cycles with triple bonds}

Next, we present a constraint term that targets the generation of molecules that include cycles with covalent bonds of order three. Starting from a probabilistic representation of the weight matrix $p(\boldsymbol{W})$, this constraint is specified by two terms: (i) a function to inspect membership in a cycle, and (ii) an indicator function for triple bonds.

A path is defined as a sequence of unique connected edges and vertices. If the pair of nodes $(v_m, v_n)$ is a member of a cycle, there should be at least two paths that connect $v_m$ and $v_n$. One path only contains the edge $\varepsilon_{m,n}$ and the other is a path of a size larger than $1$. We can use this to formulate a function for membership in a cycle. Given a graph $\mathcal{G}$, we look if there is a path other than $\varepsilon_{m,n}$ that connects $v_n$ and $v_m$. This can be achieved by inspecting graph $\mathcal{G}-\varepsilon_{m,n}$, obtained by eliminating the edge $\varepsilon_{m,n}$ from $\mathcal{G}$. Note that we do not need to check if $\varepsilon_{m,n}$ exists in the graph $\mathcal{G}$ since the indicator function for triple bonds does that for us in the next step.

A walk from node $v_m$ to node $v_n$ is a sequence of connected edges and vertices that starts from $v_m$ and ends at $v_n$. Let $\boldsymbol{A}_{\varepsilon_{m,n}}$ represent the adjacency matrix of the graph $\mathcal{G}-\varepsilon_{m,n}$. The number of walks of size $k$ between $v_m$ and $v_n$ is indicated by $(\boldsymbol{A}_{\varepsilon_{m,n}}^k)_{m,n}=(\boldsymbol{A}_{\varepsilon_{m,n}}^k)_{n,m}$. Thus, the total number of the walks connecting $v_n$ and $v_m$ is computed by $(\sum_k\boldsymbol{A}_{\varepsilon_{m,n}}^k)_{m,n}$. If this element is zero for the sum up to $k=N-1$, there is no path of a size larger than $1$ in graph $\mathcal{G}$ connecting nodes $v_n$ and $v_m$. Otherwise, the element is non-zero. Since $m\neq n$, inspecting connections between $v_n$ and $v_m$ only involves the off-diagonal elements of the summation matrix. As $k=0$ only affects the diagonal elements, which do not impact the constraint, we can write this sum as geometric series and express it in its closed form and define
\begin{equation}
    \boldsymbol{B}_{\varepsilon_{m,n}} := (\boldsymbol{I}-\frac{1}{N}\boldsymbol{A}_{\varepsilon_{m,n}})^{-1}, 
    \label{eq:cyc_ind}
\end{equation}
as the function to study membership in a cycle. The factor $1/N$ is introduced to ensure the convergence of the geometric series, since $|(1/N)\boldsymbol{A}_{\varepsilon_{m,n}}|<1$. We show the derivation in Appendix~\ref{app:Proof}.

Suppose that $\mathcal{I}_{III}$ represents an indicator function for the triple bonds in graph $\mathcal{G}$, such that $\mathcal{I}_{III}(W_{m,n})$ is non-zero if $W_{m,n}=III$ and $\mathcal{I}_{III}(W_{m,n})=0$, otherwise. For the graph not to have cycles with triple bonds, each edge should satisfy the following constraint:
\begin{equation}
	\mathcal{I}_{III}(W_{m,n}) (B_{\varepsilon_{m,n}})_{m, n}\leq 0, \quad \forall m\neq n.
	\label{eq:trip_cyc_non_diff}
\end{equation}
Simply put, if $\mathcal{I}_{III}(W_{m,n})$ is non-zero, edge $\varepsilon_{m,n}$ should not belong to a cycle; hence, $(B_{\varepsilon_{m,n}})_{m, n}=0$. On the other hand, if $\varepsilon_{m,n}$ belongs to a cycle, $(B_{\varepsilon_{m,n}})_{m, n}$ is non-zero, in which case $\mathcal{I}_{III}(W_{m,n})$ should be zero.

Given a probabilistic weight matrix $p_{\boldsymbol{\theta}}(\boldsymbol{W}^{(l)}|\boldsymbol{z}^{(l)})$, which is mapped from a sample $\boldsymbol{z}^{(l)}$ drawn from $p(\boldsymbol{z})$, we formulate a differentiable variant of the constraint defined in Eq.~(\ref{eq:trip_cyc_non_diff}). In Eq.~(\ref{eq:prob_adj}), a probabilistic adjacency matrix $\bar{\boldsymbol{A}}$ for the graph $\mathcal{G}$ was defined. Similarly, we define a probabilistic adjacency matrix $\bar{\boldsymbol{A}}_{\varepsilon_{m,n}}$ for the graph $\mathcal{G}-\varepsilon_{m,n}$ by setting elements $\bar{A}_{m,n}$ and $\bar{A}_{n,m}$ to zero. We substitute this into Eq.~(\ref{eq:cyc_ind}) to define $\bar{\boldsymbol{B}}_{\varepsilon_{m,n}} := (\boldsymbol{I}-\frac1N\bar{\boldsymbol{A}}_{\varepsilon_{m,n}})^{-1}$. On the other hand, we can define a probabilistic version of $\mathcal{I}_{III}$ as
\begin{equation}
    \bar{D}_{m,n}:=p_{\boldsymbol{\theta}}(W_{m,n}=III|\boldsymbol{z}).
    \label{eq:prob_trip}
\end{equation}
Using these matrices, we can reformulate Eq.~(\ref{eq:trip_cyc_non_diff}) and obtain the constraint
\begin{equation}
	\mathcal{C}^{(2)}(\boldsymbol{\theta}, \boldsymbol{z})=\sum^N_{m=1}\sum^N_{\substack{n=1\\n> m}}\bar{D}_{m, n} (\bar{B}_{\varepsilon_{m,n}})_{m, n}.
	\label{eq:trip_cyc_diff}
\end{equation}

\section{\label{sec:UQ}Model Uncertainty}

Bayesian inference has been used for quantifying uncertainty in VAE models~\cite{schoberl2019predictive}. Quantifying uncertainties in model parameters would result in error bars over the predicted physicochemical properties. Given a set of $K$ i.i.d. observations $\mathscr{G}=\left\{\mathcal{G}^{(1)}, \ldots, \mathcal{G}^{(K)}\right\}$ sampled from $p_{target}(\mathcal{G})$, we are interested in finding a model $p_{\boldsymbol{\theta}}(\mathcal{G})$ parameterized by $\boldsymbol{\theta}$ that closely approximates $p_{target}(\mathcal{G})$. 
The parameters $\boldsymbol{\theta}$ of such a model can be estimated through Eq.~(\ref{eq:MLE}). However, the computation of the marginal log-likelihood requires an intractable integration. To overcome this, Eq.~(\ref{eq:ELBO}) defines a lower-bound on the marginal log-likelihood by introducing an auxiliary density $q_{\boldsymbol{\phi}}$ parameterized by $\boldsymbol{\phi}$. Therefore, we find the Maximum Likelihood Estimate~(MLE) for parameters $\boldsymbol{\theta}$ and $\boldsymbol{\phi}$ by maximizing the lower-bound on the marginal log-likelihood
\begin{equation}
    \boldsymbol{\theta}_{\text{MLE}}, \boldsymbol{\phi}_{\text{MLE}}=\argmax_{\boldsymbol{\theta},\boldsymbol{\phi}}\mathscr{L}_{ELBO}(\boldsymbol{\theta},\boldsymbol{\phi};\mathscr{G})=\argmax_{\boldsymbol{\theta},\boldsymbol{\phi}}\sum_{k=1}^K\mathscr{L}_{ELBO}(\boldsymbol{\theta}, \boldsymbol{\phi};\mathcal{G}^{(k)}).
\end{equation}

The limited number of data in $\mathscr{G}$ results in uncertainties in the learned parameters. By employing predictive posterior in Eq.~(\ref{eq:BP}), one can quantify the epistemic uncertainties in the model. The procedure is summarized in the following steps:
\begin{itemize}
    \item Draw a posterior sample $\boldsymbol{\theta}^j\sim p(\boldsymbol{\theta}|\mathscr{G})$.
    \item Obtain predictive samples $\bar{\mathcal{G}}_j^{(t)}$, with $t = 1,\dots, T$, given $\boldsymbol{\theta}^j$.
\end{itemize}

As noted above, this requires a full posterior over model parameters $p(\boldsymbol{\theta}|\mathscr{G})$. Finding this posterior is computationally impractical. One common way is to approximate the posterior distribution $p(\boldsymbol{\theta}|\mathscr{G})$ with a Gaussian distribution using the Laplace method. This method requires the computation of the Hessian of the log-posterior. In some problems, this normal distribution gives a poor approximation to the full posterior of model parameters. For instance, we observed that  models with $\boldsymbol{\theta}$ drawn from the approximated Gaussian distribution suffer from extremely low validity of the sampled molecules. Additionally, the computation of the Hessian can be overwhelmingly expensive. Taking these into consideration, Newton and Raftery~\cite{newton1991weighted, newton1994approximate} proposed the weighted likelihood bootstrap~(WLB) to simulate approximately from the posterior distribution. This method is a direct extension of Rubin's Bayesian bootstrap~\cite{rubin1981bayesian}.

Given the dataset $\mathscr{G}=\{\mathcal{G}^{(1)}, \dots, \mathcal{G}^{(K)}\}$, the bootstrap method~\cite{efron1992bootstrap} generates multiple samples $\tilde{\mathscr{G}}_1,\dots, \tilde{\mathscr{G}}_B$ by sampling from $\mathscr{G}$, with replacement. In classical bootstrap, the sampling weight $\pi_k$ associated with data $\mathcal{G}^{(k)}$ is drawn from the discrete set $\left\{0, \frac{1}{K}, \ldots, \frac{K}{K}\right\}$, where the numerator is the number of times $n_k$ that data $\mathcal{G}^{(k)}$ is in the resampled dataset and the denominator is the size of the dataset $K=|\tilde{\mathscr{G}}|$. Rubin~\cite{rubin1981bayesian} presented a Bayesian analog for bootstrap by treating the sampling weight $\boldsymbol{\pi}$ as an unknown variable drawn from a posterior distribution over $\boldsymbol{\pi}$. By imposing an improper, non-informative, Dirichlet prior distribution~\cite{haldane1948precision}
\begin{equation}
    \label{eq:prior}
    p(\boldsymbol{\pi}) =\mathcal{D}ir(\boldsymbol{\pi};\boldsymbol{\alpha}), \quad \text{with}\ \boldsymbol{\alpha} = [0,\dots,0]\in \mathbb{R}^K,
\end{equation}
over $\boldsymbol{\pi}$ and observing sample $\mathscr{G}$, the Bayes rule can be used to derive the posterior distribution over sampling weights
\begin{align}
    \label{eq:weightBayes}
    \begin{split}
        p(\boldsymbol{\pi}|\mathscr{G}) &\propto p(\mathscr{G}|\boldsymbol{\pi})p(\boldsymbol{\pi})\\
        & \propto \prod_k\pi_k^{{n_k}}\prod_k\pi_k^{\alpha_k-1}\\
        & \propto \prod_k\pi_k^{{n_k+\alpha_k-1}}.
    \end{split}
\end{align}
Bootstrap methods work under the assumption that all distinct values in $\mathscr{G}$ have been observed~\cite{rubin1981bayesian}, i. e., $n_1=\dots=n_K=1$. From Eqs.~(\ref{eq:prior}) and (\ref{eq:weightBayes}), the posterior over bootstrap sampling weights follows a Dirichlet distribution over the bounded finite-dimensional space
\begin{equation}
    \label{eq:weightpost}
    p(\boldsymbol{\pi}|\mathscr{G})=\mathcal{D}ir(\boldsymbol{\pi};\boldsymbol{\alpha}'), \quad \text{with}\ \boldsymbol{\alpha}' = [1,\dots,1]\in \mathbb{R}^K.
\end{equation}
A resampled dataset $\tilde{\mathscr{G}}$ can be denoted by the original dataset $\mathscr{G}$ and the associated resampling weights $\boldsymbol{\pi}$ as $\tilde{\mathscr{G}} = (\mathscr{G}, \boldsymbol{\pi})$.

In the problems where the solution for model parameters can be computed using MLE, Newton and Raftery~\cite{newton1994approximate} proposed the  maximization of a weighted likelihood
\begin{equation}
    \label{eq:MWLE}
    \boldsymbol{\theta}_{\text{MWLE}} (\boldsymbol{\pi}) = \argmax_{\boldsymbol{\theta}} \sum_{k=1}^{K} \pi_k\log p(\mathcal{G}^{(k)}|\boldsymbol{\theta}).
\end{equation}
Since
\begin{equation}
    \log p(\mathcal{G}^{(k)}|\boldsymbol{\theta})\geq\mathscr{L}_{ELBO}(\boldsymbol{\theta}, \boldsymbol{\phi};\mathcal{G}^{(k)}),
\end{equation}
and because $\pi_k$ has a positive value, we can define a lower-bound on the weighted marginal log-likelihood. Therefore, $\boldsymbol{\theta}_{\text{MWLE}}$ can be computed by 
\begin{equation}
    \boldsymbol{\theta}_{\text{MWLE}}, \boldsymbol{\phi}_{\text{MWLE}}=\argmax_{\boldsymbol{\theta},\boldsymbol{\phi}}\sum_{k=1}^K\pi_k\mathscr{L}_{ELBO}(\boldsymbol{\theta}, \boldsymbol{\phi};\mathcal{G}^{(k)}).
\end{equation}

Newton and Raftery~\cite{newton1994approximate} simulate approximately from the posterior distribution over $\boldsymbol{\theta}$ by repeated sampling from the posterior distribution $p(\boldsymbol{\pi}|\mathscr{G})$ and maximizing a weighted likelihood to calculate $\boldsymbol{\theta}_{\text{MWLE}}$. The method can be summarized as
\begin{itemize}
    \item Draw a posterior sample $\boldsymbol{\pi}\sim p(\boldsymbol{\pi}|\mathscr{G})$.
    \item Calculate $\boldsymbol{\theta}_{\text{MWLE}}$ from weighted sample $\tilde{\mathscr{G}}=(\mathscr{G}, \boldsymbol{\pi})$.
\end{itemize}

Fushiki~\cite{fushiki2010bayesian} takes advantage of this approximation to propose a Bayesian bootstrap predictive distribution
\begin{equation}
    \label{eq:BBPD}
    p_{B B}(\bar{\mathcal{G}} \mid \mathscr{G})=\int p\left(\bar{\mathcal{G}} \mid \boldsymbol{\theta}_{\text{MWLE}}(\boldsymbol{\pi})\right) p(\boldsymbol{\pi} \mid \mathscr{G}) d \boldsymbol{\pi}.
\end{equation}
A Monte Carlo (MC) estimate of the predictive distribution 
\begin{equation}
    p(\bar{\mathcal{G}}|\mathscr{G})=\frac{1}{B}\sum_{b=1}^{B}p(\bar{\mathcal{G}}|\boldsymbol{\theta}_{\text{MWLE}}(\boldsymbol{\pi}^{b})),
\end{equation}
is obtained by drawing $B$ sampling weights $\boldsymbol{\pi^b}$ of size $K$ from the posterior (Eq.~(\ref{eq:weightpost})). We use $B=25$ to estimate the credible intervals~\cite{breiman1996bagging, clyde2001bagging}. Algorithm~\ref{alg:BBPD} describes the computation of credible intervals.

\begin{algorithm}[H]
	\caption{Estimation of credible intervals. \label{alg:BBPD}}
	\begin{algorithmic}
	\STATE \textbf{Input} $B$ the number of bootstrap samples to be drawn and $T$ the number of predictive samples.
    \FOR{$b=1,\dots,B$}
        \STATE Draw a posterior sample $\boldsymbol{\pi}^b\sim p(\boldsymbol{\pi}|\mathscr{G})$ (Eq.~(\ref{eq:weightpost})).
        \STATE Calculate $\boldsymbol{\theta}^{b}_{\text{MWLE}}$ from weighted sample $(\mathscr{G}, \boldsymbol{\pi}^b)$ (Eq.~(\ref{eq:MWLE})).
        \STATE Obtain predictive samples $\bar{\mathcal{G}}^{(t)}_{b}$, with $t = 1,\dots, T$, given $\boldsymbol{\theta}^{b}_{\text{MWLE}}$ (Eq.~(\ref{eq:BBPD})).
        \STATE Estimate properties $\hat{a}(\boldsymbol{\theta}^{b}_{\text{MWLE}})=\frac{1}{T}\sum_{t=1}^{T}a(\bar{\mathcal{G}}^{(t)}_{b})$, given the predictive samples $\bar{\mathcal{G}}^{(t)}_{b}$, with $t = 1,\dots, T$.
    \ENDFOR
    \STATE Estimate credible intervals from $\hat{a}(\boldsymbol{\theta}^{1:B}_{\text{MWLE}})$.
	\end{algorithmic}
\end{algorithm}

\section{\label{sec:results}Results}

In this section, we provide details on the training of the encoder and graph and signal decoders. The latent representation of the molecular dataset is obtained, and the generative model is used to produce realistic molecules. We assess the model in terms of the generated molecules' chemical validity, their novelty, and their uniqueness. Further, we look at the frequency of various structural features of these molecules. We inspect the smoothness of the discovered latent space by sampling molecules from a latent grid and comparing the generated molecules and constructing their latent property maps. Then, we provide probabilistic estimates of molecular properties for conditional and unconditional generative tasks. Our focus is on computing these statistics using a limited number of training data points and studying the induced uncertainty in the model.

We use a small subset of a molecular dataset to train the model, which includes learning parameters of mappings from the molecular graph to its latent representation and vice versa. We perform our experiments using a $J = 30$ dimensional latent space. We trained the model with samples from the QM9 database~\cite{ramakrishnan2014quantum, ruddigkeit2012enumeration}, which consists of $133885$ small drug-like organic molecules. These molecules are constructed of a maximum of $9$ heavy atoms, including carbon, oxygen, nitrogen, and fluorine. To visualize the latent space, we use a subset of the molecules from the test set. As the $J-$dimensional latent space cannot directly be visualized on the $2-$dimensional plane, we perform Principal Component Analysis (PCA) for the illustrations shown in this work. We display these representations using various molecular properties and structural features.

We use a number of molecular properties to show the performance of the model. These include   physicochemical properties like polar surface area (PSA)~\cite{ertl2000fast}, which is a measure of the polarity of a molecule and is the sum of the surface areas of all polar atoms in the molecule; molecular weight (MolWt), which is the sum of atomic weights for the atoms of the molecule, where the atomic weights are the weighted average of atomic isotopes based on their abundance in nature; and octanol-water partition coefficient (LogP), which amounts for the lipophilicity of the molecule.

\subsection{\label{subsec:scat}Scattering Layer}

First, we start by looking at the latent space generated with the graph scattering layers. The constructed feature maps compose the hidden layer input to the FCN layers. Here, the encoding network uses $M=4$ layers of scattering transform with a filter-bank of $\mathcal{J}=12$ kernels. This leads to a total of $\sum_{m=0}^{M-1}\mathcal{J}^m=1885$ coefficients per signal dimension, where the signal includes a one-hot vector of atom types.

\begin{figure}[H] 
    \setlength{\unitlength}{0.01\textwidth} 
    \begin{picture}(100,26)
    \put(2, 0){\includegraphics[width=0.32\textwidth]{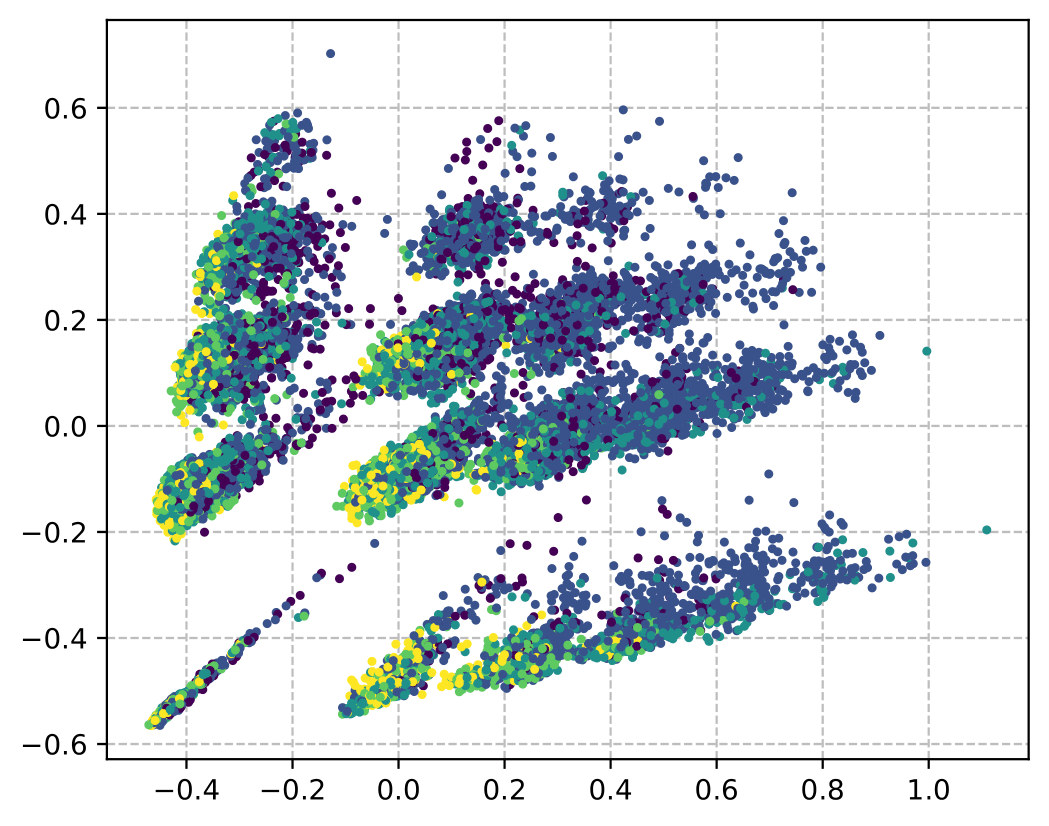}}
    \put(35, 0){\includegraphics[width=0.32\textwidth]{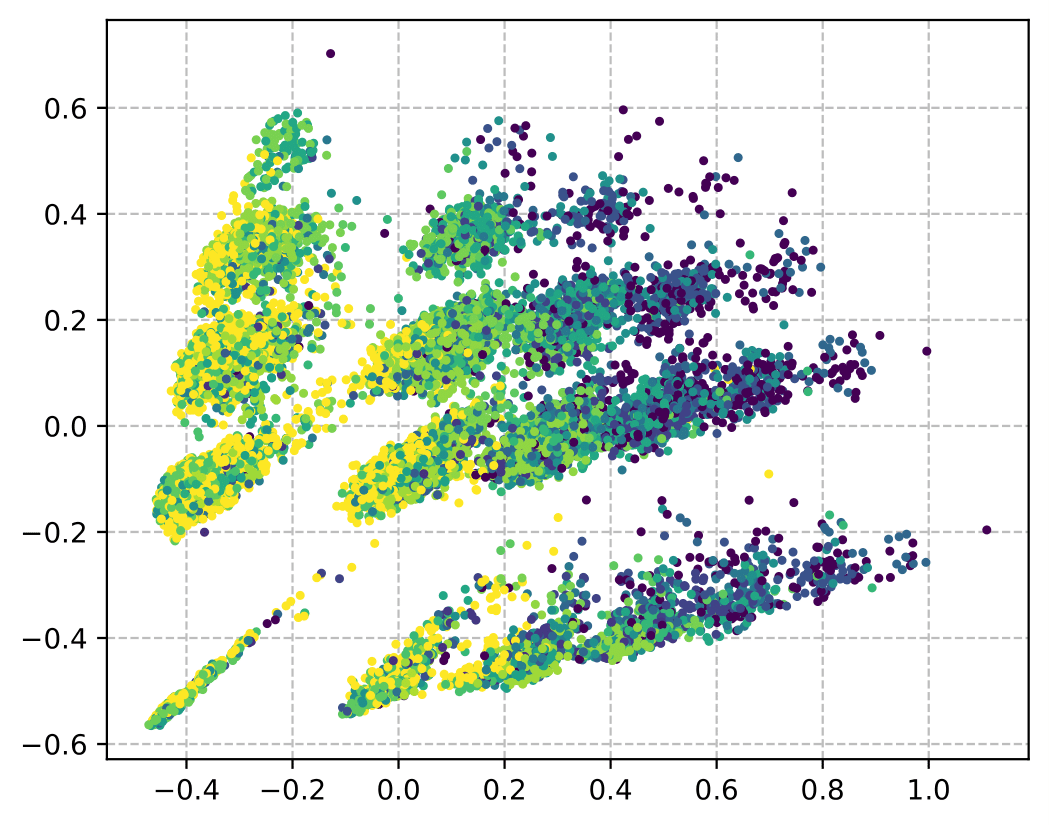}}
    \put(68, 0){\includegraphics[width=0.32\textwidth]{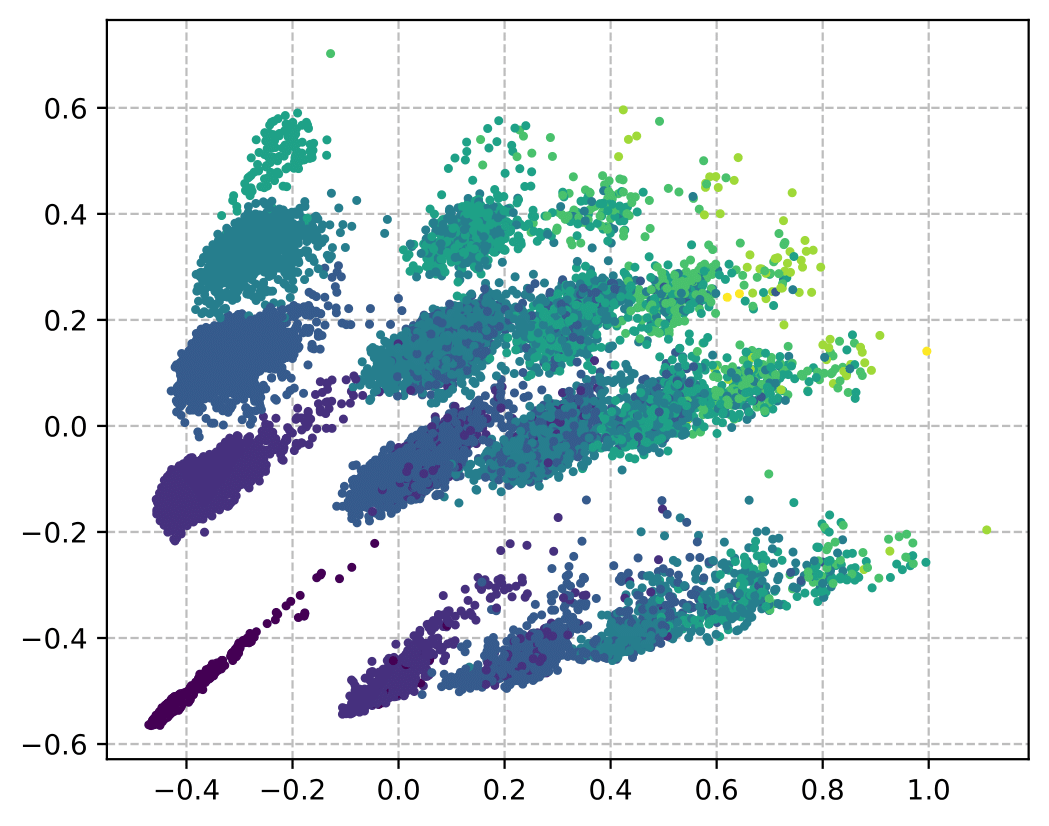}}
    \put(5.85,22.5){\footnotesize  (a)}
    \put(38.85,22.5){\footnotesize  (b)}
    \put(71.85,22.5){\footnotesize  (c)}
    \end{picture}
    \caption{Feature space generated using the  graph scattering coefficients $\mathcal{A}(\mathcal{G})$ from molecular graphs $\mathcal{G}$ for $20000$ molecules from the QM9 dataset. The illustration is projected on two principal axes using PCA and is colored by different structural features, including (a) number of rings, (b) fraction of carbons with SP3 hybridization, and (c) number of hydrogen-bond acceptor atoms. \label{fig:scat_lat}}
\end{figure}
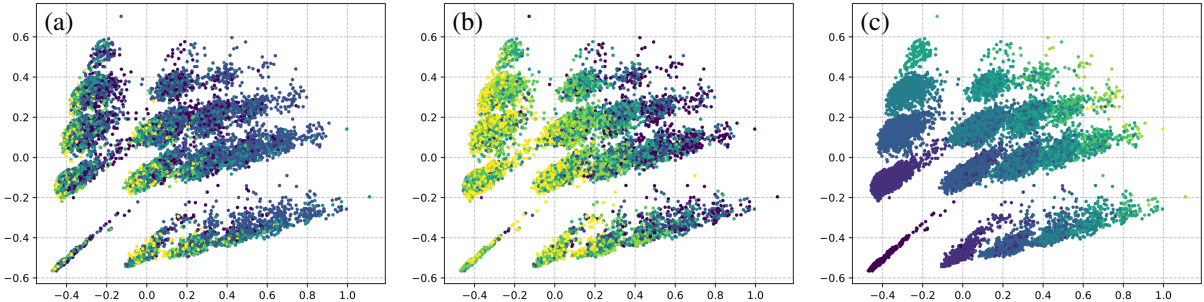

To visualize this space, we use PCA to project the high-dimensional feature space onto a $2-$dimensional space. Given that the problem consists of $5$ node types, the total size of the input to layer $n_{\boldsymbol{\phi}}^{(1)}$ (Fig.~\ref{fig:ModelArch}) is $1885\times5=9425$, as indicated in Table~\ref{tab:model_spec}. Fig.~\ref{fig:scat_lat} uses two principal axes to illustrate the feature space generated by the scattering coefficients. Each point represents a single molecule and is colored by its characteristics. It is common for the molecular features to map molecules with similar structures close together. This motivates us to inspect the latent space by coloring representations in Fig.~\ref{fig:scat_lat} based on different structural features.

\begin{table*}
\rule[-8pt]{0pt}{12pt}
  \centering
  \caption{\label{tab:model_spec}Details of the network architecture.}
    \begin{ruledtabular}
    \begin{tabular}{ccccc}
    Layer              & Input dimension & Output dimension & Activation layer & Activation function \\ \hline
    $l^{(1)}_{\theta}$ &  $J$            &  $2\times J$     & $a^{(1)}$        & Leaky ReLU  \\
    $l^{(2)}_{\theta}$ &  $2\times J$    &  $4\times J$     & $a^{(2)}$        & Leaky ReLU  \\
    $l^{(3)}_{\theta}$ &  $4\times J$    &  $8\times J$     & $a^{(3)}$        & Leaky ReLU  \\
    $l^{(4)}_{\theta}$ &  $8\times J$    &  $288$           & $a^{(4)}$        & Leaky ReLU  \\
    $-$                &  $-$            &  $-$             & $a^{(5)}$        & Leaky ReLU  \\
    $l^{(6)}_{\theta}$ &  $324+J$        &  $45$            & $a^{(6)}_f$      & Leaky ReLU  \\
    $n^{(1)}_{\phi}$   &  $5\times\sum_{m=0}^{M-1}\mathcal{J}^m$          &  $5\times\sum_{m=0}^{M-1}\mathcal{J}^m$           & $-$              & $-$         \\
    $l^{(1)}_{\phi}$   &  $5\times\sum_{m=0}^{M-1}\mathcal{J}^m$          &  $400$           & $-$              & $-$         \\
    $n^{(2)}_{\phi}$   &  $400$          &  $400$           & $\tilde{a}^{(1)}$& ReLU        \\
    $l^{(2)}_{\phi}$   &  $400$          &  $J$             & $-$              & $-$         \\
    $l^{(3)}_{\phi}$   &  $400$          &  $J$             & $-$              & $-$         
    \end{tabular}
    \end{ruledtabular}
\end{table*}

\subsection{\label{subsec:model}Generative model}
In this section, we study the constrained and the base models in terms of the quality of the sampled molecules, their predicted chemical space, and the latent space organization in terms of different structural features and molecular properties. We employ the architecture described in Section~\ref{sec:model} and illustrated in Fig.~\ref{fig:ModelArch}. The dimensionality of the fully-connected layers and the type of the activation layers are summarized in Table~\ref{tab:model_spec}.

\begin{figure}[h] 
    \setlength{\unitlength}{0.01\textwidth} 
    \begin{picture}(100,65)
    \put(18,0){\includegraphics[width=0.32\textwidth]{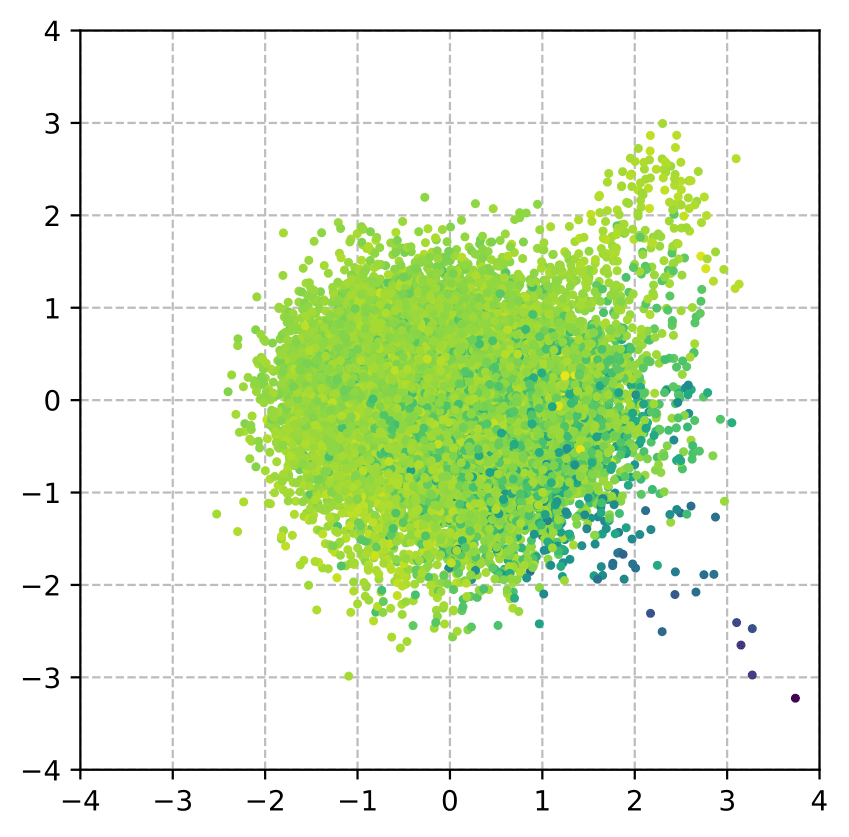}}     
    \put(51,0){\includegraphics[width=0.32\textwidth]{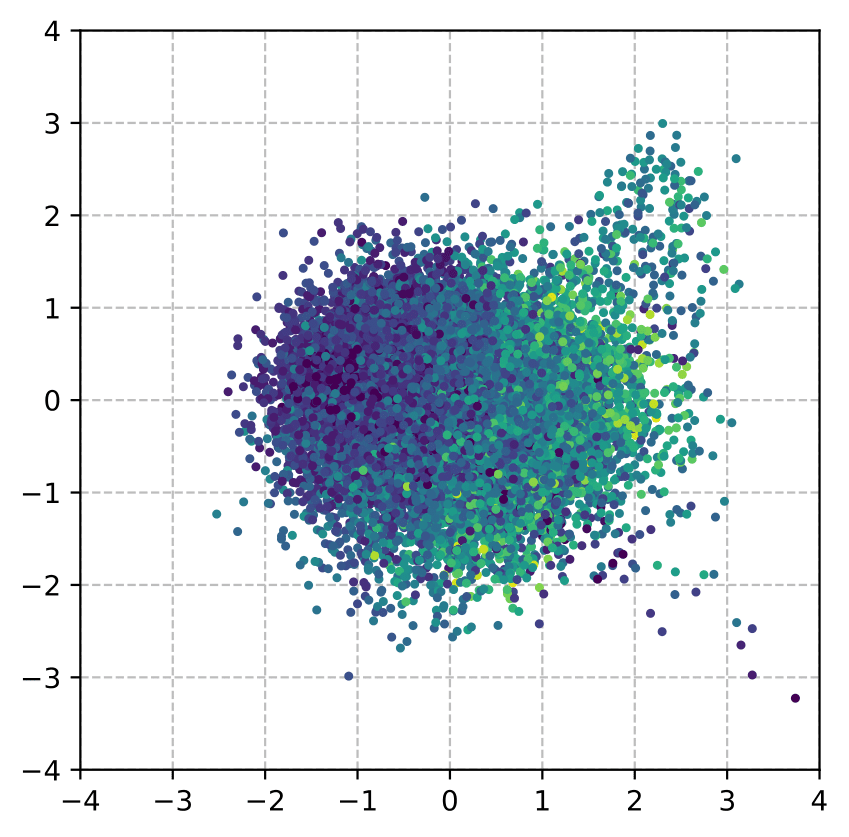}}
    \put(18,32.5){\includegraphics[width=0.32\textwidth]{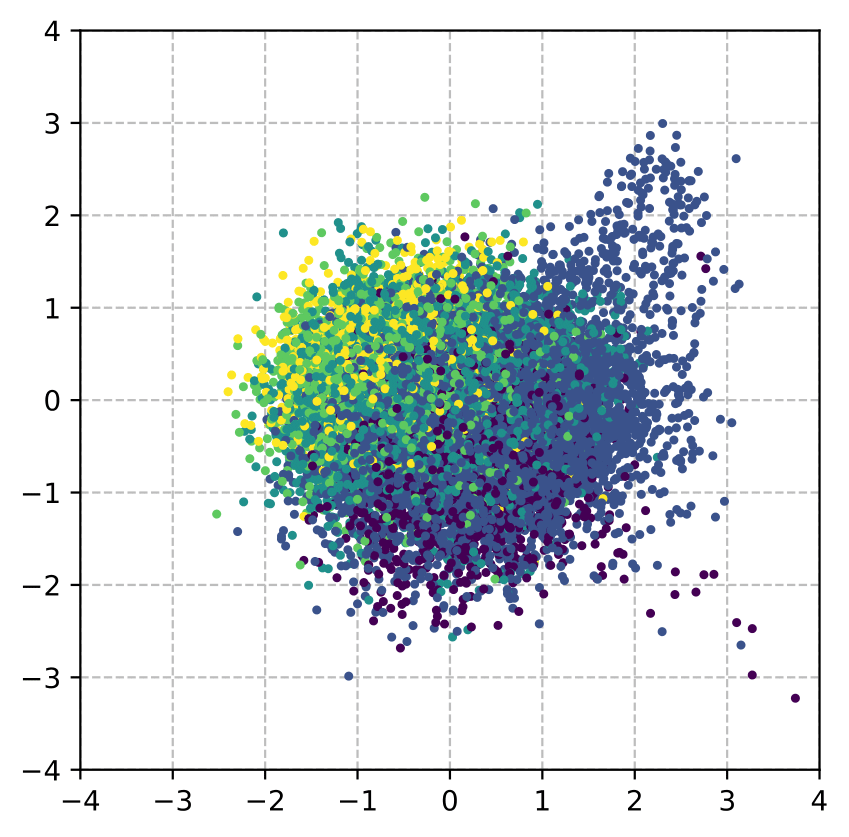}}     
    \put(51,32.5){\includegraphics[width=0.32\textwidth]{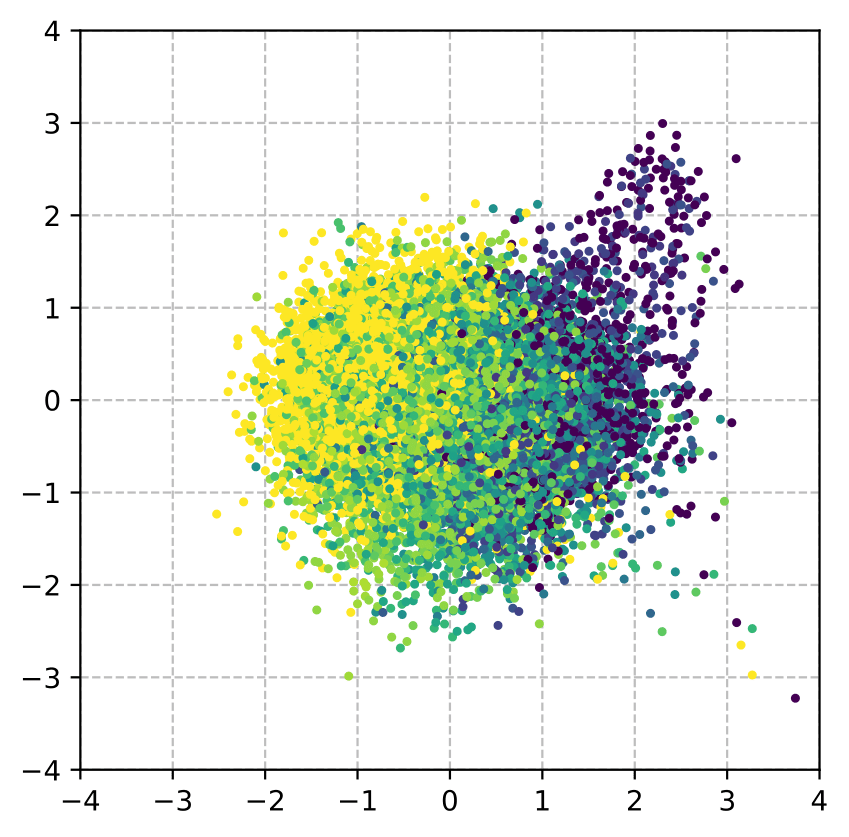}}
    \put(21.5,60.7){\footnotesize  (a)}
    \put(54.5,60.7){\footnotesize  (b)}
    \put(21.5,28.2){\footnotesize  (c)}
    \put(54.5,28.2){\footnotesize  (d)}
    \end{picture}
    \caption{Feature space generated by mapping $20000$ test set molecules using the hybrid scattering encoder $q_{\boldsymbol{\phi}}(\boldsymbol{z}|\mathcal{G})$. The model is trained using a small-sized dataset $\mathscr{G}$ with $K=600$ datapoints. Here, the graph signals $\boldsymbol{f}$ consist of atom type information. Latent representations are presented on the plane of principal axes and are colored by different structural features and properties: (a) Number of rings, (b) fraction of carbons with SP3 hybridization, (c) Molecular weight  [Da], and (d) Polar surface area [\AA$^2$]. \label{fig:latent}}
\end{figure}
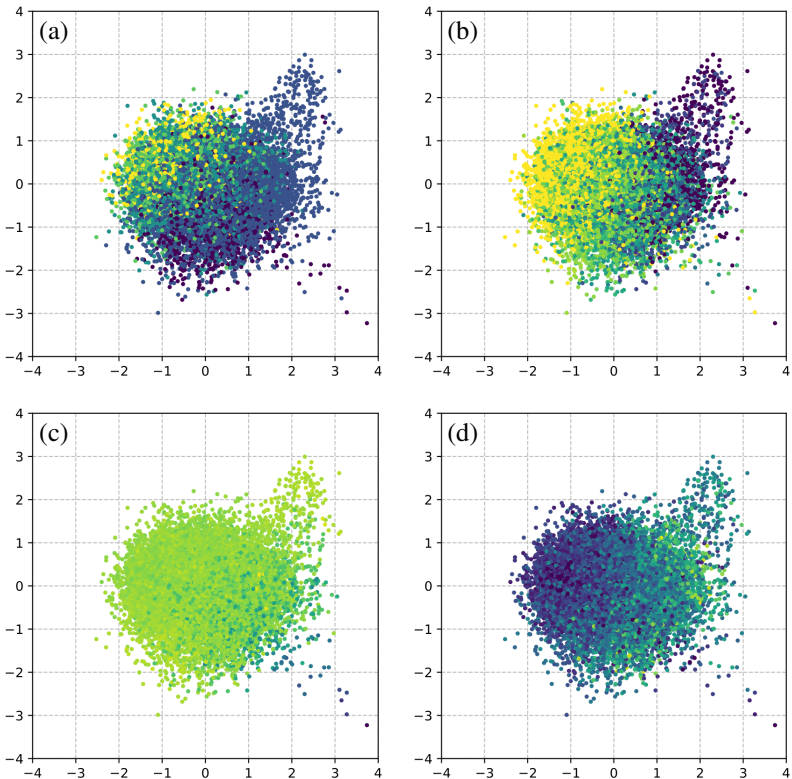

In Fig.~\ref{fig:latent}, we have trained the base model using $K=600$ training data and encoded $20000$ molecular graphs of the test set to the feature space and used two principal axes to represent it on a $2-D$ plane. In such a feature space, it is expected that the molecular representations with similar structures and physicochemical properties are close to each other. In Fig.~\ref{fig:latent}, we take a look at a few examples. We start by illustrating the latent representations in terms of the number of rings. We observe, in Fig.~\ref{fig:latent}(a), that as we move from the bottom to the top left corner in the plane of the principal axes, the number of rings increases. Fig.~\ref{fig:latent}(b) shows a similar pattern regarding the fraction of carbon atoms with SP3 hybridization in the molecule. In this figure, as we move from the upper right to the upper left region, the fraction increases. To put these into perspective, we compare them with the latent space organization in terms of the molecular weight in Fig.~\ref{fig:latent}(c). It is observed that as we move from the upper left side to the lower right side on the PCA projection plane, the molecular weight as well as the size of the molecules, decrease. Lastly, we visualize the latent space in terms of polar surface area in Fig.~\ref{fig:latent}(d). We see that the molecules with latent representation on the upper right region have higher polar surface area values than the molecules in the left side. As this primarily is affected by the number of oxygen and nitrogen atoms, we expect the molecules on the upper right corner to have more nitrogen and oxygen atoms than those on the left side.

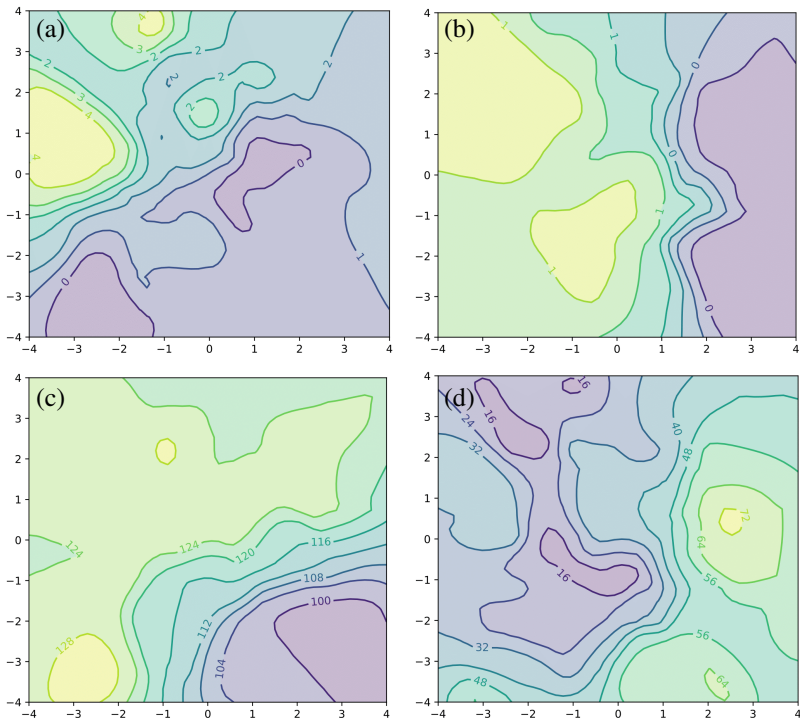
\begin{figure}[h] 
    \setlength{\unitlength}{0.01\textwidth} 
    \begin{picture}(100,60)
    \put(18,0){\includegraphics[trim=0 0 170 0, clip, width=0.32\textwidth]{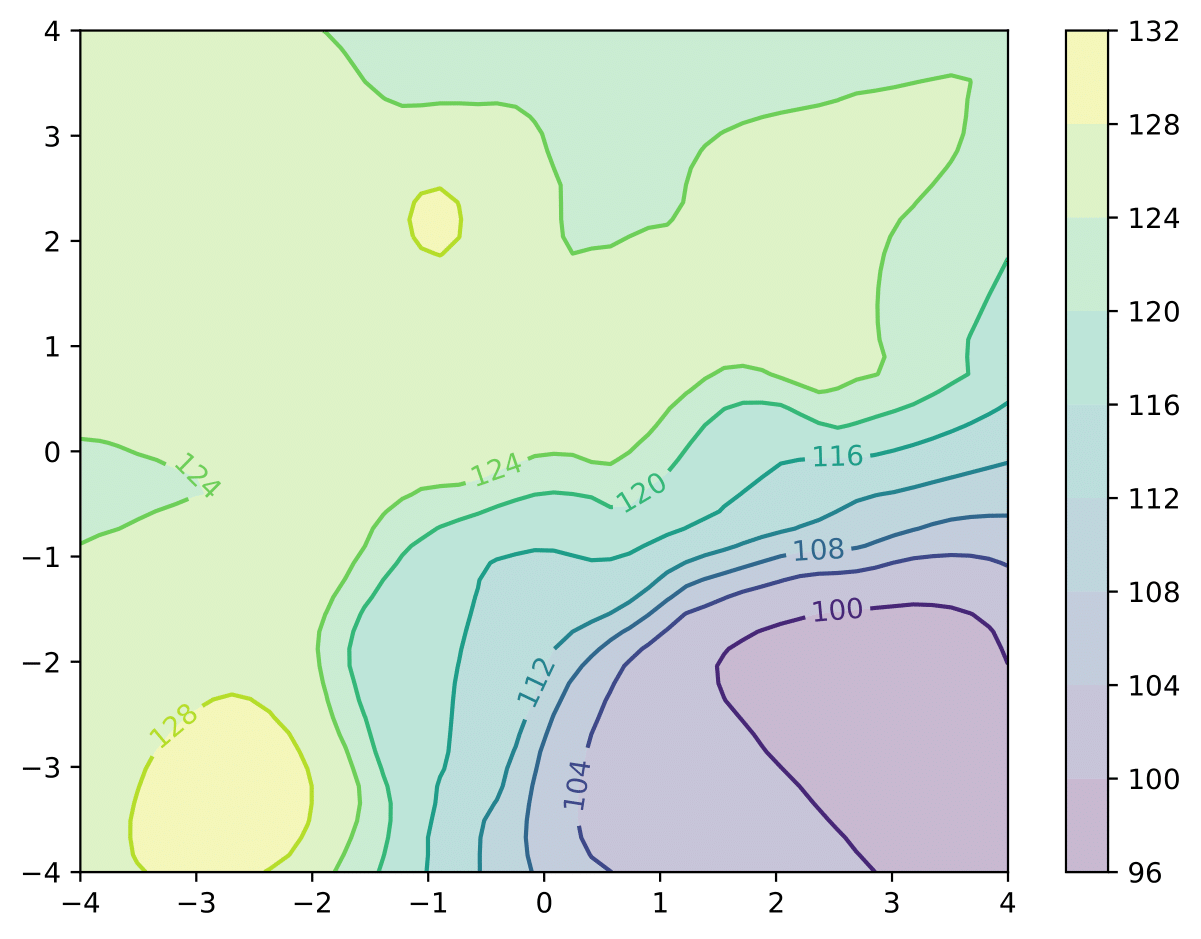}}     
    \put(51,0){\includegraphics[trim=0 0 170 0, clip, width=0.317\textwidth]{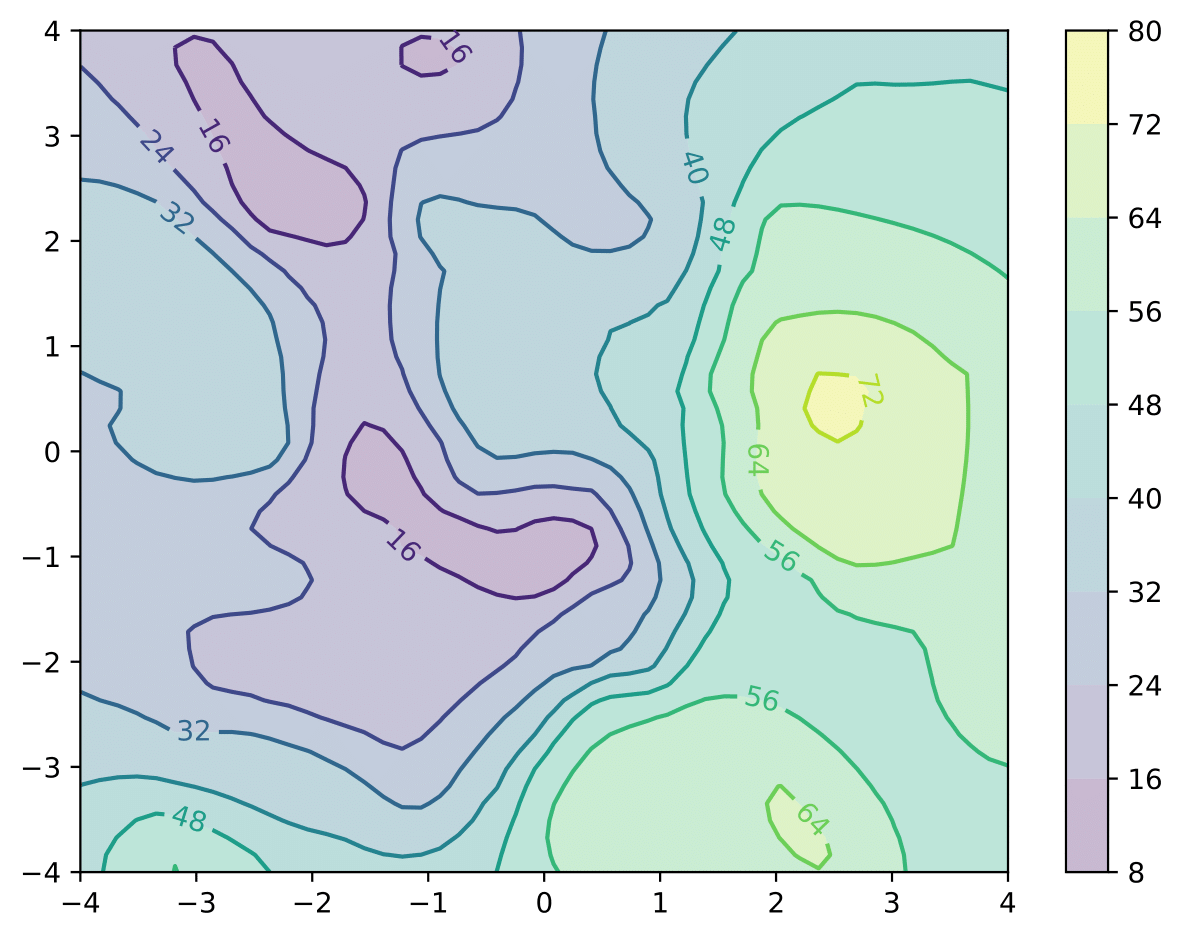}}
    \put(18,29.5){\includegraphics[trim=0 0 170 0, clip, width=0.32\textwidth]{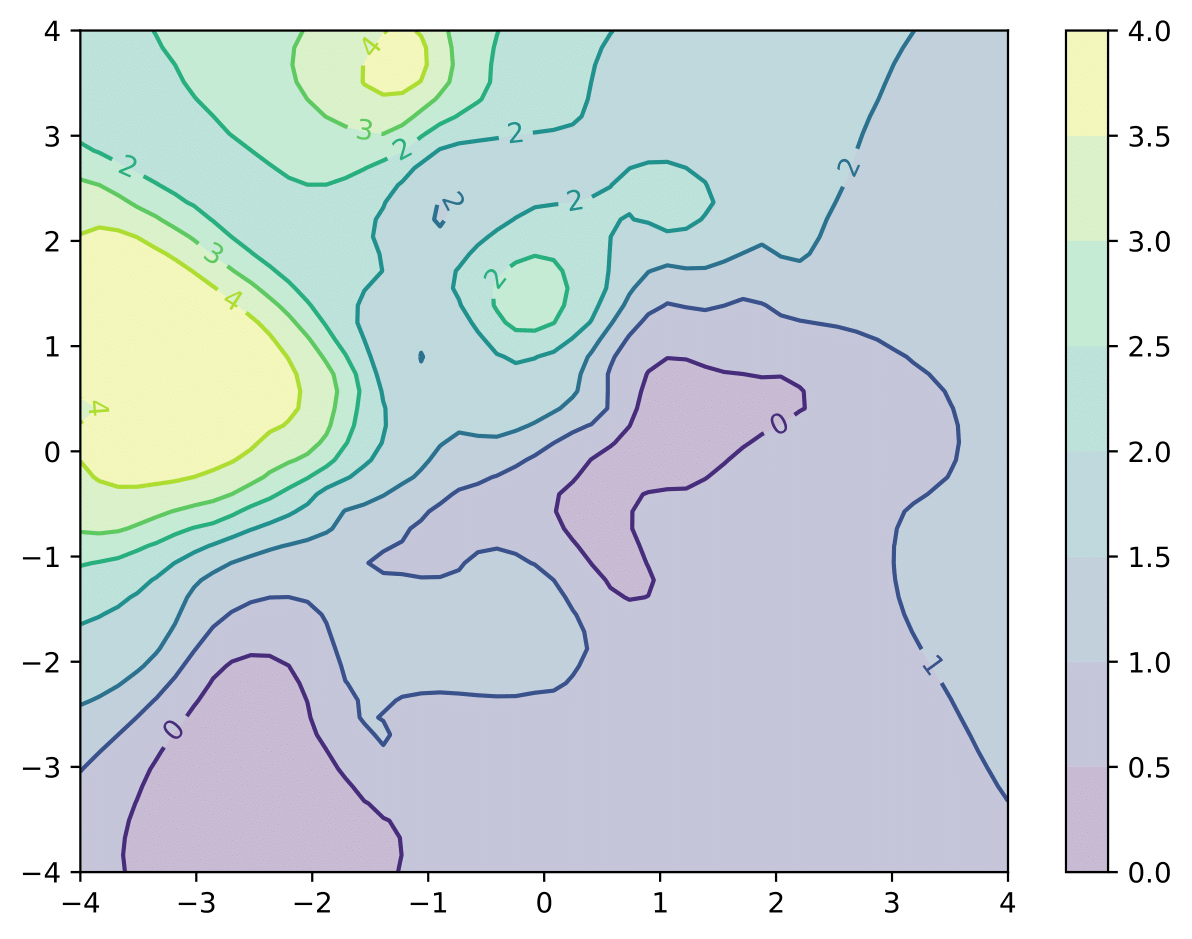}} 
    \put(51,29.5){\includegraphics[trim=0 0 170 0, clip, width=0.324\textwidth]{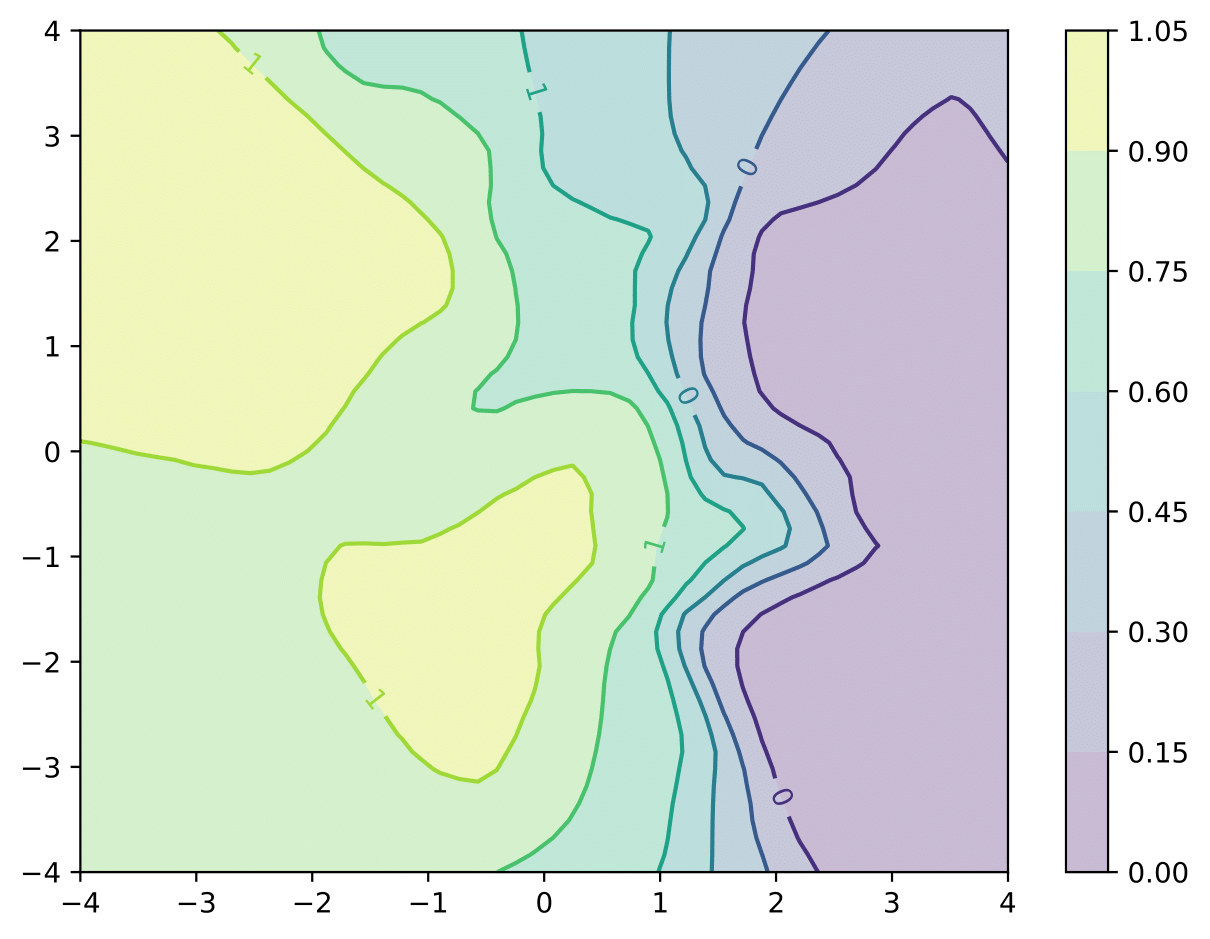}}
    \put(21.,55.9){\footnotesize (a)}
    \put(54,55.9){\footnotesize (b)}
    \put(21,26){\footnotesize (c)}
    \put(54,26){\footnotesize (d)}
    \end{picture}
    \caption{Feature space map of the structural features and physicochemical properties generated by graph scattering VAE. We have used $K=600$ training data to obtain the model. Contours are colored by the  (a) Number of rings, (b) Fraction of carbons with SP3 hybridization, (c) Molecular weight [Da], and (d) Polar surface area [\AA$^2$].} \label{fig:latent_map}
\end{figure}

We further analyze the feature space by constructing a latent contour map in Fig.~\ref{fig:latent_map}. In high-dimension, normally distributed data is concentrated on a spherical shell. As a result, linear interpolation between two points may represent regions of low probability~\cite{gomez2018automatic}. On the other hand, constructing a grid by randomly choosing $2$ orthonormal vectors from the latent space as basis may result in a large number of molecular structures that are not chemically valid~\cite{simonovsky2018graphvae}. In this work, we use the covariance matrix of the latent representation of the training data to find the orthonormal bases. To this end, we construct a projection matrix by performing eigen-decomposition of the covariance matrix and choosing the eigenvectors corresponding to the two largest eigenvalues as our bases. We construct a $2-$dimensional grid and project it onto the $J-$dimensional latent space using the transpose of this projection matrix. We then pass the $J-$dimensional grid points to the decoding network $p_{\boldsymbol{\theta}}(\mathcal{G}|\boldsymbol{z})$, which outputs probabilistic representations of the molecular graphs in return. A point estimate from a probabilistic graph is achieved by node-wise and edge-wise $\argmax$ function over the node and edge probabilities in the graph. As discussed later in this section, there is no guarantee that the generated molecules have a chemically valid combination of atoms and bonds. As a result, some grid points might correspond to chemically invalid molecules, for which the properties cannot be estimated. After filtering the invalid ones, we estimate the property values of the remaining molecules, which give us the property maps. Note that since the grid points corresponding to the invalid molecules have been eliminated, Gaussian Process (GP) regression is used to provide a smooth contour map.

\begin{figure}[h] 
    \setlength{\unitlength}{0.01\textwidth} 
    \begin{picture}(100,76)
    \put(2,50){\includegraphics[width=0.32\textwidth]{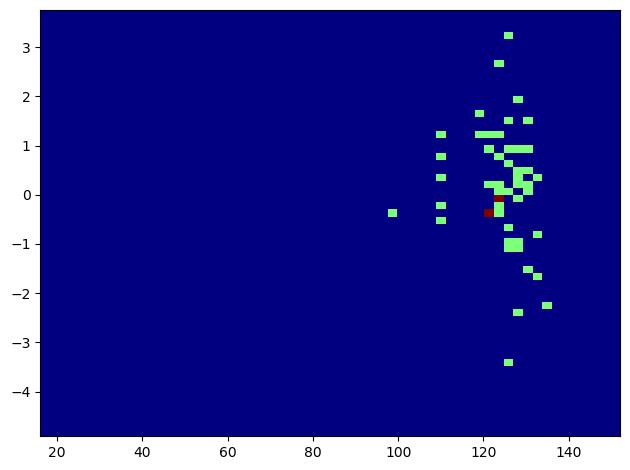}}
    \put(35,50){\includegraphics[width=0.32\textwidth]{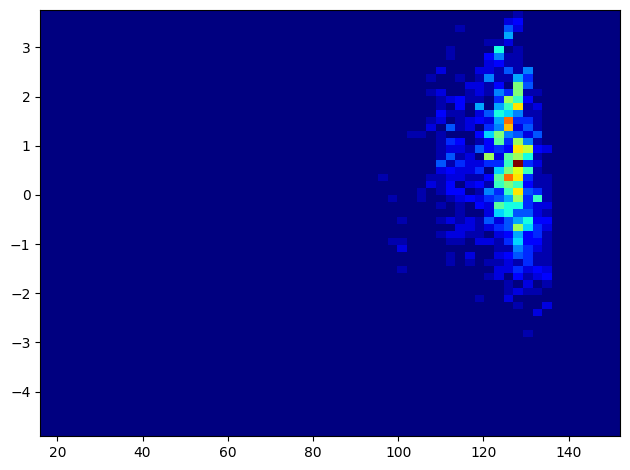}}
    \put(68,50){\includegraphics[width=0.32\textwidth]{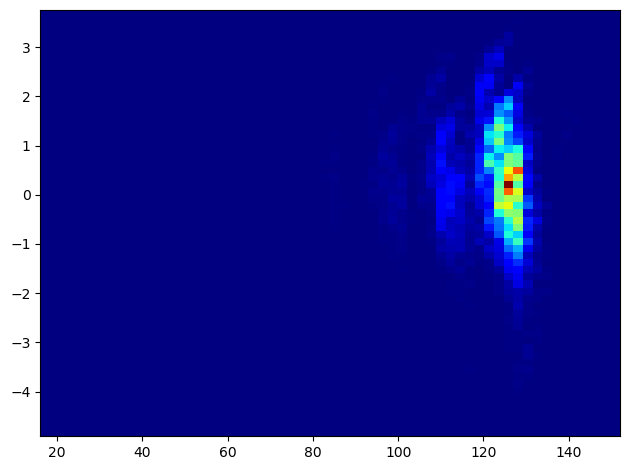}}
    \put(2,25){\includegraphics[width=0.32\textwidth]{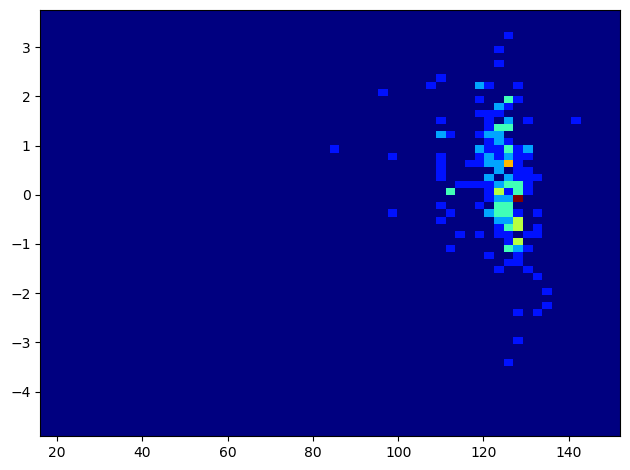}}
    \put(35,25){\includegraphics[width=0.32\textwidth]{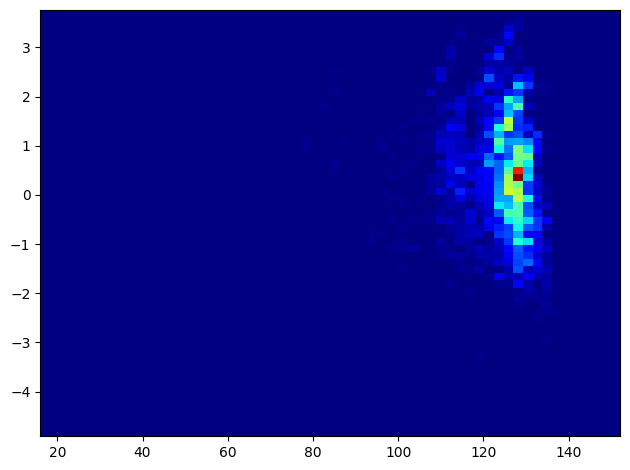}}
    \put(68,25){\includegraphics[width=0.32\textwidth]{Fig8c_prop_joint_train_QM9.png}}
    \put(2,0){\includegraphics[width=0.32\textwidth]{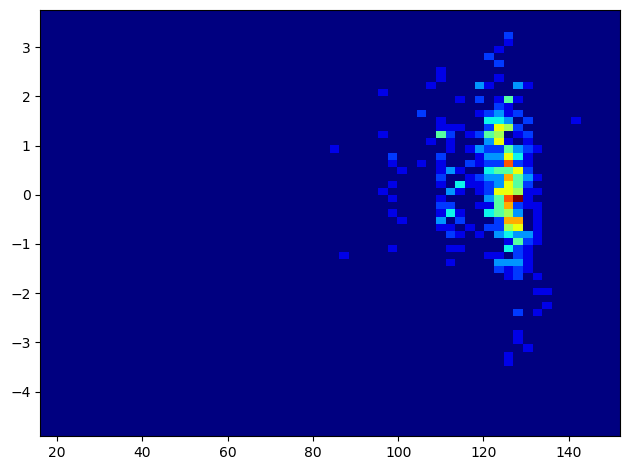}}
    \put(35,0){\includegraphics[width=0.32\textwidth]{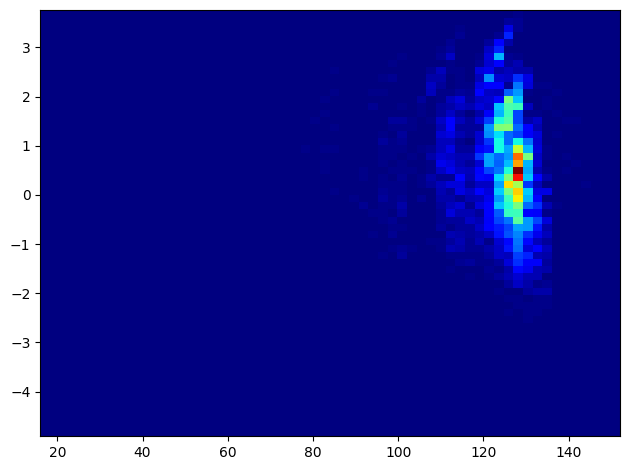}}
    \put(68,0){\includegraphics[width=0.32\textwidth]{Fig8c_prop_joint_train_QM9.png}}
    \put(0,59){\rotatebox{90}{\footnotesize $K=50$}}
    \put(0,34){\rotatebox{90}{\footnotesize $K=200$}}
    \put(0,9){\rotatebox{90}{\footnotesize $K=500$}}
    \put(13,74){\footnotesize  Training data}
    \put(48,74){\footnotesize  Prediction}
    \put(82,74){\footnotesize  Database}
    \end{picture}
    \caption{Predicted chemical space of the QM9 dataset defined by octanol-water partition coefficients on the $y-$axis based on molecular weight [Da] ($x-$axis) for various sizes of training data $K$. Plots are generated by sampling 10000 molecules. On the right column, $133885$ samples from the QM9 database are used to plot the joint map. \label{fig:joint_map}}
\end{figure}

\begin{figure}[h] 
    \setlength{\unitlength}{0.01\textwidth} 
    \begin{picture}(100,33)
    \put(0, 0){\includegraphics[width=0.32\textwidth]{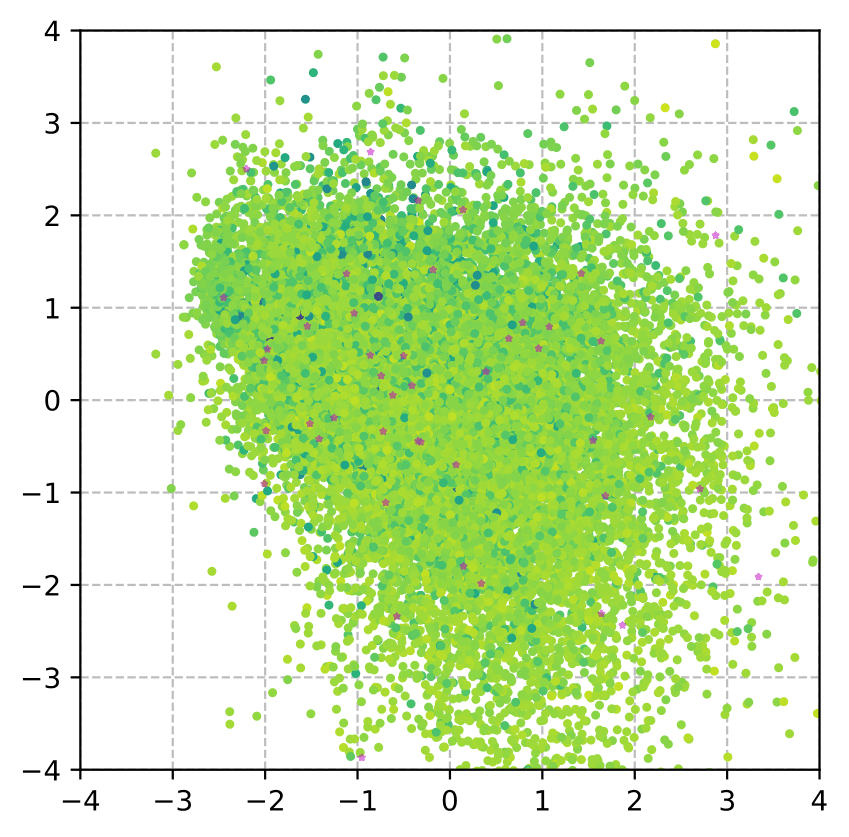}}
    \put(33, 0){\includegraphics[width=0.32\textwidth]{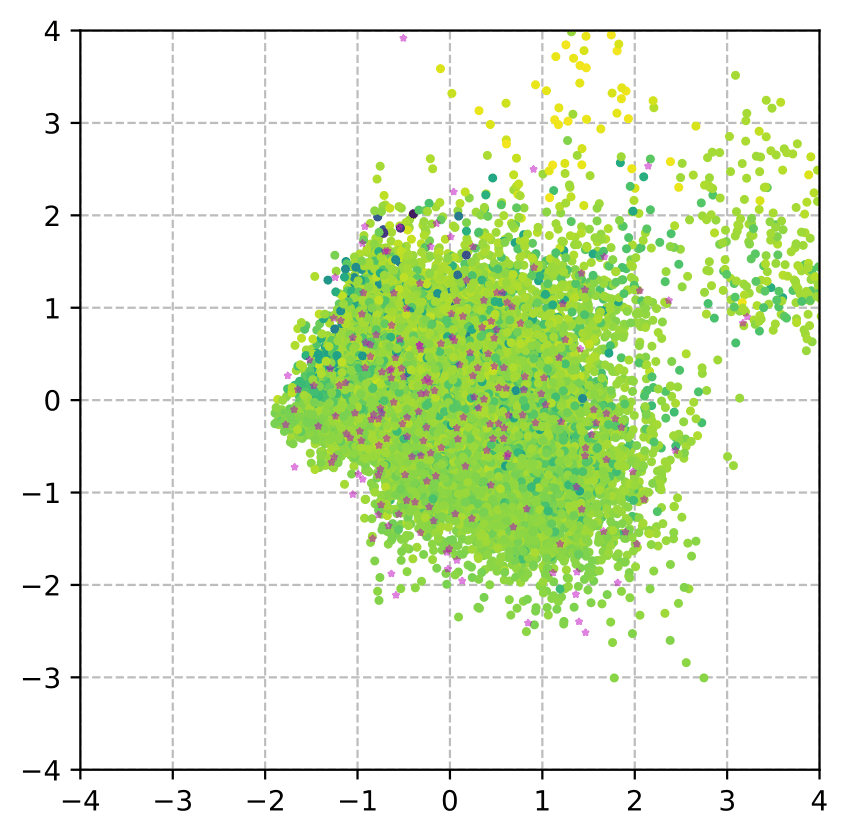}}
    \put(66, 0){\includegraphics[width=0.32\textwidth]{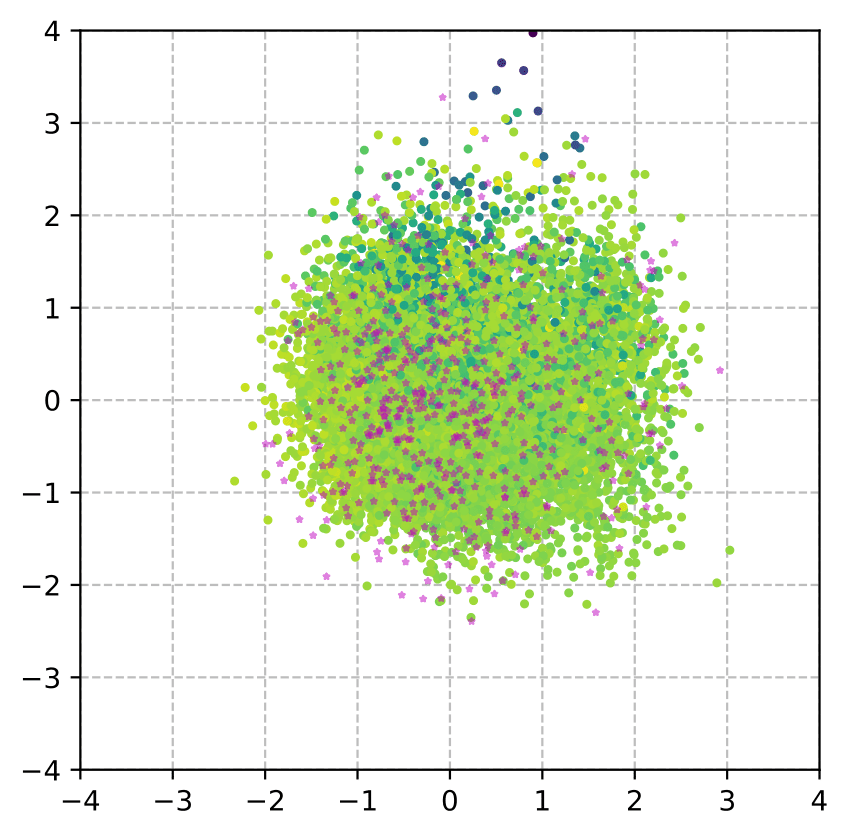}}
    \put(3.7,28.2){\footnotesize (a)}
    \put(36.7,28.2){\footnotesize (b)}
    \put(69.7,28.2){\footnotesize (c)}
    \end{picture}
    \caption{Latent space representation of graph scattering VAE trained with different training dataset sizes (a) $K = 50$, (b) $K = 200$, and (c) $K = 500$. The plots are constructed by mapping $20000$ molecular graphs from the test set using the hybrid scattering encoder $q_{\boldsymbol{\phi}}(\boldsymbol{z}|\mathcal{G})$ and projecting them on the $2-$D plane using PCA. Representations are colored by molecular weight. Training data points are shown in magenta. \label{fig:latent_}}
\end{figure}

The chemical space of the molecules defined by their physicochemical properties plays an essential role in drug design~\cite{kwon2001handbook}. For example, Lipinski's rule of five (Ro5)~\cite{lipinski1997experimental} suggests limits on the octanol-water partition coefficient, molecular weight, and hydrogen bond donors and acceptors to ensure that the drug-like properties are maintained during the development of drug molecules. As a result, the predictive model needs to give a reliable estimate of these chemical spaces. Each row in Fig.~\ref{fig:joint_map} compares the chemical space distribution of the training set constructed by the octanol-water partition coefficient and molecular weight with the one for the predicted molecules using the base model. The right column is the database's chemical space, which is constructed by $133885$ samples from the QM9 database. In Fig.~\ref{fig:joint_map}, we compare the models trained with various sizes of the training set. We observe that the model performs well with a training size as small as $K=50$ datapoints and can yield a reliable estimate of the database's chemical space. Fig.~\ref{fig:latent_} shows the latent space of the models corresponding to different training sizes, where training datapoints are marked with magenta stars.

Next, we inspect the structural features of the sampled molecules. To examine the structural features, we look at the frequency of various functional groups (FG) within the dataset and compare it with that of unique and valid molecules among $10000$ generated samples. FGs are a set of one or more atoms with particular structures and chemical properties within a molecule. When present in a molecular structure, they contribute these chemical properties to the molecule. Table~\ref{tab:FG} summarizes these results for various FGs. These include hydrocarbons, groups containing oxygen such as aldehydes, groups containing nitrogen such as amines, and halogen-based groups, e.g., halides. 

\begin{table*}
    \caption{\label{tab:FG}Frequency of functional groups (in percentage) for $133885$ molecules of the QM9 dataset and $10000$ samples generated by the model.}
    \begin{ruledtabular}
        \begin{tabular}{lrrlrr}
        Functional Group  & Dataset & Samples & Functional Group  & Dataset & Samples \\ \hline
        Acetylenic carbon & $13.28$ & $9.58$ & Ether             & $43.14$ & $40.32$ \\ 
        Aldehyde          & $10.23$ & $8.80$ & Halide            & $1.62$  & $0.60$  \\
        Alkyl carbon      & $95.21$ & $98.41$ & Hydrazone         & $0.00$  & $1.17$  \\
        Amide             & $6.81$  & $4.10$  & Hydroxyl          & $32.44$ & $44.21$ \\
        Amino acid        & $0.84$  & $0.45$  & Ketone            & $11.22$ & $7.72$  \\
        Carbonyl          & $34.74$ & $26.31$ & Nitrile           & $12.44$ & $9.23$  \\
        Carboxylic acid   & $0.85$  & $0.51$  & Primary amines    & $9.35$  & $10.35$ \\
        Ester             & $3.44$  & $2.40$  & Secondary amines  & $28.73$ & $32.39$
        \end{tabular}
    \end{ruledtabular}
\end{table*}

\begin{figure}[h] 
    \setlength{\unitlength}{0.01\textwidth} 
    \begin{picture}(100,87)
    \put(0,75){\includegraphics[width=0.16\textwidth]{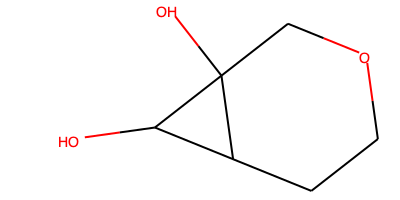}}
    \put(16,75){\includegraphics[width=0.16\textwidth]{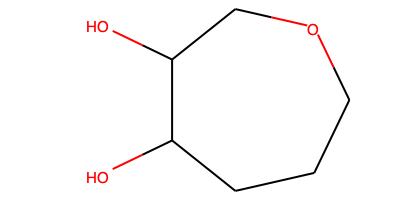}}
    \put(32,75){\includegraphics[width=0.16\textwidth]{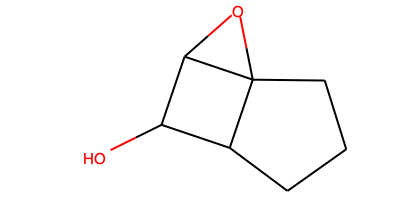}}
    \put(48,75){\includegraphics[width=0.16\textwidth]{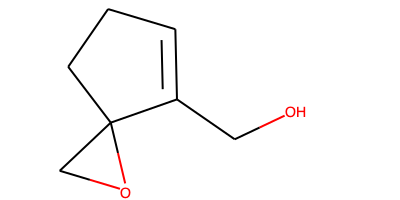}}
    \put(64,75){\includegraphics[width=0.16\textwidth]{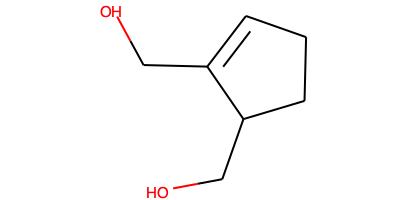}}
    \put(80,75){\includegraphics[width=0.16\textwidth]{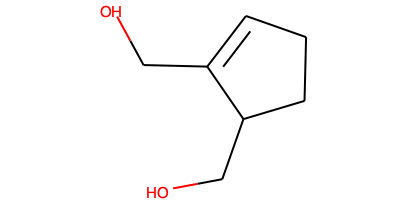}}
    
    \put(0,60){\includegraphics[width=0.16\textwidth]{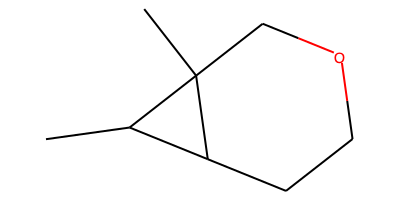}}
    \put(16,60){\includegraphics[width=0.16\textwidth]{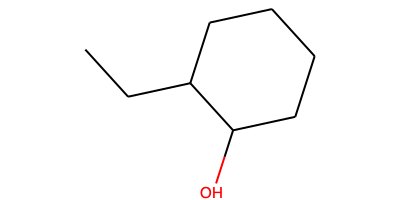}}
    \put(32,60){\includegraphics[width=0.16\textwidth]{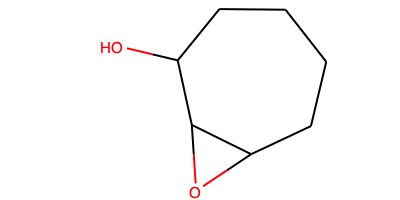}}
    \put(48,60){\includegraphics[width=0.16\textwidth]{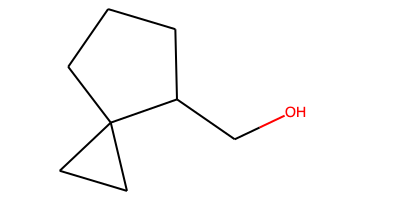}}
    \put(64,60){\includegraphics[width=0.16\textwidth]{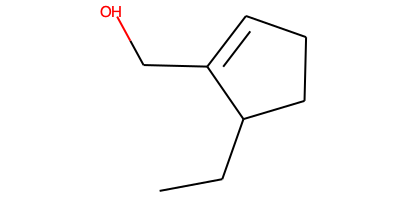}}
    \put(80,60){\includegraphics[width=0.16\textwidth]{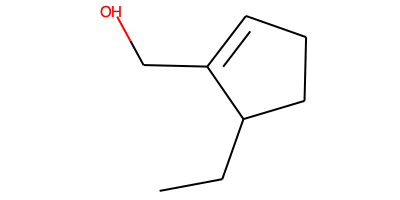}}
    
    \put(0,45){\includegraphics[width=0.16\textwidth]{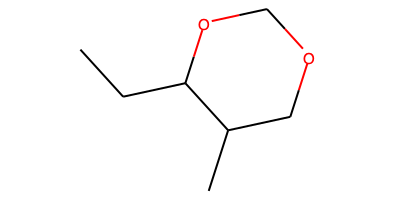}}
    \put(16,45){\includegraphics[width=0.16\textwidth]{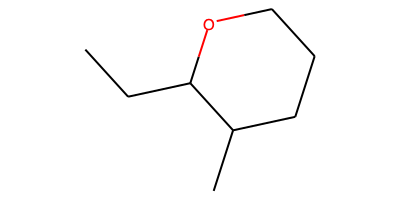}}
    \put(32,45){\includegraphics[width=0.16\textwidth]{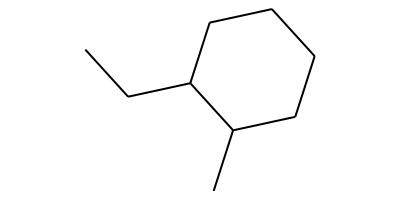}}
    \put(48,45){\includegraphics[width=0.16\textwidth]{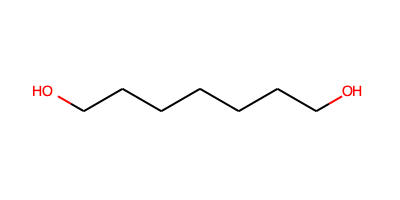}}
    \put(64,45){\includegraphics[width=0.16\textwidth]{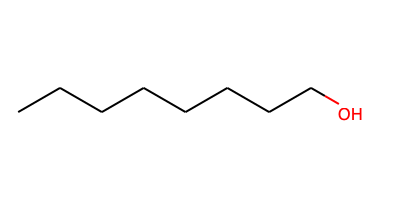}}
    \put(80,45){\includegraphics[width=0.16\textwidth]{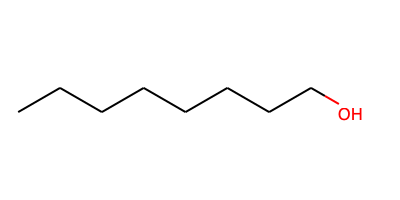}}
    
    \put(0,30){\includegraphics[width=0.16\textwidth]{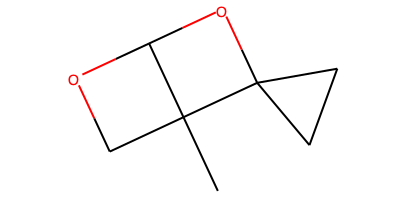}}
    \put(16,30){\includegraphics[width=0.16\textwidth]{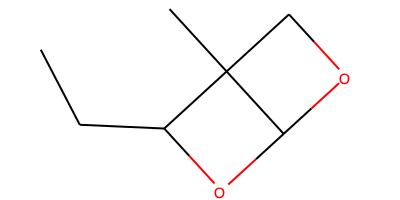}}
    \put(32,30){\includegraphics[width=0.16\textwidth]{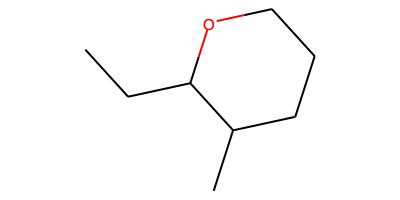}}
    \put(48,30){\includegraphics[width=0.16\textwidth]{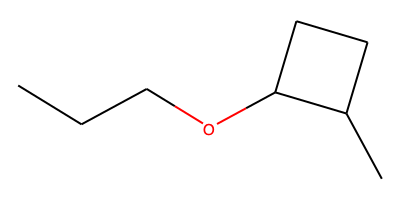}}
    \put(64,30){\includegraphics[width=0.16\textwidth]{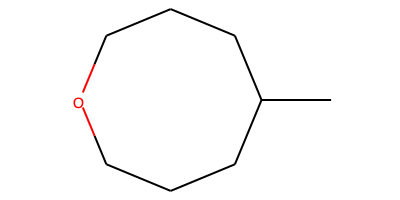}}
    \put(80,30){\includegraphics[width=0.16\textwidth]{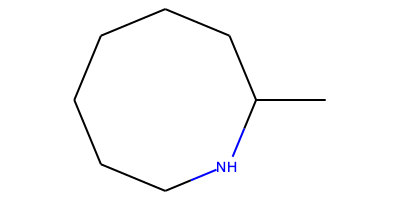}}
    
    \put(0,15){\includegraphics[width=0.16\textwidth]{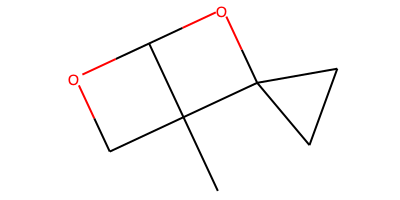}}
    \put(16,15){\includegraphics[width=0.16\textwidth]{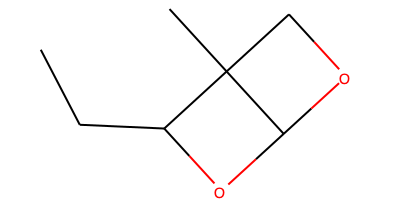}}
    \put(32,15){\includegraphics[width=0.16\textwidth]{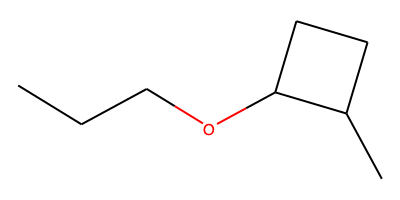}}
    \put(48,15){\includegraphics[width=0.16\textwidth]{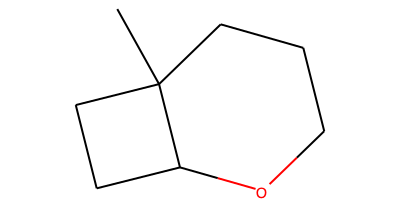}}
    \put(64,15){\includegraphics[width=0.16\textwidth]{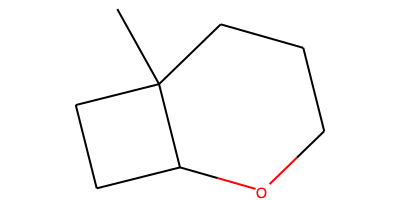}}
    \put(80,15){\includegraphics[width=0.16\textwidth]{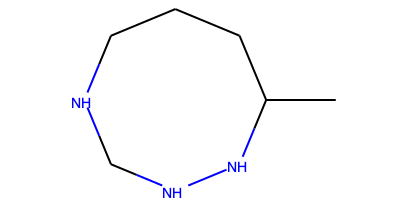}}
    
    \put(0,0){\includegraphics[width=0.16\textwidth]{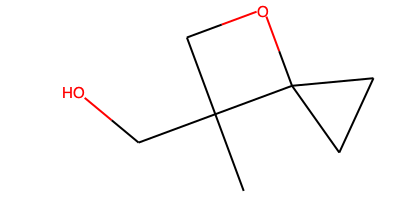}}
    \put(16,0){\includegraphics[width=0.16\textwidth]{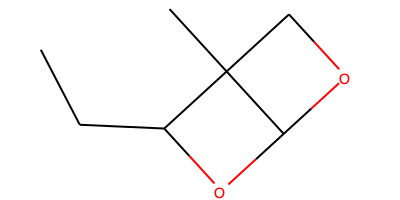}}
    \put(32,0){\includegraphics[width=0.16\textwidth]{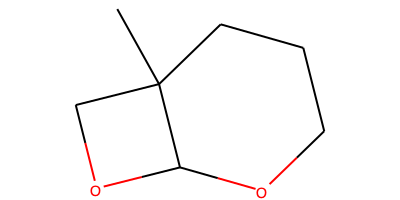}}
    \put(48,0){\includegraphics[width=0.16\textwidth]{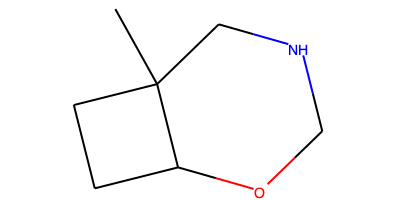}}
    \put(64,0){\includegraphics[width=0.16\textwidth]{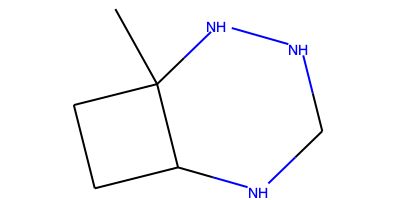}}
    \put(80,0){\includegraphics[width=0.16\textwidth]{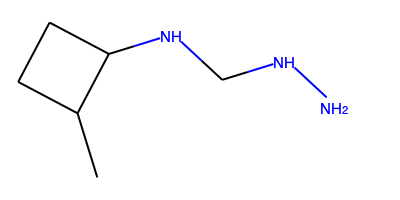}}
    
    \end{picture}
    \caption{Lewis representations of the molecules generated by mapping a set of latent space grid points to their graph representations using the decoder $p_{\boldsymbol{\theta}}(\mathcal{G}|\boldsymbol{z})$ of the constrained model.\label{fig:latent_smooth}}
\end{figure}

Next, we examine the latent space's smoothness in the constrained model by sampling grid points on a plane in this space (as described for Fig.~\ref{fig:latent_map}) and mapping them to their molecular graph representations. We train the network with a dataset $\mathscr{G}$ of size $K=600$ and use the regularization approach~\cite{ma2018constrained} to impose validity constraints on connectivity and valency of the molecule, as well as physical constraints on $3-$member cycles and cycles with triple bonds to encourage energetically more stable samples. Different constraints are defined at different levels for each node, each edge, or the whole graph. We use a single parameter $\mu$ for each type of constraint $\mathcal{C}$. We tune the four regularization parameters until the desired outputs in terms of validity and energetic stability are achieved. Comparing the generated molecules (Fig.~\ref{fig:latent_smooth}), we see a smooth transition from one molecule to the other for each row and column. Results for the constrained model are obtained with $\mu=\{0.8, 0.4, 0.8, 1000\}$ for the valence, connectivity, $3-$member cycle, and cycle with triple bond constraints, respectively.

\begin{figure}[H] 
    \setlength{\unitlength}{0.01\textwidth} 
    \begin{picture}(100,87)
    \put(2,75){\includegraphics[width=0.18\textwidth]{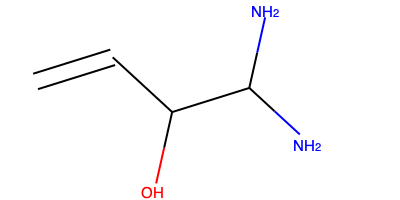}}
    \put(21,75){\includegraphics[width=0.18\textwidth]{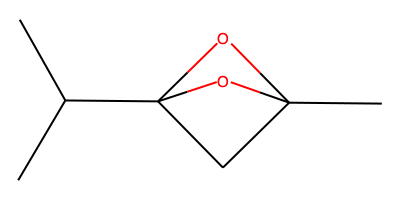}}
    \put(40,75){\includegraphics[width=0.18\textwidth]{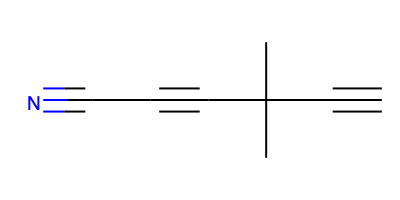}}
    \put(59,75){\includegraphics[width=0.18\textwidth]{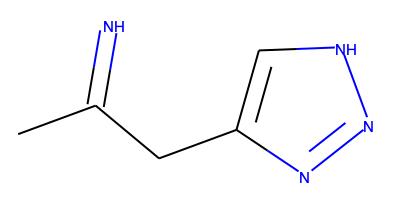}}
    \put(78,75){\includegraphics[width=0.18\textwidth]{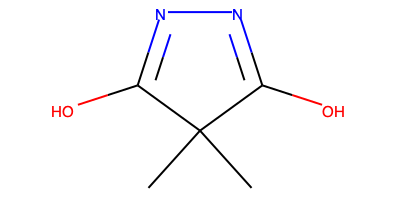}}
    
    \put(2,60){\includegraphics[width=0.18\textwidth]{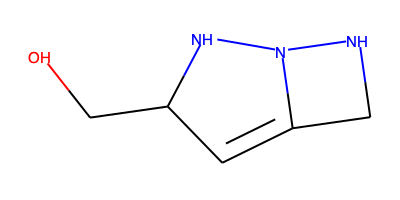}}
    \put(21,60){\includegraphics[width=0.18\textwidth]{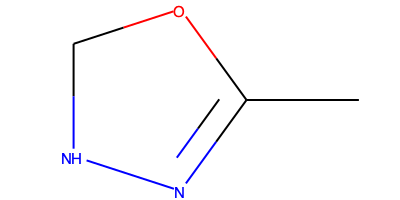}}
    \put(40,60){\includegraphics[width=0.18\textwidth]{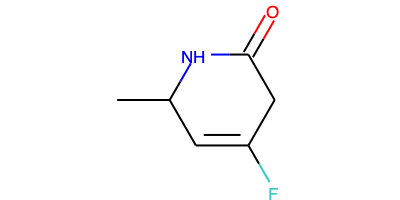}}
    \put(59,60){\includegraphics[width=0.18\textwidth]{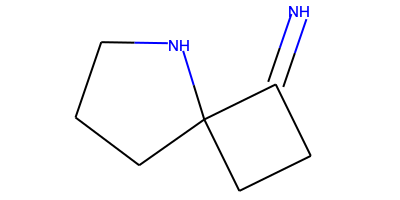}}
    \put(78,60){\includegraphics[width=0.18\textwidth]{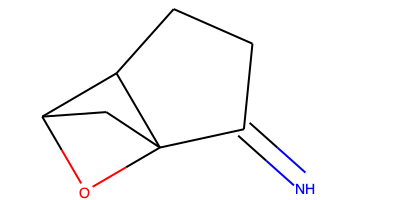}}
    
    \put(2,45){\includegraphics[width=0.18\textwidth]{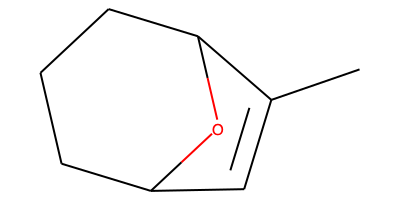}}
    \put(21,45){\includegraphics[width=0.18\textwidth]{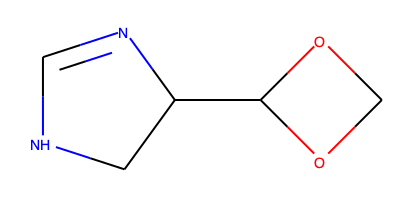}}
    \put(40,45){\includegraphics[width=0.18\textwidth]{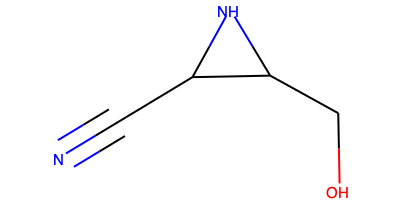}}
    \put(59,45){\includegraphics[width=0.18\textwidth]{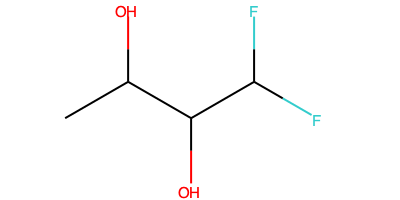}}
    \put(78,45){\includegraphics[width=0.18\textwidth]{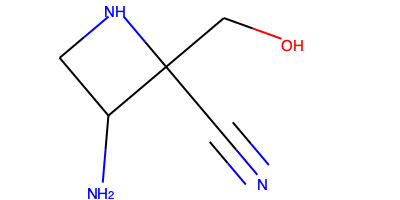}}
    
    \put(2,30){\includegraphics[width=0.18\textwidth]{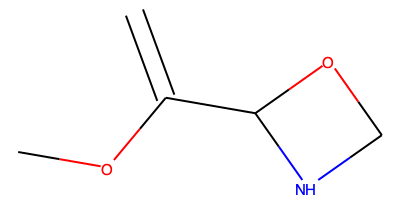}}
    \put(21,30){\includegraphics[width=0.18\textwidth]{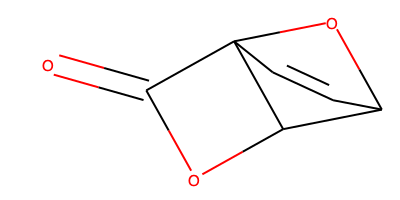}}
    \put(40,30){\includegraphics[width=0.18\textwidth]{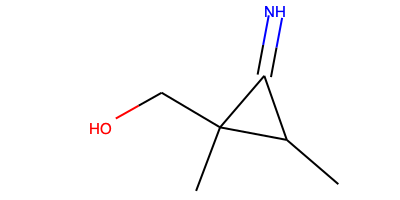}}
    \put(59,30){\includegraphics[width=0.18\textwidth]{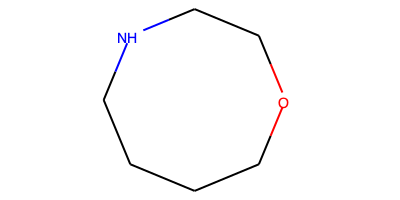}}
    \put(78,30){\includegraphics[width=0.18\textwidth]{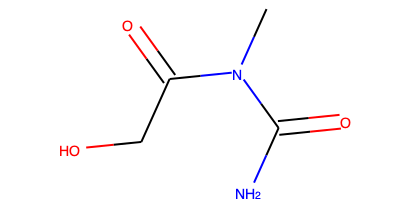}}
    
    \put(2,15){\includegraphics[width=0.18\textwidth]{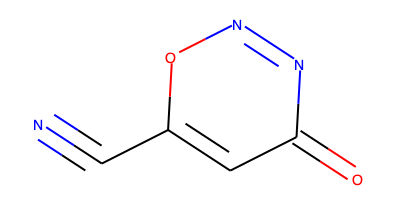}}
    \put(21,15){\includegraphics[width=0.18\textwidth]{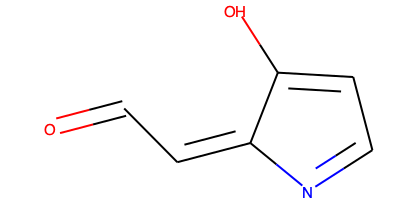}}
    \put(40,15){\includegraphics[width=0.18\textwidth]{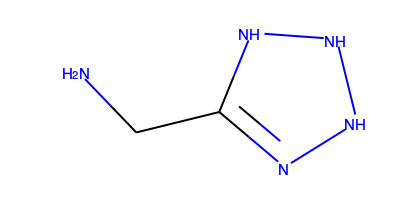}}
    \put(59,15){\includegraphics[width=0.18\textwidth]{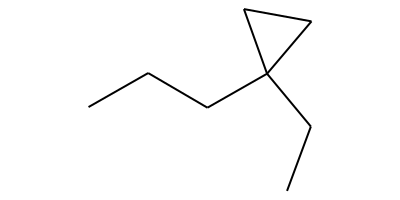}}
    \put(78,15){\includegraphics[width=0.18\textwidth]{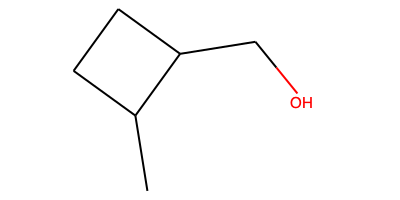}}
    
    \put(2,0){\includegraphics[width=0.18\textwidth]{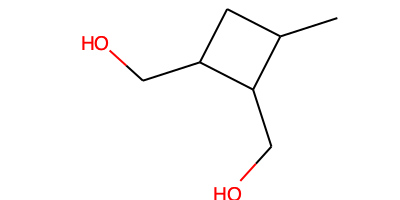}}
    \put(21,0){\includegraphics[width=0.18\textwidth]{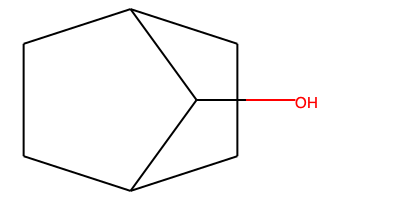}}
    \put(40,0){\includegraphics[width=0.18\textwidth]{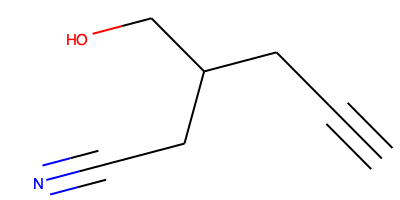}}
    \put(59,0){\includegraphics[width=0.18\textwidth]{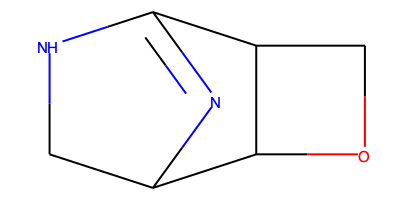}}
    \put(78,0){\includegraphics[width=0.18\textwidth]{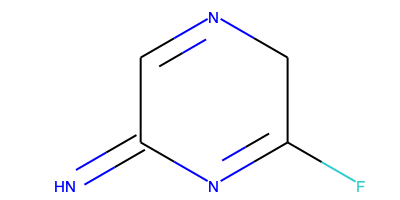}}
    
    \end{picture}
    \caption{Example of the Lewis representations of the valid molecules sampled using the constrained model with training set size $K=600$. Atom colors follow the CPK coloring convention~\cite{koltun1965space}.\label{fig:sample_mol}}
\end{figure}

As noted earlier, not all the molecular graphs sampled using the generative model have chemically valid structures. We use triple quality measures to assess the the generated molecules' quality, including validity score $H_{vld}$, uniqueness score $H_{unq}$, and novelty score $H_{vld}$. The validity score $H_{vld}$ indicates what percentage of the predicted graphs $\bar{\mathscr{G}}$ have a valid molecular structure, 
\begin{equation}
    \label{eq:val}
    H_{vld}=\frac{|valid(\bar{\mathscr{G}})|}{|\bar{\mathscr{G}}|}, 
\end{equation}
where $\bar{\mathscr{G}}=\{\bar{\mathcal{G}}^{(i)}\}_{i=1}^T$ denotes the collection of sampled molecules. Note that a valid molecule refers to a single connected graph with no violation of the atoms' valency. Fig.~\ref{fig:sample_mol} illustrates examples of valid molecules sampled using the generative model. Moreover, the uniqueness score $H_{unq}$ indicates what fraction of the molecular graphs are unique among the sampled outputs, i. e.
\begin{equation}
    \label{eq:unq}
    H_{unq}=\frac{\left|valid(\bar{\mathscr{G}}^{*})\right|}{|valid(\bar{\mathscr{G}})|}.
\end{equation}
Here, $\bar{\mathscr{G}}^{*}$ shows the set of sampled molecules. Lastly, the novelty score $H_{nvl}$ indicates if the generated molecular graphs $\bar{\mathscr{G}}$ were present in the training dataset $\mathscr{G}=\{\mathcal{G}^{(i)}\}_{i=1}^K$ or if they are novel molecules, i. e.
\begin{equation}
    \label{eq:nov}
    H_{nvl}=\frac{|valid(\bar{\mathscr{G}})|-|\mathscr{G} \cap valid(\bar{\mathscr{G}})|}{|valid(\bar{\mathscr{G}})|}.
\end{equation}

In Table~\ref{tab:qualmetric}, we examine the quality scores for $T=10000$ molecules generated by the constrained and base models with training dataset size $K=600$. We have distinguished the percentage of the molecules with different validity issues, namely connectivity and valency. Note that the number of valid molecules and molecules with each validity issue might not sum to $10000$ as some molecules might have both issues. We observe that enforcing the constraints has increased the output molecules' validity collectively and individually for each constraint. At the same time, we see a decrease in the number of cycles containing triple bonds and $3-$member cycles. We note that we observed a decrease in the uniqueness score and an increase in the novelty score using this regularization approach. As in~\cite{ma2018constrained}, we choose the regularization parameters using a trial and error approach.

\begin{table*}[h]
    \caption{\label{tab:qualmetric}Quality metrics for graph scattering variational autoencoder.}
    \begin{ruledtabular}
        \begin{tabular}{lccccccc}
        \multicolumn{1}{c}{\multirow{2}{*}{Type}} & \multicolumn{3}{c}{$H_{val}$}   & \multirow{2}{*}{\begin{tabular}[c]{@{}c@{}}$3-$member\\ cycle\end{tabular}} & \multirow{2}{*}{\begin{tabular}[c]{@{}c@{}}Triple bond\\ cycle\end{tabular}} & \multirow{2}{*}{$H_{unq}$} & \multirow{2}{*}{$H_{nvl}$} \\ \cline{2-4}
        \multicolumn{1}{c}{} & Total & Valency & Connectivity & & & & \\\hline
        Base & $77.8\%$ & $15.9\%$ & $7.2\%$ & $3910$ & $113$ & $77.7\%$ & $86.3\%$ \\
        Constrained & $89.9\%$ & $5.6\%$ & $4.6\%$ & $1745$ & $35$ & $71.1\%$ & $95.8\%$
        \end{tabular}
    \end{ruledtabular}
\end{table*}

Table~\ref{tab:qualmetricbenchmarks} further compares the present work with benchmark models. In all cases, quality metrics are computed for $10$k sampled molecules. In many examples, the need for large training data is restrictive, particularly when data generation is expensive. To that end, we have focused on the problem where only a limited number of training data is available. Nevertheless, the results compare very well with the models that take advantage of the full dataset. In uniqueness, the model performs better than the baseline models, while it is second only to MolGAN~\cite{de2018molgan} in validity score. We believe that a balance between the model's quality scores is preferred over a high score in one category and a low score in the other, as in reference~\onlinecite{de2018molgan}. In novelty, about $86$ percent of the generated samples are out of the dataset, which increases to about $96$ percent for the constrained model. Furthermore, we study the performance of the model in the large data regime using a dataset of size $K=10000$. We observe that using more data improves all the quality metrics for the sampled molecular structures. It is noteworthy that while some autoregressive models have reached perfect validity scores, they generally do so by some stoppage criteria, which cease adding components right before a molecule becomes invalid. This focuses more on enforcing validity measures during sampling rather than during training. Hence, these models are not compared here.

\begin{table}[h]
    \caption{\label{tab:qualmetricbenchmarks}Comparison of quality metrics for different molecular generative models.}
    \begin{ruledtabular}
    \begin{tabular}{lccc}
        \multicolumn{1}{c}{Method}   & $H_{val}$     & $H_{unq}$   & $H_{nvl}$  \\ \hline
        CVAE~\cite{gomez2018automatic}\footnotemark[1]         & $10.3\%$  & $67.5\%$  & $90.0\%$\\
        GVAE~\cite{kusner2017grammar}\footnotemark[1]          & $60.2\%$  & $9.3\%$   & $80.9\%$\\
        GraphVAE~\cite{simonovsky2018graphvae}\footnotemark[1] & $55.7\%$  & $76.0\%$  & $61.6\%$\\
        MolGAN~\cite{de2018molgan}\footnotemark[1]             & $98.1\%$  & $10.4\%$  & $94.2\%$\\
        GSVAE\footnotemark[2]                                                 & $77.8\%$  & $77.7\%$  & $86.3\%$\\
        GSVAE\footnotemark[3]                                                 & $87.1\%$  & $82.6\%$  & $92.7\%$
    \end{tabular}
    \end{ruledtabular}
    \footnotetext[1]{Baseline values are reported from Ref.~\onlinecite{de2018molgan}.}
    \footnotetext[2]{Trained with small dataset $K=600$}
    \footnotetext[3]{Trained with large dataset $K=10000$}
\end{table}

\subsection{\label{subsec:UQ_res}Uncertainty of the model}

In this section, we use the model developed in Section~\ref{sec:UQ} to compute the predictive estimates of physicochemical molecular properties. We begin by selecting properties that can be readily estimated from the generated molecules using an empirical or exact relationship. These include i) the octanol-water partition coefficient, which is computed using an atomic contribution scheme~\cite{wildman1999prediction}, and ii) polar surface area~\cite{ertl2000fast}. We sample molecules from the distribution $p_{\boldsymbol{\theta}}(\mathcal{G})$ and compute predictive estimates of their properties. The property values for each sample $\bar{\mathcal{G}}$ are computed using RDKit~\cite{landrum2016rdkit}, which is an open-source cheminformatics software.

\begin{figure}[h]
    \centering
    \subfigure[$K=50$]{
      \includegraphics[width=0.32\textwidth]{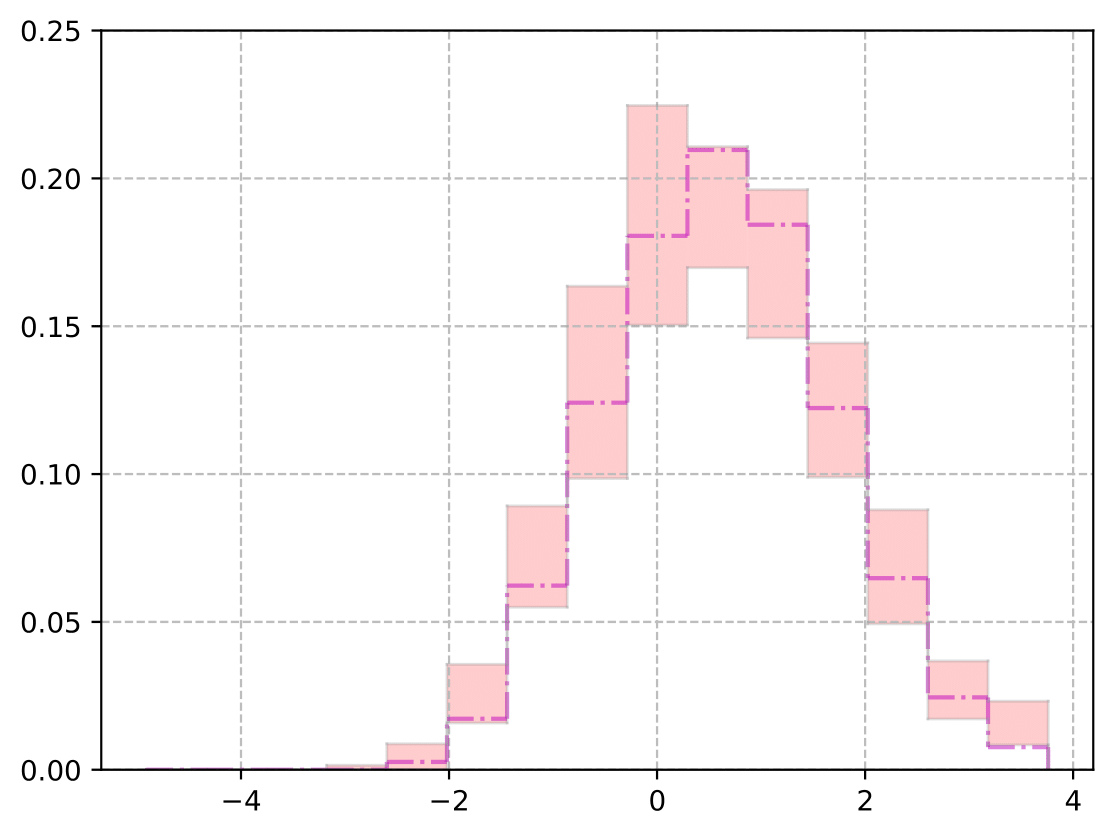}}
    \subfigure[$K=200$]{
      \includegraphics[width=0.32\textwidth]{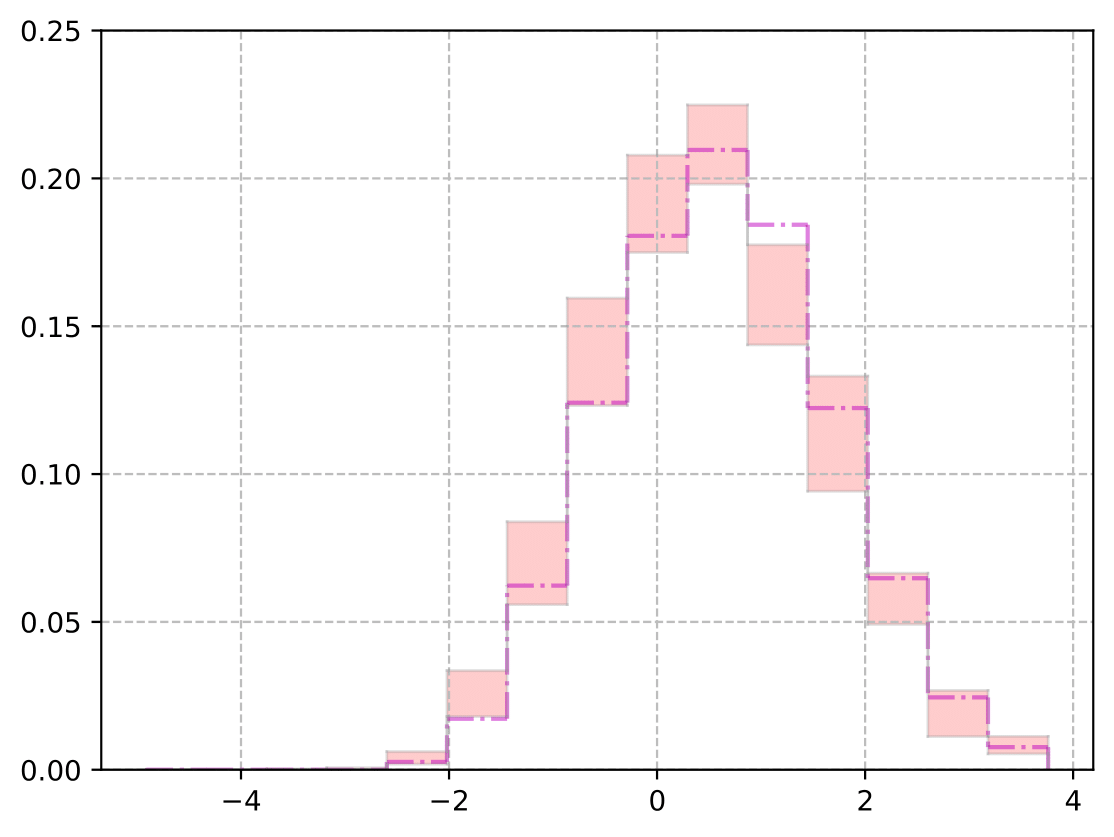}}
    \subfigure[$K=500$]{
      \includegraphics[width=0.32\textwidth]{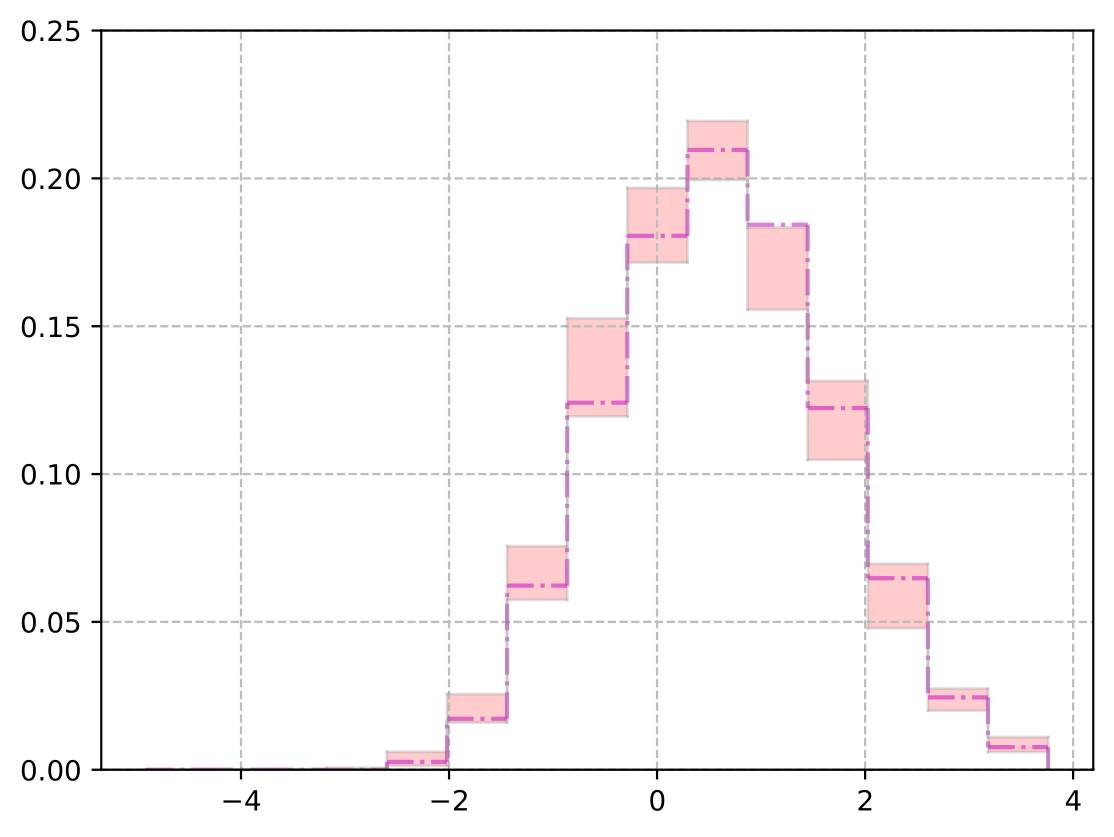}}
    \caption{Predicted octanol-water partition coefficients with latent space dimension $J=30$ for various sizes $K$ of the training dataset from the QM9 database. The reference solution, indicated in magenta ({\color{magenta}\dashdotted}), is estimated by sampling from the model $p_{\boldsymbol{\theta}}(\mathcal{G})$ trained by $K=10000$ molecules. The shaded area represents the $1\%-99\%$ credible interval, reflecting the induced epistemic uncertainty from the limited amount of the training data.} \label{fig:UQ_logP}
\end{figure}

\begin{figure}[h]
    \centering
    \subfigure[$K=50$]{
      \includegraphics[width=0.32\textwidth]{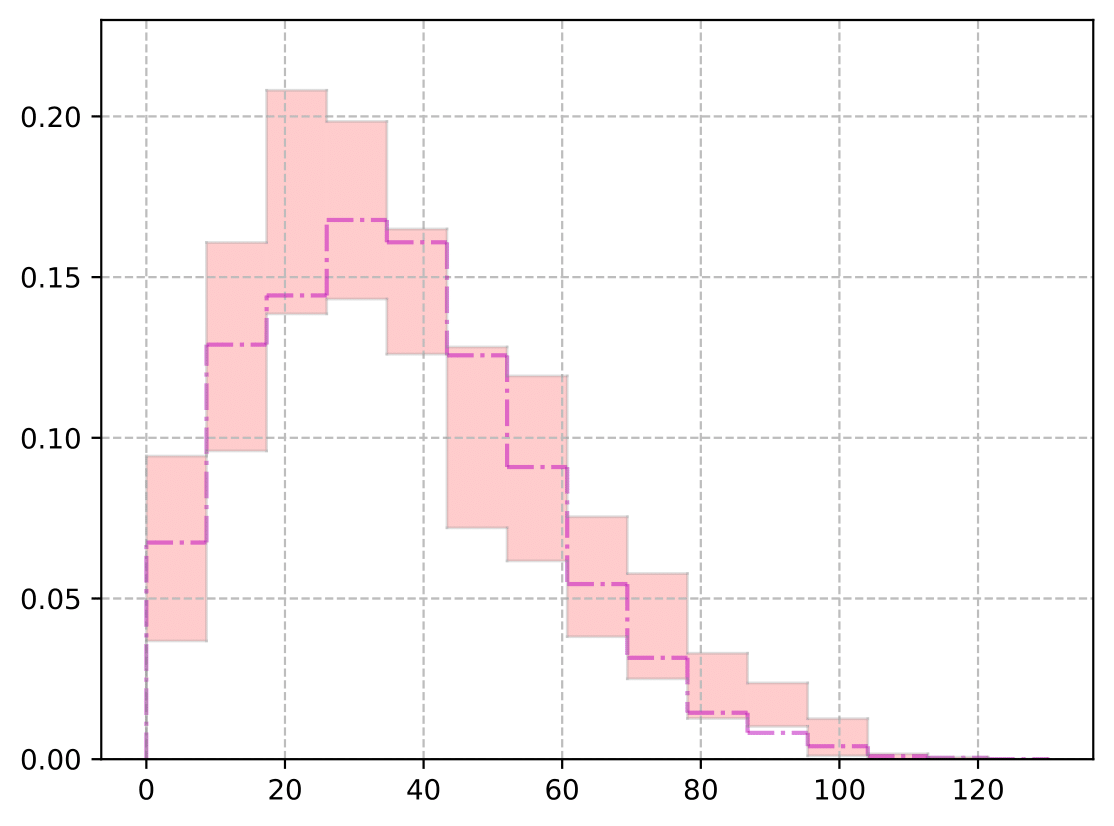}}
    \subfigure[$K=200$]{
      \includegraphics[width=0.32\textwidth]{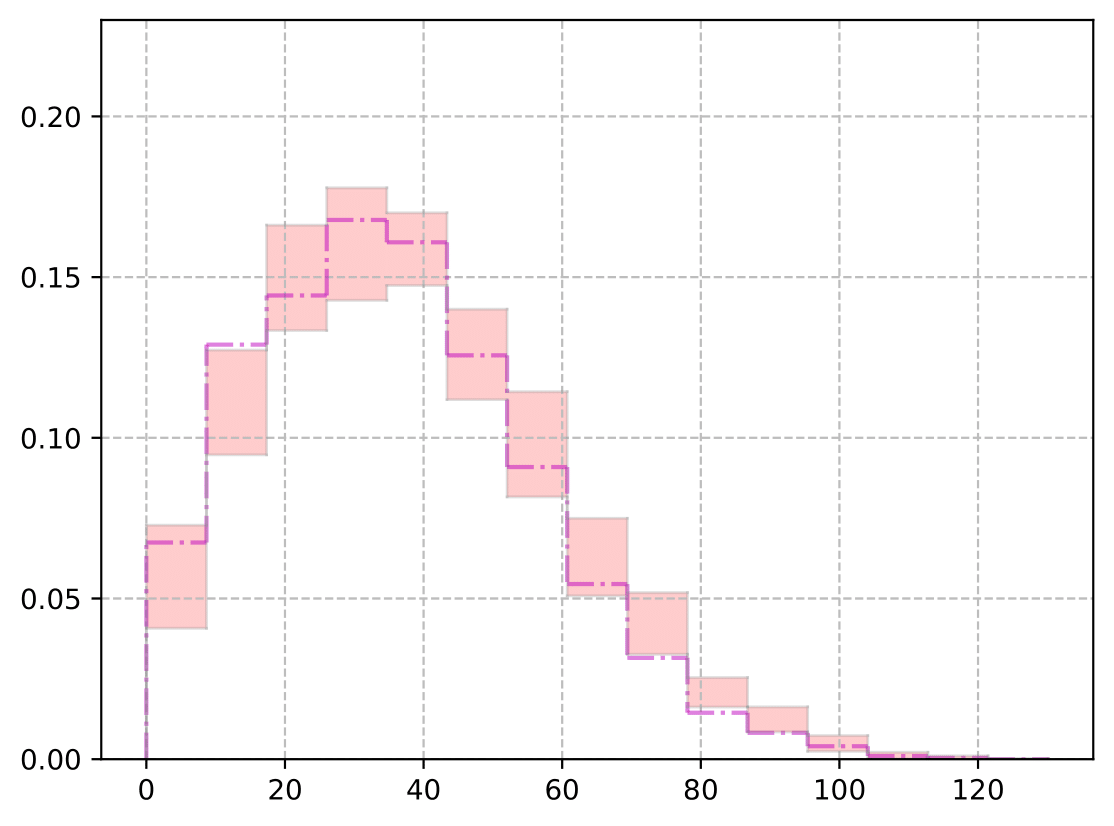}}
    \subfigure[$K=500$]{
      \includegraphics[width=0.32\textwidth]{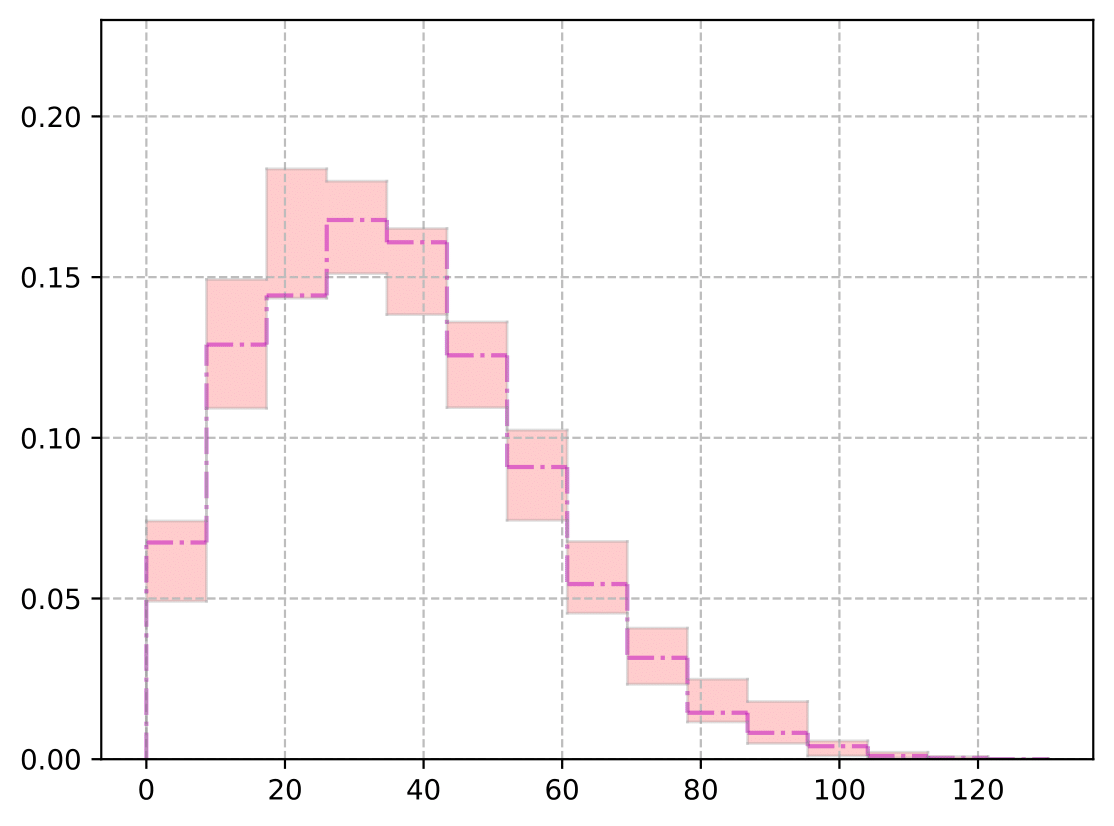}}
    \caption{Predicted polar surface area [\AA$^2$] with $J=30$ dimensional latent space for various training dataset sizes $K$ from the QM9 database. The reference solution ({\color{magenta}\dashdotted}) is estimated by samples from $p_{\boldsymbol{\theta}}(\mathcal{G})$ trained with $K=10000$ molecules. The shaded area represents the $1\%-99\%$ credible interval, which indicates the epistemic uncertainty induced by the limited training size $K$.} \label{fig:UQ_PSA}
\end{figure}

Furthermore, we can use samples from the predictive distribution to estimate credible intervals over physicochemical molecular properties, as detailed in Algorithm~\ref{alg:BBPD}. Figs.~\ref{fig:UQ_logP} and~\ref{fig:UQ_PSA} illustrate the computed error bars over the octanol-water partition coefficient and polar surface area, respectively. In these figures, we have trained the model with three different small training dataset sizes $K=\{50, 200, 500\}$ and simulated approximately from the posterior distribution using WLB to construct the credible intervals. To this end, we sample $10000$ molecules from the model and use the valid molecules to plot the error bars. We compare this to the reference solution computed using $K=10000$ training data. We observe that the reference solution falls within the shaded credible interval, representing the model's epistemic uncertainty. The probabilistic confidence over estimated properties increases as we increase the size of the training dataset.

\subsection{\label{subsec:cnd_dsgn}Conditional design}

One of the popular applications of generative models, particularly in drug discovery, is to design molecules with desired properties. In this section, we examine the performance of the model for a conditional generative task. Additionally, we provide error bars around the probabilistic estimates of the properties. These error bars reflect the epistemic uncertainty in the model.

In a conditional generative task, instead of training the model only on the molecular graphs, we simultaneously train the model on the molecules and their corresponding properties. To that end, the encoding network $q_{\boldsymbol{\phi}}$ represents the conditional distribution of the latent variable $\boldsymbol{z}$ given the graph $\mathcal{G}$ and its property labels $\boldsymbol{y}$. In other words, Eq.~(\ref{eq:var_approx}) takes the form
\begin{equation}
    \label{eq:var_approx_cond}
	q_{\boldsymbol{\phi}}(\boldsymbol{z}|\mathcal{G}, \boldsymbol{y})=\mathcal{N}(\boldsymbol{z};\boldsymbol{\mu}_{\boldsymbol{\phi}}(\mathcal{G}, \boldsymbol{y}), \boldsymbol{S}_{\boldsymbol{\phi}}(\mathcal{G}, \boldsymbol{y})).
\end{equation}

Labels $\boldsymbol{y}$ are graph-level properties, while the input signals $\boldsymbol{f}$ to the scattering layers contain node-level values; that is, the type of the atoms on each node $v_i$. To accommodate that, we add the property label as an additional signal dimension to every node on the graph. This increases the size of the signal on each node from $5$ (one-hot vector of atom types) to $5+Y$, where $Y$ denotes the number of molecular properties or size of the label vector $|\boldsymbol{y}|$.

Similarly, the decoder $p_{\boldsymbol{\theta}}$ takes a sample $\boldsymbol{z}$ from the latent space combined with molecular property labels $\boldsymbol{y}$ and maps them to a molecular representation $\mathcal{G}$. Hence, we can reformulate Eq.~(\ref{eq:graph_model}) as
\begin{equation}
    \label{eq:graph_model_cond}
	p_{\theta}(\mathcal{G}|\boldsymbol{z}, \boldsymbol{y})=p_{\theta}(\boldsymbol{W}, \boldsymbol{f}|\boldsymbol{z}, \boldsymbol{y})=p_{\theta}(\boldsymbol{f}|\boldsymbol{z}, \boldsymbol{W}, \boldsymbol{y})p_{\theta}(\boldsymbol{W}|\boldsymbol{z}, \boldsymbol{y}).
\end{equation}
Using Eqs.~(\ref{eq:var_approx_cond}) and~(\ref{eq:graph_model_cond}), we reformulate the evidence lower-bound (Eq.~(\ref{eq:ELBO})) for training the model for the conditional design
\begin{align}
    \label{eq:ELBO_dsgn}
    \begin{split}
        \mathscr{L}_{ELBO}(\boldsymbol{\theta}, \boldsymbol{\phi} ; \mathscr{G}, \boldsymbol{y})
        =&\sum_{i=1}^{K}\mathbb{E}_{q_{\boldsymbol{\phi}}(\boldsymbol{z}^{(i)} | \mathcal{G}^{(i)}, y^{(i)})}\left[\log p_{\boldsymbol{\theta}}(\mathcal{G}^{(i)} | \boldsymbol{z}^{(i)}, y^{(i)})\right]\\
        &-\sum_{i=1}^{K}\mathrm{D_{KL}}\left[q_{\boldsymbol{\phi}}(\boldsymbol{z}^{(i)} | \mathcal{G}^{(i)}, y^{(i)}) \| p_{\boldsymbol{\theta}}(\boldsymbol{z}^{(i)})\right].
    \end{split}
\end{align}
In Table~\ref{tab:model_spec_cond}, we summarize the required changes to the architecture for the conditional generative model. The rest of the layers have the same dimensionality as stated in Table~\ref{tab:model_spec}. 

In this model, new molecules are generated by sampling from $p(\boldsymbol{z})$ and $p(\boldsymbol{y})=\mathcal{N}(\boldsymbol{y};\boldsymbol{\mu}_{\boldsymbol{y}}, \boldsymbol{S}_{\boldsymbol{y}})$, where $\boldsymbol{\mu}_{\boldsymbol{y}}$ and $\boldsymbol{S}_{\boldsymbol{y}}$ are computed from the training data, and mapping them to the molecular graph domain using Eq.~(\ref{eq:graph_model_cond}). Given a target value $y_i$, we sample the rest of the property values from $p(y_{l,l\neq i}|y_{i})$.

\begin{figure}[h]
    \centering
    \subfigure[$K=50$]{
      \includegraphics[width=0.32\textwidth]{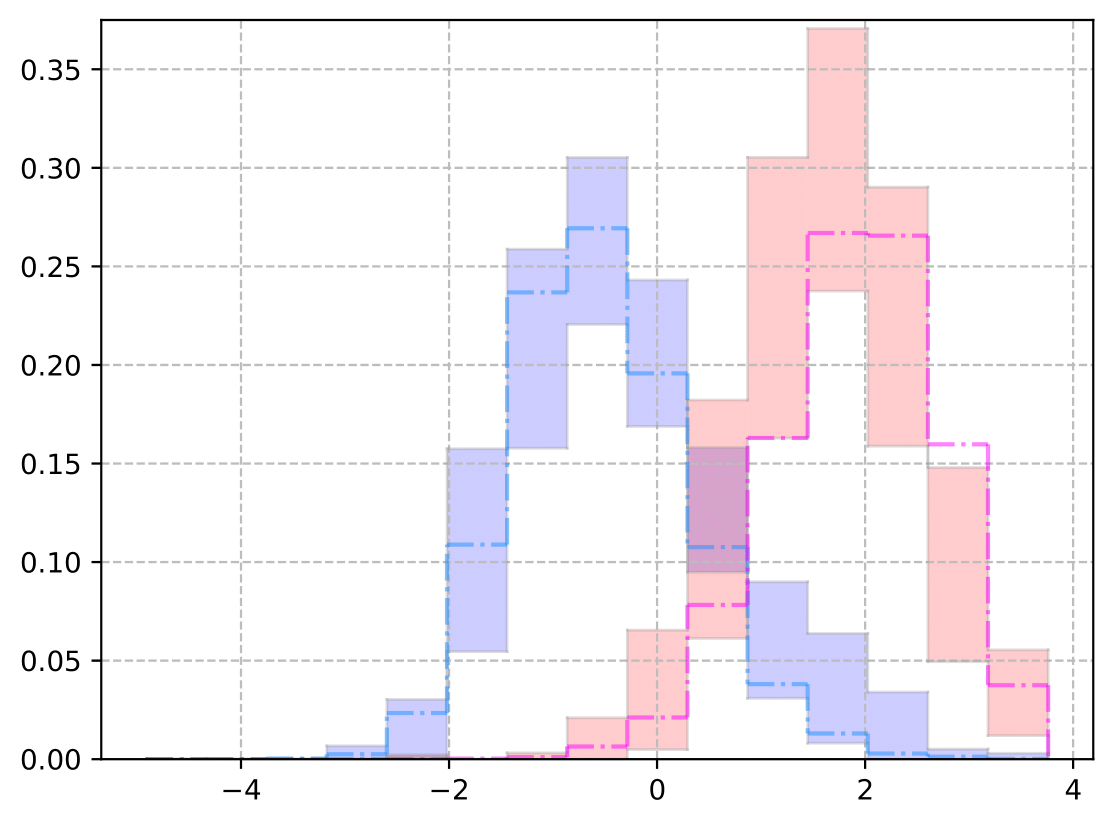}}
    \subfigure[$K=200$]{
      \includegraphics[width=0.32\textwidth]{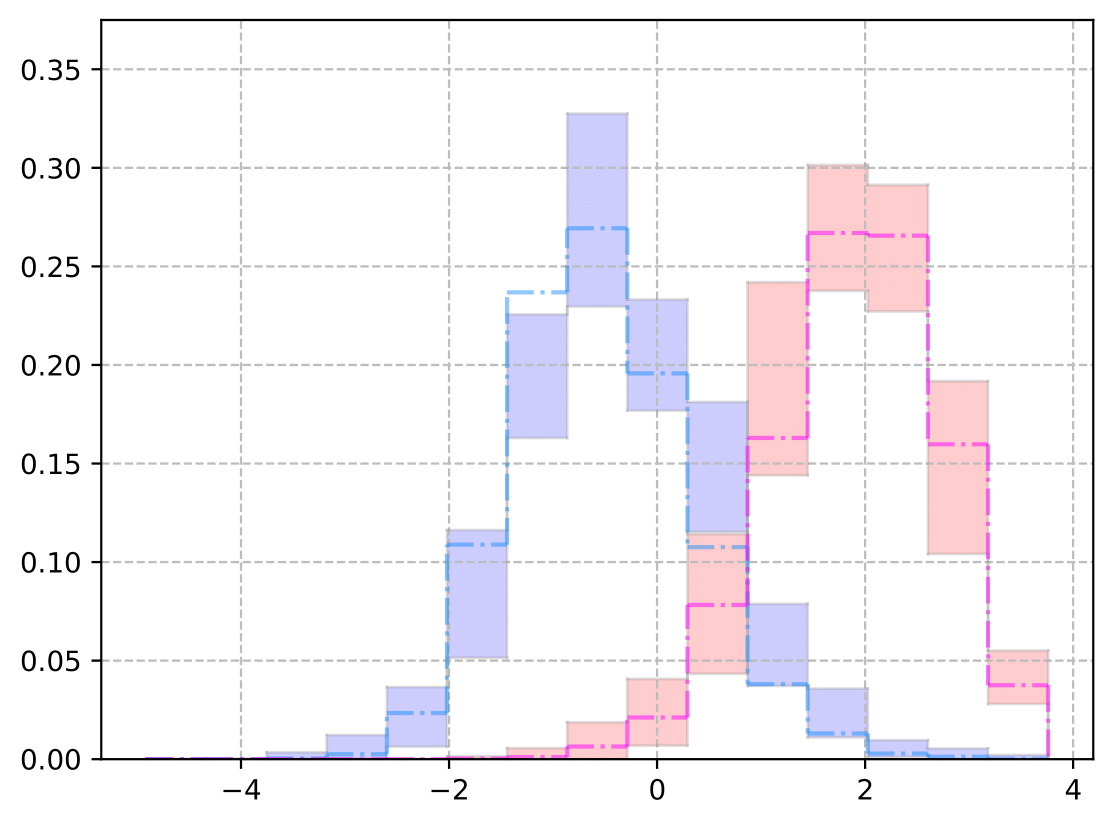}}
    \subfigure[$K=500$]{
      \includegraphics[width=0.32\textwidth]{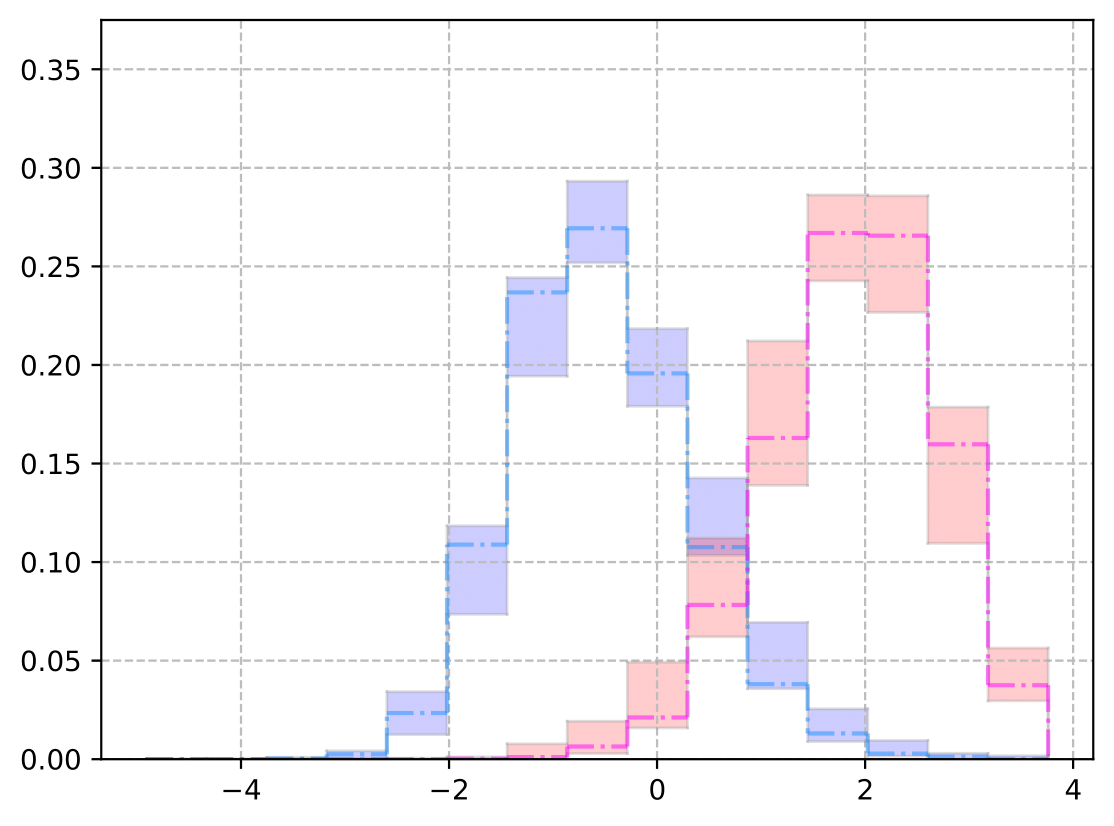}}
    \caption{Predicted octanol-water partition coefficient for the conditional design. We use a $30-$dimensional latent space and train the model with various training sizes $K=\{50, 200, 500\}$ from the QM9 database. The reference solutions, indicated in blue ({\color{ProcessBlue}\dashdotted}) for $y=-1$ and magenta ({\color{magenta}\dashdotted}) for $y=2$, are estimated by training the model with $K = 10000$ molecules. The $1\%-99\%$ credible interval is represented as error bars around the reference solution and captures the model's uncertainty.} \label{fig:UQ_cond_LogP}
\end{figure}

\begin{figure}[h]
    \centering
    \subfigure[$K=50$]{
      \includegraphics[width=0.32\textwidth]{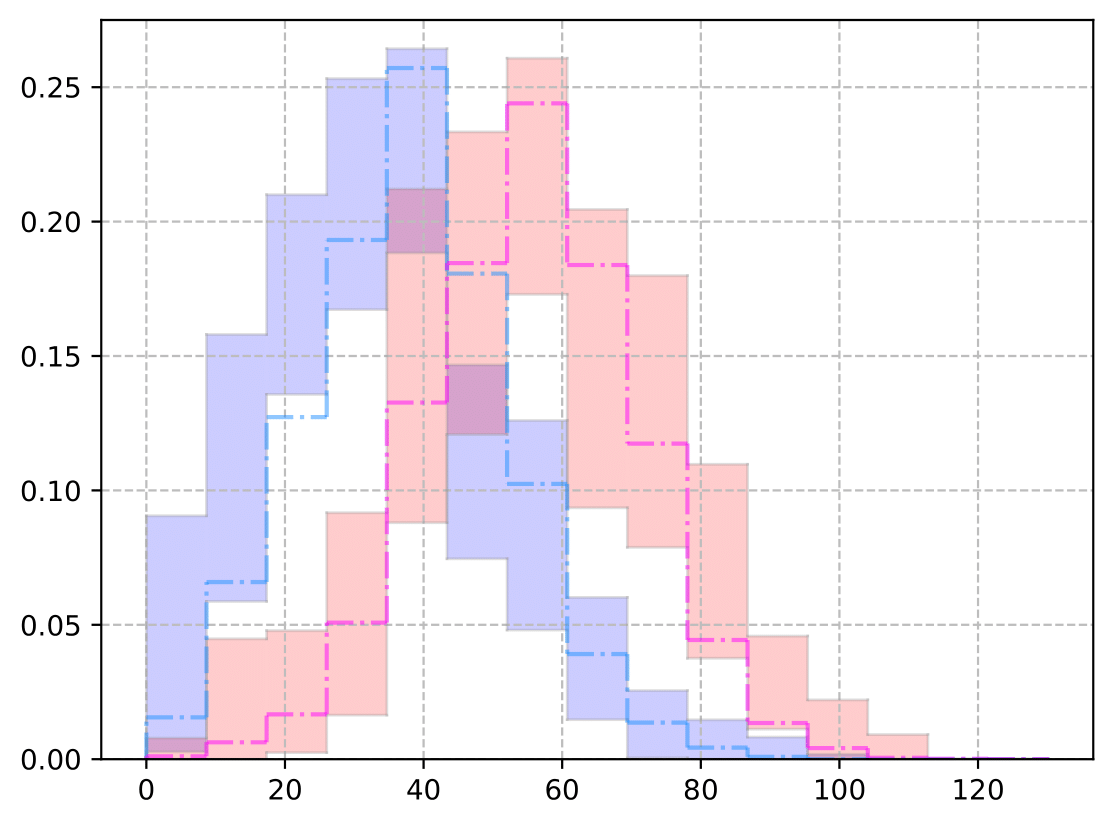}}
    \subfigure[$K=200$]{
      \includegraphics[width=0.32\textwidth]{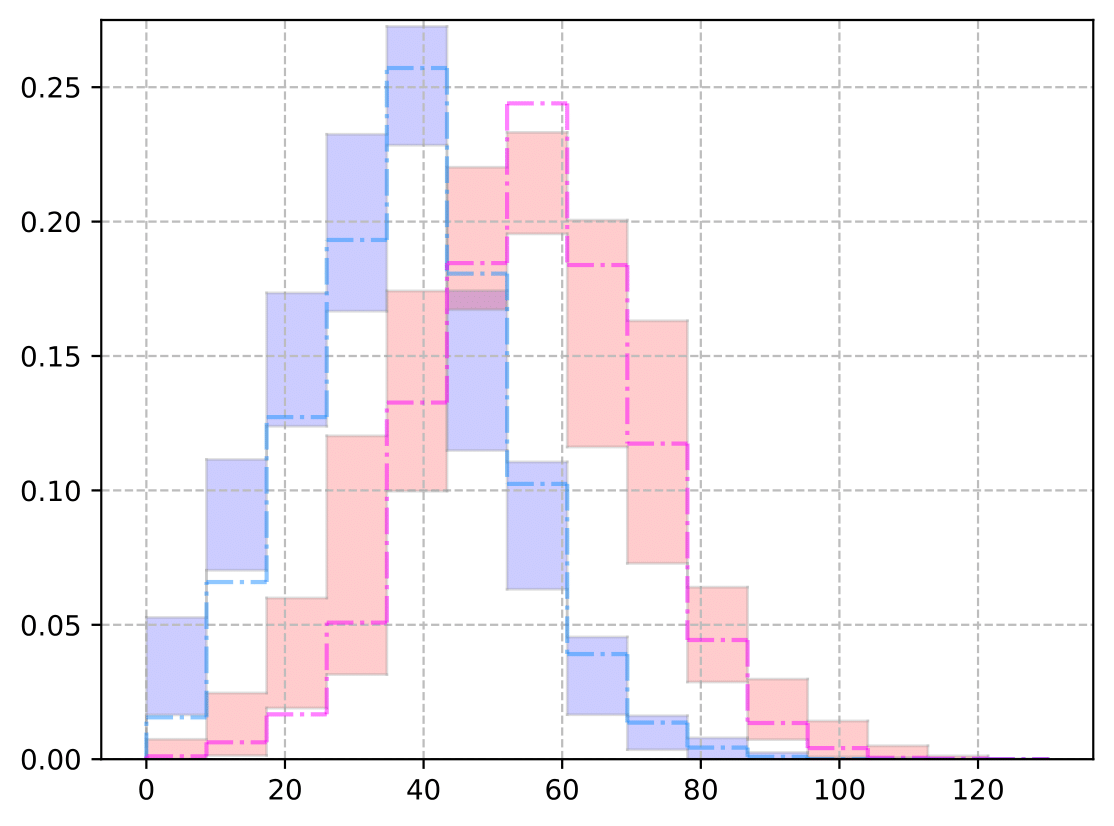}}
    \subfigure[$K=500$]{
      \includegraphics[width=0.32\textwidth]{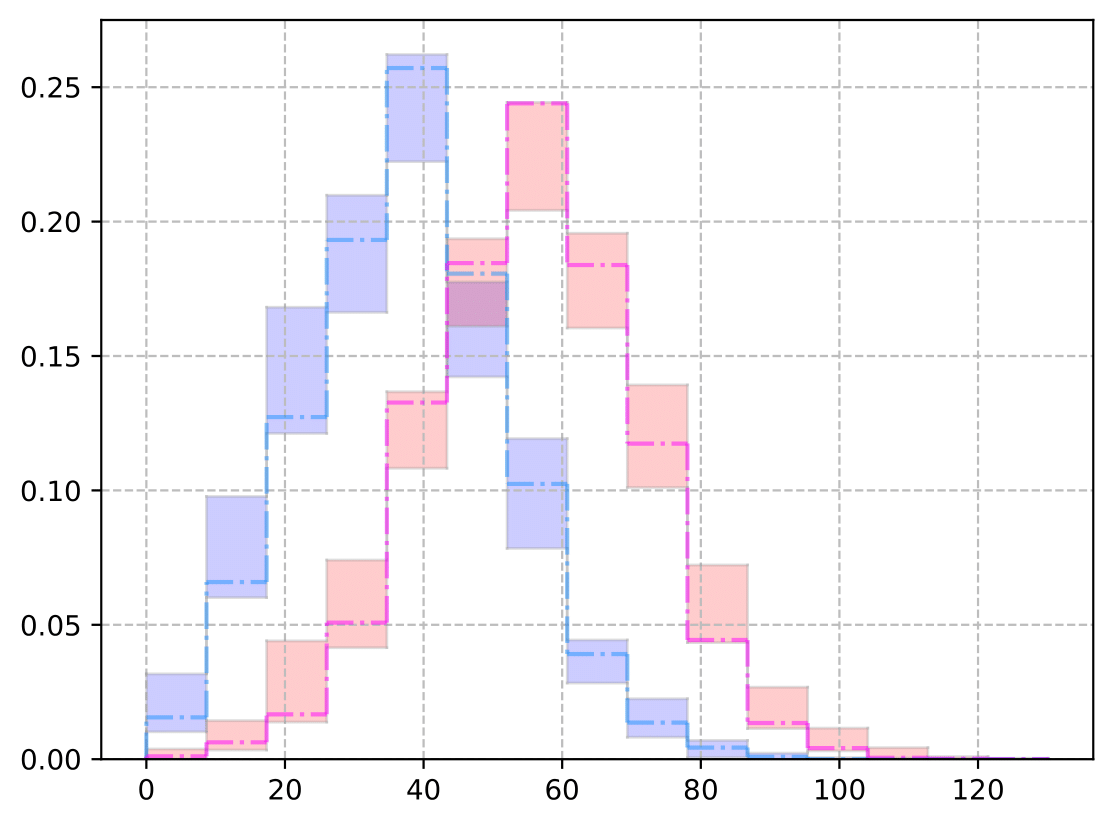}}
    \caption{Predicted polar surface area [\AA$^2$] for the case of conditional design with latent space dimension $J = 30$ for various sizes $K$ of the training dataset $\mathscr{G}$ from the QM9 database. Similar to the previous example, the reference solutions (blue ({\color{ProcessBlue}\dashdotted}) for $y=40$ and magenta ({\color{magenta}\dashdotted}) for $y=60$) are obtained by training with $K=10000$ samples. The shaded area represents the $1\%-99\%$ credible interval, reflecting the induced epistemic uncertainty from the lack of training information.} \label{fig:UQ_cond_PSA}
\end{figure}

Figs.~\ref{fig:UQ_cond_LogP} and~\ref{fig:UQ_cond_PSA} show the conditional design for two different target values of the octanol-water partition coefficient and polar surface area, respectively. In these figures, estimated properties are accompanied by the shaded area which represents the probabilistic confidence over them. As more data becomes available, our confidence over the predictive estimates of the properties grows.

\begin{table*}[h]
\rule[-8pt]{0pt}{12pt}
  \centering
  \caption{\label{tab:model_spec_cond}Network architecture modifications for the conditional generative task.}
    \begin{ruledtabular}
    \begin{tabular}{ccccc}
    Layer              & Input dimension & Output dimension & Activation layer & Activation function \\ \hline
    $l^{(1)}_{\theta}$ &  $J+Y$            &  $2\times (J+Y)$     & $a^{(1)}$        & Leaky ReLU  \\
    $l^{(2)}_{\theta}$ &  $2\times (J+Y)$    &  $4\times (J+Y)$     & $a^{(2)}$        & Leaky ReLU  \\
    $l^{(3)}_{\theta}$ &  $4\times (J+Y)$    &  $8\times (J+Y)$     & $a^{(3)}$        & Leaky ReLU  \\
    $l^{(4)}_{\theta}$ &  $8\times (J+Y)$    &  $288$           & $a^{(4)}$        & Leaky ReLU  \\
    $l^{(6)}_{\theta}$ &  $324+J+Y$        &  $45$            & $a^{(6)}_f$      & Leaky ReLU  \\
    $n^{(1)}_{\phi}$   &  $(5+Y)\times\sum_{m=0}^{M-1}\mathcal{J}^m$ &  $(5+Y)\times\sum_{m=0}^{M-1}\mathcal{J}^m$ & $-$              & $-$         \\
    $l^{(1)}_{\phi}$   &  $(5+Y)\times\sum_{m=0}^{M-1}\mathcal{J}^m$  &  $400$           & $-$              & $-$
    \end{tabular}
    \end{ruledtabular}
\end{table*}

\section{\label{sec:conclude}Conclusions}

In this work, we presented a generative model for constructing molecular graphs based on a VAE framework. Particular emphasis was placed in the case where limited training data is available. The variational approximation used was based on a hybrid graph scattering network, which consists of layers of graph scattering transform followed by fully-connected layers.
We constructed scattering layers using adaptive spectral filters, which are tailored to the training dataset. This increases the discriminatory power of the encoding network by reducing the correlation between the neighboring kernels.

For the decoding network, we use a one-shot graph generation model that returns a probabilistic 
representation of the molecule in terms of the distributions over different atoms and bond orders. 
The model first computes the probability of the bond orders and then computes the atom features' conditional probability given the underlying topology of the molecules. We further use a VAE regularization approach to impose constraints that enhance the sampled molecules' validity and 
energetic stability. The latter includes constraints on $3-$member cycles and cycles with triple bonds, which are strained chemical structures and thus, energetically less stable.

We provided visualization of the latent space using different structural and physicochemical properties of the molecules. We further assessed the model's performance in terms of the quality metrics and satisfaction of the constraints. We compared the generated molecules with the results of benchmark models in three categories: chemical validity, uniqueness, and novelty. Results show that when the latent space's dimensionality was set to $J=30$, the present model compares very well with benchmark models, despite those models being trained with approximately $100k$ training data points. 

Here, the regularization parameters for each set of constraints are approximated with a single value. For instance, the valence capacity constraint is defined over each node in the graph. However, we approximate the regularization parameter for all the nodes with a single parameter $\mu$. This simplification may limit the improvement of the quality. Using individual parameters may enhance the quality of the results.

Further, the model's predictive capabilities were investigated through the chemical space constructed from sampled molecules with various small-sized training datasets. Additionally, the frequency of various functional groups in the output samples is surveyed and compared to the QM9 database. We computed predictive estimates of the physicochemical properties by sampling molecules from the predictive model. 

To account for the use of limited training data in the problem, we use the Bayesian bootstrap predictive distribution to quantify the confidence in the predictive estimates of the properties. A posterior distribution over the resampling scheme's weights can be derived by imposing a prior on the sampling weights. Samples from this posterior distribution are directly passed to the model to yield maximum weighted likelihood estimates of the model parameters. This Bayesian formalism provides error bars over the predictive estimates. These error bars quantify the epistemic uncertainties in the learned parameters of the decoding network $p_{\boldsymbol{\theta}}(\mathcal{G}|\boldsymbol{z})$.

Lastly, a conditional generative task was developed to design molecules with given target property values. In this task, the model was trained along with the property labels in a supervised fashion, and then, molecules were sampled conditioned on the target property values. Tests were performed for three sizes of limited training data, and the uncertainty of the models was assessed. 

\appendix

\section{\label{app:Proof}Proof of Eq.~(\ref{eq:cyc_ind})}

Suppose $\boldsymbol{A}$ represents the adjacency matrix of graph $\mathcal{G}$ of size $N$. A walk of size $k$ from node $v_m$ to node $v_n$ is a sequence of $k$ connected edges between the two nodes. The total number of size $k$ walks between $v_m$ and $v_n$ is computed by $(\boldsymbol{A}^k)_{m,n}$. If there exists a unique sequence of edges and nodes between nodes $v_m$ and $v_n$, there should at least be a walk of size less than or equal to $N-1$ between them. In other words, element $(m,n)$ of matrix
\begin{equation}
    \boldsymbol{C}:=\sum_{k=1}^{N-1} \boldsymbol{A}^k,
    \label{eq:sum_powers}
\end{equation}
should be non-zero. Alternatively, to reduce the computations, we can modify Eq.~(\ref{eq:sum_powers}) as a geometric series and compute it using a closed form expression. Since inspecting connectivity of $v_m$ and $v_n$ only concerns the off-diagonal elements in Eq.~(\ref{eq:sum_powers}), we can add $\boldsymbol{I}$ to this sum without loss of generality. Also, we can neglect the fact that we only need to sum Eq.~(\ref{eq:sum_powers}) up to $N-1$ and write it in the form of geometric series 
\begin{equation}
    \boldsymbol{C}:=\sum_{k=0}^{\infty} \boldsymbol{A}^k.
    \label{eq:geometric}
\end{equation}
This geometric series has the closed form 
\begin{equation}
    \boldsymbol{C}=(\boldsymbol{I}-\boldsymbol{A})^{-1}
    \label{eq:geometric_closed}
\end{equation}
if $|\boldsymbol{A}|<1$. But this may not hold. Since we only inspect whether the elements of $\boldsymbol{C}$ are zero or not, and since $|(1/N)\boldsymbol{A}|<1$, we can substitute $\boldsymbol{A}$ by $(1/N)\boldsymbol{A}$ in Eq.~(\ref{eq:geometric}) to guarantee convergence of the geometric series
\begin{equation}
    \boldsymbol{B}:=\sum_{k=0}^{\infty} (\frac1N\boldsymbol{A})^k=(\boldsymbol{I}-\frac1N\boldsymbol{A})^{-1}.
\end{equation}

\vspace*{0.1in}

\noindent \textbf{DATA AVAILABILITY}

The PyTorch-based implementation, training/test datasets and script files to generate all the results in this paper are publicly available at \url{https://github.com/zabaras/GSVAE}.

\nocite{*}
\bibliography{aipsamp} 

\begin{thebibliography}{71}%
\makeatletter
\providecommand \@ifxundefined [1]{%
 \@ifx{#1\undefined}
}%
\providecommand \@ifnum [1]{%
 \ifnum #1\expandafter \@firstoftwo
 \else \expandafter \@secondoftwo
 \fi
}%
\providecommand \@ifx [1]{%
 \ifx #1\expandafter \@firstoftwo
 \else \expandafter \@secondoftwo
 \fi
}%
\providecommand \natexlab [1]{#1}%
\providecommand \enquote  [1]{``#1''}%
\providecommand \bibnamefont  [1]{#1}%
\providecommand \bibfnamefont [1]{#1}%
\providecommand \citenamefont [1]{#1}%
\providecommand \href@noop [0]{\@secondoftwo}%
\providecommand \href [0]{\begingroup \@sanitize@url \@href}%
\providecommand \@href[1]{\@@startlink{#1}\@@href}%
\providecommand \@@href[1]{\endgroup#1\@@endlink}%
\providecommand \@sanitize@url [0]{\catcode `\\12\catcode `\$12\catcode
  `\&12\catcode `\#12\catcode `\^12\catcode `\_12\catcode `\%12\relax}%
\providecommand \@@startlink[1]{}%
\providecommand \@@endlink[0]{}%
\providecommand \url  [0]{\begingroup\@sanitize@url \@url }%
\providecommand \@url [1]{\endgroup\@href {#1}{\urlprefix }}%
\providecommand \urlprefix  [0]{URL }%
\providecommand \Eprint [0]{\href }%
\providecommand \doibase [0]{http://dx.doi.org/}%
\providecommand \selectlanguage [0]{\@gobble}%
\providecommand \bibinfo  [0]{\@secondoftwo}%
\providecommand \bibfield  [0]{\@secondoftwo}%
\providecommand \translation [1]{[#1]}%
\providecommand \BibitemOpen [0]{}%
\providecommand \bibitemStop [0]{}%
\providecommand \bibitemNoStop [0]{.\EOS\space}%
\providecommand \EOS [0]{\spacefactor3000\relax}%
\providecommand \BibitemShut  [1]{\csname bibitem#1\endcsname}%
\let\auto@bib@innerbib\@empty
\bibitem [{\citenamefont {Sanchez-Lengeling}\ and\ \citenamefont
  {Aspuru-Guzik}(2018)}]{sanchez2018inverse}%
  \BibitemOpen
  \bibfield  {author} {\bibinfo {author} {\bibfnamefont {B.}~\bibnamefont
  {Sanchez-Lengeling}}\ and\ \bibinfo {author} {\bibfnamefont {A.}~\bibnamefont
  {Aspuru-Guzik}},\ }\bibfield  {title} {\enquote {\bibinfo {title} {Inverse
  molecular design using machine learning: Generative models for matter
  engineering},}\ }\href@noop {} {\bibfield  {journal} {\bibinfo  {journal}
  {Science}\ }\textbf {\bibinfo {volume} {361}},\ \bibinfo {pages} {360--365}
  (\bibinfo {year} {2018})}\BibitemShut {NoStop}%
\bibitem [{\citenamefont {Drews}(2000)}]{drews2000drug}%
  \BibitemOpen
  \bibfield  {author} {\bibinfo {author} {\bibfnamefont {J.}~\bibnamefont
  {Drews}},\ }\bibfield  {title} {\enquote {\bibinfo {title} {Drug discovery: a
  historical perspective},}\ }\href@noop {} {\bibfield  {journal} {\bibinfo
  {journal} {Science}\ }\textbf {\bibinfo {volume} {287}},\ \bibinfo {pages}
  {1960--1964} (\bibinfo {year} {2000})}\BibitemShut {NoStop}%
\bibitem [{\citenamefont {Hoppe}\ and\ \citenamefont
  {Sariciftci}(2004)}]{hoppe2004organic}%
  \BibitemOpen
  \bibfield  {author} {\bibinfo {author} {\bibfnamefont {H.}~\bibnamefont
  {Hoppe}}\ and\ \bibinfo {author} {\bibfnamefont {N.~S.}\ \bibnamefont
  {Sariciftci}},\ }\bibfield  {title} {\enquote {\bibinfo {title} {Organic
  solar cells: An overview},}\ }\href@noop {} {\bibfield  {journal} {\bibinfo
  {journal} {Journal of materials research}\ }\textbf {\bibinfo {volume}
  {19}},\ \bibinfo {pages} {1924--1945} (\bibinfo {year} {2004})}\BibitemShut
  {NoStop}%
\bibitem [{\citenamefont {Jorgensen}(2009)}]{jorgensen2009efficient}%
  \BibitemOpen
  \bibfield  {author} {\bibinfo {author} {\bibfnamefont {W.~L.}\ \bibnamefont
  {Jorgensen}},\ }\bibfield  {title} {\enquote {\bibinfo {title} {Efficient
  drug lead discovery and optimization},}\ }\href@noop {} {\bibfield  {journal}
  {\bibinfo  {journal} {Accounts of chemical research}\ }\textbf {\bibinfo
  {volume} {42}},\ \bibinfo {pages} {724--733} (\bibinfo {year}
  {2009})}\BibitemShut {NoStop}%
\bibitem [{\citenamefont {Elton}\ \emph {et~al.}(2019)\citenamefont {Elton},
  \citenamefont {Boukouvalas}, \citenamefont {Fuge},\ and\ \citenamefont
  {Chung}}]{elton2019deep}%
  \BibitemOpen
  \bibfield  {author} {\bibinfo {author} {\bibfnamefont {D.~C.}\ \bibnamefont
  {Elton}}, \bibinfo {author} {\bibfnamefont {Z.}~\bibnamefont {Boukouvalas}},
  \bibinfo {author} {\bibfnamefont {M.~D.}\ \bibnamefont {Fuge}}, \ and\
  \bibinfo {author} {\bibfnamefont {P.~W.}\ \bibnamefont {Chung}},\ }\bibfield
  {title} {\enquote {\bibinfo {title} {Deep learning for molecular design—a
  review of the state of the art},}\ }\href@noop {} {\bibfield  {journal}
  {\bibinfo  {journal} {Molecular Systems Design \& Engineering}\ }\textbf
  {\bibinfo {volume} {4}},\ \bibinfo {pages} {828--849} (\bibinfo {year}
  {2019})}\BibitemShut {NoStop}%
\bibitem [{\citenamefont {Faber}\ \emph {et~al.}(2017)\citenamefont {Faber},
  \citenamefont {Hutchison}, \citenamefont {Huang}, \citenamefont {Gilmer},
  \citenamefont {Schoenholz}, \citenamefont {Dahl}, \citenamefont {Vinyals},
  \citenamefont {Kearnes}, \citenamefont {Riley},\ and\ \citenamefont
  {Von~Lilienfeld}}]{faber2017prediction}%
  \BibitemOpen
  \bibfield  {author} {\bibinfo {author} {\bibfnamefont {F.~A.}\ \bibnamefont
  {Faber}}, \bibinfo {author} {\bibfnamefont {L.}~\bibnamefont {Hutchison}},
  \bibinfo {author} {\bibfnamefont {B.}~\bibnamefont {Huang}}, \bibinfo
  {author} {\bibfnamefont {J.}~\bibnamefont {Gilmer}}, \bibinfo {author}
  {\bibfnamefont {S.~S.}\ \bibnamefont {Schoenholz}}, \bibinfo {author}
  {\bibfnamefont {G.~E.}\ \bibnamefont {Dahl}}, \bibinfo {author}
  {\bibfnamefont {O.}~\bibnamefont {Vinyals}}, \bibinfo {author} {\bibfnamefont
  {S.}~\bibnamefont {Kearnes}}, \bibinfo {author} {\bibfnamefont {P.~F.}\
  \bibnamefont {Riley}}, \ and\ \bibinfo {author} {\bibfnamefont {O.~A.}\
  \bibnamefont {Von~Lilienfeld}},\ }\bibfield  {title} {\enquote {\bibinfo
  {title} {Prediction errors of molecular machine learning models lower than
  hybrid {DFT} error},}\ }\href@noop {} {\bibfield  {journal} {\bibinfo
  {journal} {Journal of chemical theory and computation}\ }\textbf {\bibinfo
  {volume} {13}},\ \bibinfo {pages} {5255--5264} (\bibinfo {year}
  {2017})}\BibitemShut {NoStop}%
\bibitem [{\citenamefont {G{\'o}mez-Bombarelli}\ \emph
  {et~al.}(2018)\citenamefont {G{\'o}mez-Bombarelli}, \citenamefont {Wei},
  \citenamefont {Duvenaud}, \citenamefont {Hern{\'a}ndez-Lobato}, \citenamefont
  {S{\'a}nchez-Lengeling}, \citenamefont {Sheberla}, \citenamefont
  {Aguilera-Iparraguirre}, \citenamefont {Hirzel}, \citenamefont {Adams},\ and\
  \citenamefont {Aspuru-Guzik}}]{gomez2018automatic}%
  \BibitemOpen
  \bibfield  {author} {\bibinfo {author} {\bibfnamefont {R.}~\bibnamefont
  {G{\'o}mez-Bombarelli}}, \bibinfo {author} {\bibfnamefont {J.~N.}\
  \bibnamefont {Wei}}, \bibinfo {author} {\bibfnamefont {D.}~\bibnamefont
  {Duvenaud}}, \bibinfo {author} {\bibfnamefont {J.~M.}\ \bibnamefont
  {Hern{\'a}ndez-Lobato}}, \bibinfo {author} {\bibfnamefont {B.}~\bibnamefont
  {S{\'a}nchez-Lengeling}}, \bibinfo {author} {\bibfnamefont {D.}~\bibnamefont
  {Sheberla}}, \bibinfo {author} {\bibfnamefont {J.}~\bibnamefont
  {Aguilera-Iparraguirre}}, \bibinfo {author} {\bibfnamefont {T.~D.}\
  \bibnamefont {Hirzel}}, \bibinfo {author} {\bibfnamefont {R.~P.}\
  \bibnamefont {Adams}}, \ and\ \bibinfo {author} {\bibfnamefont
  {A.}~\bibnamefont {Aspuru-Guzik}},\ }\bibfield  {title} {\enquote {\bibinfo
  {title} {Automatic chemical design using a data-driven continuous
  representation of molecules},}\ }\href@noop {} {\bibfield  {journal}
  {\bibinfo  {journal} {ACS central science}\ }\textbf {\bibinfo {volume}
  {4}},\ \bibinfo {pages} {268--276} (\bibinfo {year} {2018})}\BibitemShut
  {NoStop}%
\bibitem [{\citenamefont {Stokes}\ \emph {et~al.}(2020)\citenamefont {Stokes},
  \citenamefont {Yang}, \citenamefont {Swanson}, \citenamefont {Jin},
  \citenamefont {Cubillos-Ruiz}, \citenamefont {Donghia}, \citenamefont
  {MacNair}, \citenamefont {French}, \citenamefont {Carfrae}, \citenamefont
  {Bloom-Ackerman} \emph {et~al.}}]{stokes2020deep}%
  \BibitemOpen
  \bibfield  {author} {\bibinfo {author} {\bibfnamefont {J.~M.}\ \bibnamefont
  {Stokes}}, \bibinfo {author} {\bibfnamefont {K.}~\bibnamefont {Yang}},
  \bibinfo {author} {\bibfnamefont {K.}~\bibnamefont {Swanson}}, \bibinfo
  {author} {\bibfnamefont {W.}~\bibnamefont {Jin}}, \bibinfo {author}
  {\bibfnamefont {A.}~\bibnamefont {Cubillos-Ruiz}}, \bibinfo {author}
  {\bibfnamefont {N.~M.}\ \bibnamefont {Donghia}}, \bibinfo {author}
  {\bibfnamefont {C.~R.}\ \bibnamefont {MacNair}}, \bibinfo {author}
  {\bibfnamefont {S.}~\bibnamefont {French}}, \bibinfo {author} {\bibfnamefont
  {L.~A.}\ \bibnamefont {Carfrae}}, \bibinfo {author} {\bibfnamefont
  {Z.}~\bibnamefont {Bloom-Ackerman}},  \emph {et~al.},\ }\bibfield  {title}
  {\enquote {\bibinfo {title} {A deep learning approach to antibiotic
  discovery},}\ }\href@noop {} {\bibfield  {journal} {\bibinfo  {journal}
  {Cell}\ }\textbf {\bibinfo {volume} {180}},\ \bibinfo {pages} {688--702}
  (\bibinfo {year} {2020})}\BibitemShut {NoStop}%
\bibitem [{\citenamefont {Weininger}(1988)}]{weininger1988smiles}%
  \BibitemOpen
  \bibfield  {author} {\bibinfo {author} {\bibfnamefont {D.}~\bibnamefont
  {Weininger}},\ }\bibfield  {title} {\enquote {\bibinfo {title} {{SMILES}, a
  chemical language and information system. 1. {Introduction} to methodology
  and encoding rules},}\ }\href@noop {} {\bibfield  {journal} {\bibinfo
  {journal} {Journal of chemical information and computer sciences}\ }\textbf
  {\bibinfo {volume} {28}},\ \bibinfo {pages} {31--36} (\bibinfo {year}
  {1988})}\BibitemShut {NoStop}%
\bibitem [{\citenamefont {Kusner}, \citenamefont {Paige},\ and\ \citenamefont
  {Hern{\'a}ndez-Lobato}(2017)}]{kusner2017grammar}%
  \BibitemOpen
  \bibfield  {author} {\bibinfo {author} {\bibfnamefont {M.~J.}\ \bibnamefont
  {Kusner}}, \bibinfo {author} {\bibfnamefont {B.}~\bibnamefont {Paige}}, \
  and\ \bibinfo {author} {\bibfnamefont {J.~M.}\ \bibnamefont
  {Hern{\'a}ndez-Lobato}},\ }\bibfield  {title} {\enquote {\bibinfo {title}
  {Grammar variational autoencoder},}\ }in\ \href@noop {} {\emph {\bibinfo
  {booktitle} {Proceedings of the 34th International Conference on Machine
  Learning-Volume 70}}}\ (\bibinfo {organization} {JMLR. org},\ \bibinfo {year}
  {2017})\ pp.\ \bibinfo {pages} {1945--1954}\BibitemShut {NoStop}%
\bibitem [{\citenamefont {Bronstein}\ \emph {et~al.}(2017)\citenamefont
  {Bronstein}, \citenamefont {Bruna}, \citenamefont {LeCun}, \citenamefont
  {Szlam},\ and\ \citenamefont {Vandergheynst}}]{bronstein2017geometric}%
  \BibitemOpen
  \bibfield  {author} {\bibinfo {author} {\bibfnamefont {M.~M.}\ \bibnamefont
  {Bronstein}}, \bibinfo {author} {\bibfnamefont {J.}~\bibnamefont {Bruna}},
  \bibinfo {author} {\bibfnamefont {Y.}~\bibnamefont {LeCun}}, \bibinfo
  {author} {\bibfnamefont {A.}~\bibnamefont {Szlam}}, \ and\ \bibinfo {author}
  {\bibfnamefont {P.}~\bibnamefont {Vandergheynst}},\ }\bibfield  {title}
  {\enquote {\bibinfo {title} {Geometric deep learning: {Going} beyond
  {Euclidean} data},}\ }\href@noop {} {\bibfield  {journal} {\bibinfo
  {journal} {IEEE Signal Processing Magazine}\ }\textbf {\bibinfo {volume}
  {34}},\ \bibinfo {pages} {18--42} (\bibinfo {year} {2017})}\BibitemShut
  {NoStop}%
\bibitem [{\citenamefont {Gilmer}\ \emph {et~al.}(2017)\citenamefont {Gilmer},
  \citenamefont {Schoenholz}, \citenamefont {Riley}, \citenamefont {Vinyals},\
  and\ \citenamefont {Dahl}}]{gilmer2017neural}%
  \BibitemOpen
  \bibfield  {author} {\bibinfo {author} {\bibfnamefont {J.}~\bibnamefont
  {Gilmer}}, \bibinfo {author} {\bibfnamefont {S.~S.}\ \bibnamefont
  {Schoenholz}}, \bibinfo {author} {\bibfnamefont {P.~F.}\ \bibnamefont
  {Riley}}, \bibinfo {author} {\bibfnamefont {O.}~\bibnamefont {Vinyals}}, \
  and\ \bibinfo {author} {\bibfnamefont {G.~E.}\ \bibnamefont {Dahl}},\
  }\bibfield  {title} {\enquote {\bibinfo {title} {Neural message passing for
  quantum chemistry},}\ }in\ \href@noop {} {\emph {\bibinfo {booktitle}
  {Proceedings of the 34th International Conference on Machine Learning-Volume
  70}}}\ (\bibinfo {organization} {JMLR. org},\ \bibinfo {year} {2017})\ pp.\
  \bibinfo {pages} {1263--1272}\BibitemShut {NoStop}%
\bibitem [{\citenamefont {De~Cao}\ and\ \citenamefont
  {Kipf}(2018)}]{de2018molgan}%
  \BibitemOpen
  \bibfield  {author} {\bibinfo {author} {\bibfnamefont {N.}~\bibnamefont
  {De~Cao}}\ and\ \bibinfo {author} {\bibfnamefont {T.}~\bibnamefont {Kipf}},\
  }\bibfield  {title} {\enquote {\bibinfo {title} {Molgan: An implicit
  generative model for small molecular graphs},}\ }\href@noop {} {\bibfield
  {journal} {\bibinfo  {journal} {arXiv preprint arXiv:1805.11973}\ } (\bibinfo
  {year} {2018})}\BibitemShut {NoStop}%
\bibitem [{\citenamefont {Simonovsky}\ and\ \citenamefont
  {Komodakis}(2018)}]{simonovsky2018graphvae}%
  \BibitemOpen
  \bibfield  {author} {\bibinfo {author} {\bibfnamefont {M.}~\bibnamefont
  {Simonovsky}}\ and\ \bibinfo {author} {\bibfnamefont {N.}~\bibnamefont
  {Komodakis}},\ }\bibfield  {title} {\enquote {\bibinfo {title} {Graphvae:
  {Towards} generation of small graphs using variational autoencoders},}\ }in\
  \href@noop {} {\emph {\bibinfo {booktitle} {International Conference on
  Artificial Neural Networks}}}\ (\bibinfo {organization} {Springer},\ \bibinfo
  {year} {2018})\ pp.\ \bibinfo {pages} {412--422}\BibitemShut {NoStop}%
\bibitem [{\citenamefont {Liu}\ \emph {et~al.}(2018)\citenamefont {Liu},
  \citenamefont {Allamanis}, \citenamefont {Brockschmidt},\ and\ \citenamefont
  {Gaunt}}]{liu2018constrained}%
  \BibitemOpen
  \bibfield  {author} {\bibinfo {author} {\bibfnamefont {Q.}~\bibnamefont
  {Liu}}, \bibinfo {author} {\bibfnamefont {M.}~\bibnamefont {Allamanis}},
  \bibinfo {author} {\bibfnamefont {M.}~\bibnamefont {Brockschmidt}}, \ and\
  \bibinfo {author} {\bibfnamefont {A.}~\bibnamefont {Gaunt}},\ }\bibfield
  {title} {\enquote {\bibinfo {title} {Constrained graph variational
  autoencoders for molecule design},}\ }in\ \href@noop {} {\emph {\bibinfo
  {booktitle} {Advances in neural information processing systems}}}\ (\bibinfo
  {year} {2018})\ pp.\ \bibinfo {pages} {7795--7804}\BibitemShut {NoStop}%
\bibitem [{\citenamefont {Dinh}, \citenamefont {Krueger},\ and\ \citenamefont
  {Bengio}(2014)}]{dinh2014nice}%
  \BibitemOpen
  \bibfield  {author} {\bibinfo {author} {\bibfnamefont {L.}~\bibnamefont
  {Dinh}}, \bibinfo {author} {\bibfnamefont {D.}~\bibnamefont {Krueger}}, \
  and\ \bibinfo {author} {\bibfnamefont {Y.}~\bibnamefont {Bengio}},\
  }\bibfield  {title} {\enquote {\bibinfo {title} {Nice: Non-linear independent
  components estimation},}\ }\href@noop {} {\bibfield  {journal} {\bibinfo
  {journal} {arXiv preprint arXiv:1410.8516}\ } (\bibinfo {year}
  {2014})}\BibitemShut {NoStop}%
\bibitem [{\citenamefont {Madhawa}\ \emph {et~al.}(2019)\citenamefont
  {Madhawa}, \citenamefont {Ishiguro}, \citenamefont {Nakago},\ and\
  \citenamefont {Abe}}]{madhawa2019graphnvp}%
  \BibitemOpen
  \bibfield  {author} {\bibinfo {author} {\bibfnamefont {K.}~\bibnamefont
  {Madhawa}}, \bibinfo {author} {\bibfnamefont {K.}~\bibnamefont {Ishiguro}},
  \bibinfo {author} {\bibfnamefont {K.}~\bibnamefont {Nakago}}, \ and\ \bibinfo
  {author} {\bibfnamefont {M.}~\bibnamefont {Abe}},\ }\bibfield  {title}
  {\enquote {\bibinfo {title} {Graphnvp: An invertible flow model for
  generating molecular graphs},}\ }\href@noop {} {\bibfield  {journal}
  {\bibinfo  {journal} {arXiv preprint arXiv:1905.11600}\ } (\bibinfo {year}
  {2019})}\BibitemShut {NoStop}%
\bibitem [{\citenamefont {Jastrz{\k{e}}bski}, \citenamefont {Le{\'s}niak},\
  and\ \citenamefont {Czarnecki}(2016)}]{jastrzkebski2016learning}%
  \BibitemOpen
  \bibfield  {author} {\bibinfo {author} {\bibfnamefont {S.}~\bibnamefont
  {Jastrz{\k{e}}bski}}, \bibinfo {author} {\bibfnamefont {D.}~\bibnamefont
  {Le{\'s}niak}}, \ and\ \bibinfo {author} {\bibfnamefont {W.~M.}\ \bibnamefont
  {Czarnecki}},\ }\bibfield  {title} {\enquote {\bibinfo {title} {Learning to
  smile (s)},}\ }\href@noop {} {\bibfield  {journal} {\bibinfo  {journal}
  {arXiv preprint arXiv:1602.06289}\ } (\bibinfo {year} {2016})}\BibitemShut
  {NoStop}%
\bibitem [{\citenamefont {Segler}\ \emph {et~al.}(2018)\citenamefont {Segler},
  \citenamefont {Kogej}, \citenamefont {Tyrchan},\ and\ \citenamefont
  {Waller}}]{segler2018generating}%
  \BibitemOpen
  \bibfield  {author} {\bibinfo {author} {\bibfnamefont {M.~H.}\ \bibnamefont
  {Segler}}, \bibinfo {author} {\bibfnamefont {T.}~\bibnamefont {Kogej}},
  \bibinfo {author} {\bibfnamefont {C.}~\bibnamefont {Tyrchan}}, \ and\
  \bibinfo {author} {\bibfnamefont {M.~P.}\ \bibnamefont {Waller}},\ }\bibfield
   {title} {\enquote {\bibinfo {title} {Generating focused molecule libraries
  for drug discovery with recurrent neural networks},}\ }\href@noop {}
  {\bibfield  {journal} {\bibinfo  {journal} {ACS central science}\ }\textbf
  {\bibinfo {volume} {4}},\ \bibinfo {pages} {120--131} (\bibinfo {year}
  {2018})}\BibitemShut {NoStop}%
\bibitem [{\citenamefont {Son}\ \emph {et~al.}(2019)\citenamefont {Son},
  \citenamefont {Trivedi}, \citenamefont {Pan}, \citenamefont {Anderson},\ and\
  \citenamefont {Kondor}}]{hy2019covariant}%
  \BibitemOpen
  \bibfield  {author} {\bibinfo {author} {\bibfnamefont {H.~T.}\ \bibnamefont
  {Son}}, \bibinfo {author} {\bibfnamefont {S.}~\bibnamefont {Trivedi}},
  \bibinfo {author} {\bibfnamefont {H.}~\bibnamefont {Pan}}, \bibinfo {author}
  {\bibfnamefont {B.}~\bibnamefont {Anderson}}, \ and\ \bibinfo {author}
  {\bibfnamefont {R.}~\bibnamefont {Kondor}},\ }\bibfield  {title} {\enquote
  {\bibinfo {title} {Covariant compositional networks for learning graphs},}\
  }in\ \href@noop {} {\emph {\bibinfo {booktitle} {Proceedings of the 15th
  International Workshop on Mining and Learning with Graphs (MLG)}}}\ (\bibinfo
  {year} {2019})\BibitemShut {NoStop}%
\bibitem [{\citenamefont {Chen}\ \emph {et~al.}(2019)\citenamefont {Chen},
  \citenamefont {Ye}, \citenamefont {Zuo}, \citenamefont {Zheng},\ and\
  \citenamefont {Ong}}]{chen2019graph}%
  \BibitemOpen
  \bibfield  {author} {\bibinfo {author} {\bibfnamefont {C.}~\bibnamefont
  {Chen}}, \bibinfo {author} {\bibfnamefont {W.}~\bibnamefont {Ye}}, \bibinfo
  {author} {\bibfnamefont {Y.}~\bibnamefont {Zuo}}, \bibinfo {author}
  {\bibfnamefont {C.}~\bibnamefont {Zheng}}, \ and\ \bibinfo {author}
  {\bibfnamefont {S.~P.}\ \bibnamefont {Ong}},\ }\bibfield  {title} {\enquote
  {\bibinfo {title} {Graph networks as a universal machine learning framework
  for molecules and crystals},}\ }\href@noop {} {\bibfield  {journal} {\bibinfo
   {journal} {Chemistry of Materials}\ }\textbf {\bibinfo {volume} {31}},\
  \bibinfo {pages} {3564--3572} (\bibinfo {year} {2019})}\BibitemShut {NoStop}%
\bibitem [{\citenamefont {Coley}\ \emph {et~al.}(2019)\citenamefont {Coley},
  \citenamefont {Jin}, \citenamefont {Rogers}, \citenamefont {Jamison},
  \citenamefont {Jaakkola}, \citenamefont {Green}, \citenamefont {Barzilay},\
  and\ \citenamefont {Jensen}}]{coley2019graph}%
  \BibitemOpen
  \bibfield  {author} {\bibinfo {author} {\bibfnamefont {C.~W.}\ \bibnamefont
  {Coley}}, \bibinfo {author} {\bibfnamefont {W.}~\bibnamefont {Jin}}, \bibinfo
  {author} {\bibfnamefont {L.}~\bibnamefont {Rogers}}, \bibinfo {author}
  {\bibfnamefont {T.~F.}\ \bibnamefont {Jamison}}, \bibinfo {author}
  {\bibfnamefont {T.~S.}\ \bibnamefont {Jaakkola}}, \bibinfo {author}
  {\bibfnamefont {W.~H.}\ \bibnamefont {Green}}, \bibinfo {author}
  {\bibfnamefont {R.}~\bibnamefont {Barzilay}}, \ and\ \bibinfo {author}
  {\bibfnamefont {K.~F.}\ \bibnamefont {Jensen}},\ }\bibfield  {title}
  {\enquote {\bibinfo {title} {A graph-convolutional neural network model for
  the prediction of chemical reactivity},}\ }\href@noop {} {\bibfield
  {journal} {\bibinfo  {journal} {Chemical science}\ }\textbf {\bibinfo
  {volume} {10}},\ \bibinfo {pages} {370--377} (\bibinfo {year}
  {2019})}\BibitemShut {NoStop}%
\bibitem [{\citenamefont {Thomas}\ \emph {et~al.}(2018)\citenamefont {Thomas},
  \citenamefont {Smidt}, \citenamefont {Kearnes}, \citenamefont {Yang},
  \citenamefont {Li}, \citenamefont {Kohlhoff},\ and\ \citenamefont
  {Riley}}]{thomas2018tensor}%
  \BibitemOpen
  \bibfield  {author} {\bibinfo {author} {\bibfnamefont {N.}~\bibnamefont
  {Thomas}}, \bibinfo {author} {\bibfnamefont {T.}~\bibnamefont {Smidt}},
  \bibinfo {author} {\bibfnamefont {S.}~\bibnamefont {Kearnes}}, \bibinfo
  {author} {\bibfnamefont {L.}~\bibnamefont {Yang}}, \bibinfo {author}
  {\bibfnamefont {L.}~\bibnamefont {Li}}, \bibinfo {author} {\bibfnamefont
  {K.}~\bibnamefont {Kohlhoff}}, \ and\ \bibinfo {author} {\bibfnamefont
  {P.}~\bibnamefont {Riley}},\ }\bibfield  {title} {\enquote {\bibinfo {title}
  {Tensor field networks: Rotation-and translation-equivariant neural networks
  for 3d point clouds},}\ }\href@noop {} {\bibfield  {journal} {\bibinfo
  {journal} {arXiv preprint arXiv:1802.08219}\ } (\bibinfo {year}
  {2018})}\BibitemShut {NoStop}%
\bibitem [{\citenamefont {Weiler}\ \emph {et~al.}(2018)\citenamefont {Weiler},
  \citenamefont {Geiger}, \citenamefont {Welling}, \citenamefont {Boomsma},\
  and\ \citenamefont {Cohen}}]{weiler20183d}%
  \BibitemOpen
  \bibfield  {author} {\bibinfo {author} {\bibfnamefont {M.}~\bibnamefont
  {Weiler}}, \bibinfo {author} {\bibfnamefont {M.}~\bibnamefont {Geiger}},
  \bibinfo {author} {\bibfnamefont {M.}~\bibnamefont {Welling}}, \bibinfo
  {author} {\bibfnamefont {W.}~\bibnamefont {Boomsma}}, \ and\ \bibinfo
  {author} {\bibfnamefont {T.~S.}\ \bibnamefont {Cohen}},\ }\bibfield  {title}
  {\enquote {\bibinfo {title} {3d steerable {CNNs}: Learning rotationally
  equivariant features in volumetric data},}\ }in\ \href@noop {} {\emph
  {\bibinfo {booktitle} {Advances in Neural Information Processing Systems}}}\
  (\bibinfo {year} {2018})\ pp.\ \bibinfo {pages} {10381--10392}\BibitemShut
  {NoStop}%
\bibitem [{\citenamefont {Kondor}(2018)}]{kondor2018n}%
  \BibitemOpen
  \bibfield  {author} {\bibinfo {author} {\bibfnamefont {R.}~\bibnamefont
  {Kondor}},\ }\bibfield  {title} {\enquote {\bibinfo {title} {N-body networks:
  {A} covariant hierarchical neural network architecture for learning atomic
  potentials},}\ }\href@noop {} {\bibfield  {journal} {\bibinfo  {journal}
  {arXiv preprint arXiv:1803.01588}\ } (\bibinfo {year} {2018})}\BibitemShut
  {NoStop}%
\bibitem [{\citenamefont {Jin}, \citenamefont {Barzilay},\ and\ \citenamefont
  {Jaakkola}(2018)}]{jin2018junction}%
  \BibitemOpen
  \bibfield  {author} {\bibinfo {author} {\bibfnamefont {W.}~\bibnamefont
  {Jin}}, \bibinfo {author} {\bibfnamefont {R.}~\bibnamefont {Barzilay}}, \
  and\ \bibinfo {author} {\bibfnamefont {T.}~\bibnamefont {Jaakkola}},\
  }\bibfield  {title} {\enquote {\bibinfo {title} {Junction tree variational
  autoencoder for molecular graph generation},}\ }\href@noop {} {\bibfield
  {journal} {\bibinfo  {journal} {arXiv preprint arXiv:1802.04364}\ } (\bibinfo
  {year} {2018})}\BibitemShut {NoStop}%
\bibitem [{\citenamefont {You}\ \emph {et~al.}(2018)\citenamefont {You},
  \citenamefont {Liu}, \citenamefont {Ying}, \citenamefont {Pande},\ and\
  \citenamefont {Leskovec}}]{you2018graph}%
  \BibitemOpen
  \bibfield  {author} {\bibinfo {author} {\bibfnamefont {J.}~\bibnamefont
  {You}}, \bibinfo {author} {\bibfnamefont {B.}~\bibnamefont {Liu}}, \bibinfo
  {author} {\bibfnamefont {Z.}~\bibnamefont {Ying}}, \bibinfo {author}
  {\bibfnamefont {V.}~\bibnamefont {Pande}}, \ and\ \bibinfo {author}
  {\bibfnamefont {J.}~\bibnamefont {Leskovec}},\ }\bibfield  {title} {\enquote
  {\bibinfo {title} {Graph convolutional policy network for goal-directed
  molecular graph generation},}\ }in\ \href@noop {} {\emph {\bibinfo
  {booktitle} {Advances in neural information processing systems}}}\ (\bibinfo
  {year} {2018})\ pp.\ \bibinfo {pages} {6410--6421}\BibitemShut {NoStop}%
\bibitem [{\citenamefont {Kipf}\ and\ \citenamefont
  {Welling}(2016)}]{kipf2016variational}%
  \BibitemOpen
  \bibfield  {author} {\bibinfo {author} {\bibfnamefont {T.~N.}\ \bibnamefont
  {Kipf}}\ and\ \bibinfo {author} {\bibfnamefont {M.}~\bibnamefont {Welling}},\
  }\bibfield  {title} {\enquote {\bibinfo {title} {Variational graph
  auto-encoders},}\ }\href@noop {} {\bibfield  {journal} {\bibinfo  {journal}
  {NIPS Workshop on Bayesian Deep Learning}\ } (\bibinfo {year}
  {2016})}\BibitemShut {NoStop}%
\bibitem [{\citenamefont {Mansimov}\ \emph {et~al.}(2019)\citenamefont
  {Mansimov}, \citenamefont {Mahmood}, \citenamefont {Kang},\ and\
  \citenamefont {Cho}}]{mansimov2019molecular}%
  \BibitemOpen
  \bibfield  {author} {\bibinfo {author} {\bibfnamefont {E.}~\bibnamefont
  {Mansimov}}, \bibinfo {author} {\bibfnamefont {O.}~\bibnamefont {Mahmood}},
  \bibinfo {author} {\bibfnamefont {S.}~\bibnamefont {Kang}}, \ and\ \bibinfo
  {author} {\bibfnamefont {K.}~\bibnamefont {Cho}},\ }\bibfield  {title}
  {\enquote {\bibinfo {title} {Molecular geometry prediction using a deep
  generative graph neural network},}\ }\href@noop {} {\bibfield  {journal}
  {\bibinfo  {journal} {Scientific reports}\ }\textbf {\bibinfo {volume} {9}},\
  \bibinfo {pages} {1--13} (\bibinfo {year} {2019})}\BibitemShut {NoStop}%
\bibitem [{\citenamefont {Gebauer}, \citenamefont {Gastegger},\ and\
  \citenamefont {Sch{\"u}tt}(2019)}]{gebauer2019symmetry}%
  \BibitemOpen
  \bibfield  {author} {\bibinfo {author} {\bibfnamefont {N.}~\bibnamefont
  {Gebauer}}, \bibinfo {author} {\bibfnamefont {M.}~\bibnamefont {Gastegger}},
  \ and\ \bibinfo {author} {\bibfnamefont {K.}~\bibnamefont {Sch{\"u}tt}},\
  }\bibfield  {title} {\enquote {\bibinfo {title} {Symmetry-adapted generation
  of 3d point sets for the targeted discovery of molecules},}\ }in\ \href@noop
  {} {\emph {\bibinfo {booktitle} {Advances in Neural Information Processing
  Systems}}}\ (\bibinfo {year} {2019})\ pp.\ \bibinfo {pages}
  {7566--7578}\BibitemShut {NoStop}%
\bibitem [{\citenamefont {Mallat}(2012)}]{mallat2012group}%
  \BibitemOpen
  \bibfield  {author} {\bibinfo {author} {\bibfnamefont {S.}~\bibnamefont
  {Mallat}},\ }\bibfield  {title} {\enquote {\bibinfo {title} {Group invariant
  scattering},}\ }\href@noop {} {\bibfield  {journal} {\bibinfo  {journal}
  {Communications on Pure and Applied Mathematics}\ }\textbf {\bibinfo {volume}
  {65}},\ \bibinfo {pages} {1331--1398} (\bibinfo {year} {2012})}\BibitemShut
  {NoStop}%
\bibitem [{\citenamefont {Mallat}(2016)}]{mallat2016understanding}%
  \BibitemOpen
  \bibfield  {author} {\bibinfo {author} {\bibfnamefont {S.}~\bibnamefont
  {Mallat}},\ }\bibfield  {title} {\enquote {\bibinfo {title} {Understanding
  deep convolutional networks},}\ }\href@noop {} {\bibfield  {journal}
  {\bibinfo  {journal} {Philosophical Transactions of the Royal Society A:
  Mathematical, Physical and Engineering Sciences}\ }\textbf {\bibinfo {volume}
  {374}},\ \bibinfo {pages} {20150203} (\bibinfo {year} {2016})}\BibitemShut
  {NoStop}%
\bibitem [{\citenamefont {Oyallon}\ \emph {et~al.}(2018)\citenamefont
  {Oyallon}, \citenamefont {Zagoruyko}, \citenamefont {Huang}, \citenamefont
  {Komodakis}, \citenamefont {Lacoste-Julien}, \citenamefont {Blaschko},\ and\
  \citenamefont {Belilovsky}}]{oyallon2018scattering}%
  \BibitemOpen
  \bibfield  {author} {\bibinfo {author} {\bibfnamefont {E.}~\bibnamefont
  {Oyallon}}, \bibinfo {author} {\bibfnamefont {S.}~\bibnamefont {Zagoruyko}},
  \bibinfo {author} {\bibfnamefont {G.}~\bibnamefont {Huang}}, \bibinfo
  {author} {\bibfnamefont {N.}~\bibnamefont {Komodakis}}, \bibinfo {author}
  {\bibfnamefont {S.}~\bibnamefont {Lacoste-Julien}}, \bibinfo {author}
  {\bibfnamefont {M.}~\bibnamefont {Blaschko}}, \ and\ \bibinfo {author}
  {\bibfnamefont {E.}~\bibnamefont {Belilovsky}},\ }\bibfield  {title}
  {\enquote {\bibinfo {title} {Scattering networks for hybrid representation
  learning},}\ }\href@noop {} {\bibfield  {journal} {\bibinfo  {journal} {IEEE
  transactions on pattern analysis and machine intelligence}\ }\textbf
  {\bibinfo {volume} {41}},\ \bibinfo {pages} {2208--2221} (\bibinfo {year}
  {2018})}\BibitemShut {NoStop}%
\bibitem [{\citenamefont {Chen}, \citenamefont {Cheng},\ and\ \citenamefont
  {Mallat}(2014)}]{chen2014unsupervised}%
  \BibitemOpen
  \bibfield  {author} {\bibinfo {author} {\bibfnamefont {X.}~\bibnamefont
  {Chen}}, \bibinfo {author} {\bibfnamefont {X.}~\bibnamefont {Cheng}}, \ and\
  \bibinfo {author} {\bibfnamefont {S.}~\bibnamefont {Mallat}},\ }\bibfield
  {title} {\enquote {\bibinfo {title} {Unsupervised deep {Haar} scattering on
  graphs},}\ }in\ \href@noop {} {\emph {\bibinfo {booktitle} {Advances in
  Neural Information Processing Systems}}}\ (\bibinfo {year} {2014})\ pp.\
  \bibinfo {pages} {1709--1717}\BibitemShut {NoStop}%
\bibitem [{\citenamefont {Zou}\ and\ \citenamefont
  {Lerman}(2019)}]{zou2019graph}%
  \BibitemOpen
  \bibfield  {author} {\bibinfo {author} {\bibfnamefont {D.}~\bibnamefont
  {Zou}}\ and\ \bibinfo {author} {\bibfnamefont {G.}~\bibnamefont {Lerman}},\
  }\bibfield  {title} {\enquote {\bibinfo {title} {Graph convolutional neural
  networks via scattering},}\ }\href@noop {} {\bibfield  {journal} {\bibinfo
  {journal} {Applied and Computational Harmonic Analysis}\ } (\bibinfo {year}
  {2019})}\BibitemShut {NoStop}%
\bibitem [{\citenamefont {Gama}, \citenamefont {Ribeiro},\ and\ \citenamefont
  {Bruna}(2018)}]{gama2018diffusion}%
  \BibitemOpen
  \bibfield  {author} {\bibinfo {author} {\bibfnamefont {F.}~\bibnamefont
  {Gama}}, \bibinfo {author} {\bibfnamefont {A.}~\bibnamefont {Ribeiro}}, \
  and\ \bibinfo {author} {\bibfnamefont {J.}~\bibnamefont {Bruna}},\ }\bibfield
   {title} {\enquote {\bibinfo {title} {Diffusion scattering transforms on
  graphs},}\ }\href@noop {} {\bibfield  {journal} {\bibinfo  {journal} {arXiv
  preprint arXiv:1806.08829}\ } (\bibinfo {year} {2018})}\BibitemShut {NoStop}%
\bibitem [{\citenamefont {Coifman}\ and\ \citenamefont
  {Maggioni}(2006)}]{coifman2006diffusion}%
  \BibitemOpen
  \bibfield  {author} {\bibinfo {author} {\bibfnamefont {R.~R.}\ \bibnamefont
  {Coifman}}\ and\ \bibinfo {author} {\bibfnamefont {M.}~\bibnamefont
  {Maggioni}},\ }\bibfield  {title} {\enquote {\bibinfo {title} {Diffusion
  wavelets},}\ }\href@noop {} {\bibfield  {journal} {\bibinfo  {journal}
  {Applied and Computational Harmonic Analysis}\ }\textbf {\bibinfo {volume}
  {21}},\ \bibinfo {pages} {53--94} (\bibinfo {year} {2006})}\BibitemShut
  {NoStop}%
\bibitem [{\citenamefont {Hammond}, \citenamefont {Vandergheynst},\ and\
  \citenamefont {Gribonval}(2011)}]{hammond2011wavelets}%
  \BibitemOpen
  \bibfield  {author} {\bibinfo {author} {\bibfnamefont {D.~K.}\ \bibnamefont
  {Hammond}}, \bibinfo {author} {\bibfnamefont {P.}~\bibnamefont
  {Vandergheynst}}, \ and\ \bibinfo {author} {\bibfnamefont {R.}~\bibnamefont
  {Gribonval}},\ }\bibfield  {title} {\enquote {\bibinfo {title} {Wavelets on
  graphs via spectral graph theory},}\ }\href@noop {} {\bibfield  {journal}
  {\bibinfo  {journal} {Applied and Computational Harmonic Analysis}\ }\textbf
  {\bibinfo {volume} {30}},\ \bibinfo {pages} {129--150} (\bibinfo {year}
  {2011})}\BibitemShut {NoStop}%
\bibitem [{\citenamefont {Li}\ \emph {et~al.}(2018)\citenamefont {Li},
  \citenamefont {Vinyals}, \citenamefont {Dyer}, \citenamefont {Pascanu},\ and\
  \citenamefont {Battaglia}}]{li2018learning}%
  \BibitemOpen
  \bibfield  {author} {\bibinfo {author} {\bibfnamefont {Y.}~\bibnamefont
  {Li}}, \bibinfo {author} {\bibfnamefont {O.}~\bibnamefont {Vinyals}},
  \bibinfo {author} {\bibfnamefont {C.}~\bibnamefont {Dyer}}, \bibinfo {author}
  {\bibfnamefont {R.}~\bibnamefont {Pascanu}}, \ and\ \bibinfo {author}
  {\bibfnamefont {P.}~\bibnamefont {Battaglia}},\ }\bibfield  {title} {\enquote
  {\bibinfo {title} {Learning deep generative models of graphs},}\ }\href@noop
  {} {\bibfield  {journal} {\bibinfo  {journal} {arXiv preprint
  arXiv:1803.03324}\ } (\bibinfo {year} {2018})}\BibitemShut {NoStop}%
\bibitem [{\citenamefont {Bresson}\ and\ \citenamefont
  {Laurent}(2019)}]{bresson2019two}%
  \BibitemOpen
  \bibfield  {author} {\bibinfo {author} {\bibfnamefont {X.}~\bibnamefont
  {Bresson}}\ and\ \bibinfo {author} {\bibfnamefont {T.}~\bibnamefont
  {Laurent}},\ }\bibfield  {title} {\enquote {\bibinfo {title} {A two-step
  graph convolutional decoder for molecule generation},}\ }\href@noop {}
  {\bibfield  {journal} {\bibinfo  {journal} {arXiv preprint arXiv:1906.03412}\
  } (\bibinfo {year} {2019})}\BibitemShut {NoStop}%
\bibitem [{\citenamefont {Ma}, \citenamefont {Chen},\ and\ \citenamefont
  {Xiao}(2018)}]{ma2018constrained}%
  \BibitemOpen
  \bibfield  {author} {\bibinfo {author} {\bibfnamefont {T.}~\bibnamefont
  {Ma}}, \bibinfo {author} {\bibfnamefont {J.}~\bibnamefont {Chen}}, \ and\
  \bibinfo {author} {\bibfnamefont {C.}~\bibnamefont {Xiao}},\ }\bibfield
  {title} {\enquote {\bibinfo {title} {Constrained generation of semantically
  valid graphs via regularizing variational autoencoders},}\ }in\ \href@noop {}
  {\emph {\bibinfo {booktitle} {Advances in Neural Information Processing
  Systems}}}\ (\bibinfo {year} {2018})\ pp.\ \bibinfo {pages}
  {7113--7124}\BibitemShut {NoStop}%
\bibitem [{\citenamefont {Sch{\"o}berl}, \citenamefont {Zabaras},\ and\
  \citenamefont {Koutsourelakis}(2019)}]{schoberl2019predictive}%
  \BibitemOpen
  \bibfield  {author} {\bibinfo {author} {\bibfnamefont {M.}~\bibnamefont
  {Sch{\"o}berl}}, \bibinfo {author} {\bibfnamefont {N.}~\bibnamefont
  {Zabaras}}, \ and\ \bibinfo {author} {\bibfnamefont {P.-S.}\ \bibnamefont
  {Koutsourelakis}},\ }\bibfield  {title} {\enquote {\bibinfo {title}
  {Predictive collective variable discovery with deep {Bayesian} models},}\
  }\href@noop {} {\bibfield  {journal} {\bibinfo  {journal} {The Journal of
  chemical physics}\ }\textbf {\bibinfo {volume} {150}},\ \bibinfo {pages}
  {024109} (\bibinfo {year} {2019})}\BibitemShut {NoStop}%
\bibitem [{\citenamefont {MacKay}\ and\ \citenamefont
  {Mac~Kay}(2003)}]{mackay2003information}%
  \BibitemOpen
  \bibfield  {author} {\bibinfo {author} {\bibfnamefont {D.~J.}\ \bibnamefont
  {MacKay}}\ and\ \bibinfo {author} {\bibfnamefont {D.~J.}\ \bibnamefont
  {Mac~Kay}},\ }\href@noop {} {\emph {\bibinfo {title} {Information theory,
  inference and learning algorithms}}}\ (\bibinfo  {publisher} {Cambridge
  university press},\ \bibinfo {year} {2003})\BibitemShut {NoStop}%
\bibitem [{\citenamefont {Harris}(1989)}]{harris1989predictive}%
  \BibitemOpen
  \bibfield  {author} {\bibinfo {author} {\bibfnamefont {I.~R.}\ \bibnamefont
  {Harris}},\ }\bibfield  {title} {\enquote {\bibinfo {title} {Predictive fit
  for natural exponential families},}\ }\href@noop {} {\bibfield  {journal}
  {\bibinfo  {journal} {Biometrika}\ }\textbf {\bibinfo {volume} {76}},\
  \bibinfo {pages} {675--684} (\bibinfo {year} {1989})}\BibitemShut {NoStop}%
\bibitem [{\citenamefont {Fushiki}\ \emph {et~al.}(2005)\citenamefont
  {Fushiki}, \citenamefont {Komaki}, \citenamefont {Aihara} \emph
  {et~al.}}]{fushiki2005nonparametric}%
  \BibitemOpen
  \bibfield  {author} {\bibinfo {author} {\bibfnamefont {T.}~\bibnamefont
  {Fushiki}}, \bibinfo {author} {\bibfnamefont {F.}~\bibnamefont {Komaki}},
  \bibinfo {author} {\bibfnamefont {K.}~\bibnamefont {Aihara}},  \emph
  {et~al.},\ }\bibfield  {title} {\enquote {\bibinfo {title} {Nonparametric
  bootstrap prediction},}\ }\href@noop {} {\bibfield  {journal} {\bibinfo
  {journal} {Bernoulli}\ }\textbf {\bibinfo {volume} {11}},\ \bibinfo {pages}
  {293--307} (\bibinfo {year} {2005})}\BibitemShut {NoStop}%
\bibitem [{\citenamefont {Fushiki}(2010)}]{fushiki2010bayesian}%
  \BibitemOpen
  \bibfield  {author} {\bibinfo {author} {\bibfnamefont {T.}~\bibnamefont
  {Fushiki}},\ }\bibfield  {title} {\enquote {\bibinfo {title} {Bayesian
  bootstrap prediction},}\ }\href@noop {} {\bibfield  {journal} {\bibinfo
  {journal} {Journal of statistical planning and inference}\ }\textbf {\bibinfo
  {volume} {140}},\ \bibinfo {pages} {65--74} (\bibinfo {year}
  {2010})}\BibitemShut {NoStop}%
\bibitem [{\citenamefont {Rubin}(1981)}]{rubin1981bayesian}%
  \BibitemOpen
  \bibfield  {author} {\bibinfo {author} {\bibfnamefont {D.~B.}\ \bibnamefont
  {Rubin}},\ }\bibfield  {title} {\enquote {\bibinfo {title} {The {Bayesian}
  bootstrap},}\ }\href@noop {} {\bibfield  {journal} {\bibinfo  {journal} {The
  annals of statistics}\ ,\ \bibinfo {pages} {130--134}} (\bibinfo {year}
  {1981})}\BibitemShut {NoStop}%
\bibitem [{\citenamefont {Cayley}(1874)}]{professor1874lvii}%
  \BibitemOpen
  \bibfield  {author} {\bibinfo {author} {\bibfnamefont {P.}~\bibnamefont
  {Cayley}},\ }\bibfield  {title} {\enquote {\bibinfo {title} {Lvii. on the
  mathematical theory of isomers},}\ }\href@noop {} {\bibfield  {journal}
  {\bibinfo  {journal} {The London, Edinburgh, and Dublin Philosophical
  Magazine and Journal of Science}\ }\textbf {\bibinfo {volume} {47}},\
  \bibinfo {pages} {444--447} (\bibinfo {year} {1874})}\BibitemShut {NoStop}%
\bibitem [{\citenamefont {Sylvester}(1878)}]{sylvester1878application}%
  \BibitemOpen
  \bibfield  {author} {\bibinfo {author} {\bibfnamefont {J.~J.}\ \bibnamefont
  {Sylvester}},\ }\bibfield  {title} {\enquote {\bibinfo {title} {On an
  application of the new atomic theory to the graphical representation of the
  invariants and covariants of binary quantics, with three appendices},}\
  }\href@noop {} {\bibfield  {journal} {\bibinfo  {journal} {American Journal
  of Mathematics}\ }\textbf {\bibinfo {volume} {1}},\ \bibinfo {pages}
  {64--104} (\bibinfo {year} {1878})}\BibitemShut {NoStop}%
\bibitem [{\citenamefont {Kingma}\ and\ \citenamefont
  {Welling}(2013)}]{kingma2013auto}%
  \BibitemOpen
  \bibfield  {author} {\bibinfo {author} {\bibfnamefont {D.~P.}\ \bibnamefont
  {Kingma}}\ and\ \bibinfo {author} {\bibfnamefont {M.}~\bibnamefont
  {Welling}},\ }\bibfield  {title} {\enquote {\bibinfo {title} {Auto-encoding
  variational {Bayes}},}\ }\href@noop {} {\bibfield  {journal} {\bibinfo
  {journal} {arXiv preprint arXiv:1312.6114}\ } (\bibinfo {year}
  {2013})}\BibitemShut {NoStop}%
\bibitem [{\citenamefont {Shuman}\ \emph {et~al.}(2013)\citenamefont {Shuman},
  \citenamefont {Narang}, \citenamefont {Frossard}, \citenamefont {Ortega},\
  and\ \citenamefont {Vandergheynst}}]{shuman2013emerging}%
  \BibitemOpen
  \bibfield  {author} {\bibinfo {author} {\bibfnamefont {D.}~\bibnamefont
  {Shuman}}, \bibinfo {author} {\bibfnamefont {S.}~\bibnamefont {Narang}},
  \bibinfo {author} {\bibfnamefont {P.}~\bibnamefont {Frossard}}, \bibinfo
  {author} {\bibfnamefont {A.}~\bibnamefont {Ortega}}, \ and\ \bibinfo {author}
  {\bibfnamefont {P.}~\bibnamefont {Vandergheynst}},\ }\bibfield  {title}
  {\enquote {\bibinfo {title} {The emerging field of signal processing on
  graphs: Extending high-dimensional data analysis to networks and other
  irregular domains},}\ }\href@noop {} {\bibfield  {journal} {\bibinfo
  {journal} {IEEE Signal Processing Magazine}\ }\textbf {\bibinfo {volume}
  {3}},\ \bibinfo {pages} {83--98} (\bibinfo {year} {2013})}\BibitemShut
  {NoStop}%
\bibitem [{\citenamefont {Ortega}\ \emph {et~al.}(2018)\citenamefont {Ortega},
  \citenamefont {Frossard}, \citenamefont {Kova{\v{c}}evi{\'c}}, \citenamefont
  {Moura},\ and\ \citenamefont {Vandergheynst}}]{ortega2018graph}%
  \BibitemOpen
  \bibfield  {author} {\bibinfo {author} {\bibfnamefont {A.}~\bibnamefont
  {Ortega}}, \bibinfo {author} {\bibfnamefont {P.}~\bibnamefont {Frossard}},
  \bibinfo {author} {\bibfnamefont {J.}~\bibnamefont {Kova{\v{c}}evi{\'c}}},
  \bibinfo {author} {\bibfnamefont {J.~M.}\ \bibnamefont {Moura}}, \ and\
  \bibinfo {author} {\bibfnamefont {P.}~\bibnamefont {Vandergheynst}},\
  }\bibfield  {title} {\enquote {\bibinfo {title} {Graph signal processing:
  Overview, challenges, and applications},}\ }\href@noop {} {\bibfield
  {journal} {\bibinfo  {journal} {Proceedings of the IEEE}\ }\textbf {\bibinfo
  {volume} {106}},\ \bibinfo {pages} {808--828} (\bibinfo {year}
  {2018})}\BibitemShut {NoStop}%
\bibitem [{\citenamefont {Shuman}\ \emph {et~al.}(2015)\citenamefont {Shuman},
  \citenamefont {Wiesmeyr}, \citenamefont {Holighaus},\ and\ \citenamefont
  {Vandergheynst}}]{shuman2015spectrum}%
  \BibitemOpen
  \bibfield  {author} {\bibinfo {author} {\bibfnamefont {D.~I.}\ \bibnamefont
  {Shuman}}, \bibinfo {author} {\bibfnamefont {C.}~\bibnamefont {Wiesmeyr}},
  \bibinfo {author} {\bibfnamefont {N.}~\bibnamefont {Holighaus}}, \ and\
  \bibinfo {author} {\bibfnamefont {P.}~\bibnamefont {Vandergheynst}},\
  }\bibfield  {title} {\enquote {\bibinfo {title} {Spectrum-adapted tight graph
  wavelet and vertex-frequency frames},}\ }\href@noop {} {\bibfield  {journal}
  {\bibinfo  {journal} {IEEE Transactions on Signal Processing}\ }\textbf
  {\bibinfo {volume} {63}},\ \bibinfo {pages} {4223--4235} (\bibinfo {year}
  {2015})}\BibitemShut {NoStop}%
\bibitem [{\citenamefont {Leonardi}\ and\ \citenamefont {Van
  De~Ville}(2013)}]{leonardi2013tight}%
  \BibitemOpen
  \bibfield  {author} {\bibinfo {author} {\bibfnamefont {N.}~\bibnamefont
  {Leonardi}}\ and\ \bibinfo {author} {\bibfnamefont {D.}~\bibnamefont {Van
  De~Ville}},\ }\bibfield  {title} {\enquote {\bibinfo {title} {Tight wavelet
  frames on multislice graphs},}\ }\href@noop {} {\bibfield  {journal}
  {\bibinfo  {journal} {IEEE Transactions on Signal Processing}\ }\textbf
  {\bibinfo {volume} {61}},\ \bibinfo {pages} {3357--3367} (\bibinfo {year}
  {2013})}\BibitemShut {NoStop}%
\bibitem [{\citenamefont {Chung}\ and\ \citenamefont
  {Graham}(1997)}]{chung1997spectral}%
  \BibitemOpen
  \bibfield  {author} {\bibinfo {author} {\bibfnamefont {F.~R.}\ \bibnamefont
  {Chung}}\ and\ \bibinfo {author} {\bibfnamefont {F.~C.}\ \bibnamefont
  {Graham}},\ }\href@noop {} {\emph {\bibinfo {title} {Spectral graph
  theory}}},\ \bibinfo {number} {92}\ (\bibinfo  {publisher} {American
  Mathematical Soc.},\ \bibinfo {year} {1997})\BibitemShut {NoStop}%
\bibitem [{\citenamefont {Gama}, \citenamefont {Bruna},\ and\ \citenamefont
  {Ribeiro}(2019)}]{gama2019stability}%
  \BibitemOpen
  \bibfield  {author} {\bibinfo {author} {\bibfnamefont {F.}~\bibnamefont
  {Gama}}, \bibinfo {author} {\bibfnamefont {J.}~\bibnamefont {Bruna}}, \ and\
  \bibinfo {author} {\bibfnamefont {A.}~\bibnamefont {Ribeiro}},\ }\bibfield
  {title} {\enquote {\bibinfo {title} {Stability of graph scattering
  transforms},}\ }\href@noop {} {\bibfield  {journal} {\bibinfo  {journal}
  {arXiv preprint arXiv:1906.04784}\ } (\bibinfo {year} {2019})}\BibitemShut
  {NoStop}%
\bibitem [{\citenamefont {Newton}(1991)}]{newton1991weighted}%
  \BibitemOpen
  \bibfield  {author} {\bibinfo {author} {\bibfnamefont {M.~A.}\ \bibnamefont
  {Newton}},\ }\emph {\bibinfo {title} {The weighted likelihood bootstrap and
  an algorithm for prepivoting}},\ \href@noop {} {Ph.D. thesis},\ \bibinfo
  {school} {University of Washington} (\bibinfo {year} {1991})\BibitemShut
  {NoStop}%
\bibitem [{\citenamefont {Newton}\ and\ \citenamefont
  {Raftery}(1994)}]{newton1994approximate}%
  \BibitemOpen
  \bibfield  {author} {\bibinfo {author} {\bibfnamefont {M.~A.}\ \bibnamefont
  {Newton}}\ and\ \bibinfo {author} {\bibfnamefont {A.~E.}\ \bibnamefont
  {Raftery}},\ }\bibfield  {title} {\enquote {\bibinfo {title} {Approximate
  bayesian inference with the weighted likelihood bootstrap},}\ }\href@noop {}
  {\bibfield  {journal} {\bibinfo  {journal} {Journal of the Royal Statistical
  Society: Series B (Methodological)}\ }\textbf {\bibinfo {volume} {56}},\
  \bibinfo {pages} {3--26} (\bibinfo {year} {1994})}\BibitemShut {NoStop}%
\bibitem [{\citenamefont {Efron}(1992)}]{efron1992bootstrap}%
  \BibitemOpen
  \bibfield  {author} {\bibinfo {author} {\bibfnamefont {B.}~\bibnamefont
  {Efron}},\ }\bibfield  {title} {\enquote {\bibinfo {title} {Bootstrap
  methods: {Another} look at the jackknife},}\ }in\ \href@noop {} {\emph
  {\bibinfo {booktitle} {Breakthroughs in statistics}}}\ (\bibinfo  {publisher}
  {Springer},\ \bibinfo {year} {1992})\ pp.\ \bibinfo {pages}
  {569--593}\BibitemShut {NoStop}%
\bibitem [{\citenamefont {Haldane}(1948)}]{haldane1948precision}%
  \BibitemOpen
  \bibfield  {author} {\bibinfo {author} {\bibfnamefont {J.}~\bibnamefont
  {Haldane}},\ }\bibfield  {title} {\enquote {\bibinfo {title} {The precision
  of observed values of small frequencies},}\ }\href@noop {} {\bibfield
  {journal} {\bibinfo  {journal} {Biometrika}\ }\textbf {\bibinfo {volume}
  {35}},\ \bibinfo {pages} {297--300} (\bibinfo {year} {1948})}\BibitemShut
  {NoStop}%
\bibitem [{\citenamefont {Breiman}(1996)}]{breiman1996bagging}%
  \BibitemOpen
  \bibfield  {author} {\bibinfo {author} {\bibfnamefont {L.}~\bibnamefont
  {Breiman}},\ }\bibfield  {title} {\enquote {\bibinfo {title} {Bagging
  predictors},}\ }\href@noop {} {\bibfield  {journal} {\bibinfo  {journal}
  {Machine learning}\ }\textbf {\bibinfo {volume} {24}},\ \bibinfo {pages}
  {123--140} (\bibinfo {year} {1996})}\BibitemShut {NoStop}%
\bibitem [{\citenamefont {Clyde}\ and\ \citenamefont
  {Lee}(2001)}]{clyde2001bagging}%
  \BibitemOpen
  \bibfield  {author} {\bibinfo {author} {\bibfnamefont {M.}~\bibnamefont
  {Clyde}}\ and\ \bibinfo {author} {\bibfnamefont {H.}~\bibnamefont {Lee}},\
  }\bibfield  {title} {\enquote {\bibinfo {title} {Bagging and the {Bayesian
  Bootstrap.}}}\ }in\ \href@noop {} {\emph {\bibinfo {booktitle} {AISTATS}}}\
  (\bibinfo {year} {2001})\BibitemShut {NoStop}%
\bibitem [{\citenamefont {Ramakrishnan}\ \emph {et~al.}(2014)\citenamefont
  {Ramakrishnan}, \citenamefont {Dral}, \citenamefont {Rupp},\ and\
  \citenamefont {Von~Lilienfeld}}]{ramakrishnan2014quantum}%
  \BibitemOpen
  \bibfield  {author} {\bibinfo {author} {\bibfnamefont {R.}~\bibnamefont
  {Ramakrishnan}}, \bibinfo {author} {\bibfnamefont {P.~O.}\ \bibnamefont
  {Dral}}, \bibinfo {author} {\bibfnamefont {M.}~\bibnamefont {Rupp}}, \ and\
  \bibinfo {author} {\bibfnamefont {O.~A.}\ \bibnamefont {Von~Lilienfeld}},\
  }\bibfield  {title} {\enquote {\bibinfo {title} {Quantum chemistry structures
  and properties of 134 kilo molecules},}\ }\href@noop {} {\bibfield  {journal}
  {\bibinfo  {journal} {Scientific data}\ }\textbf {\bibinfo {volume} {1}},\
  \bibinfo {pages} {140022} (\bibinfo {year} {2014})}\BibitemShut {NoStop}%
\bibitem [{\citenamefont {Ruddigkeit}\ \emph {et~al.}(2012)\citenamefont
  {Ruddigkeit}, \citenamefont {Van~Deursen}, \citenamefont {Blum},\ and\
  \citenamefont {Reymond}}]{ruddigkeit2012enumeration}%
  \BibitemOpen
  \bibfield  {author} {\bibinfo {author} {\bibfnamefont {L.}~\bibnamefont
  {Ruddigkeit}}, \bibinfo {author} {\bibfnamefont {R.}~\bibnamefont
  {Van~Deursen}}, \bibinfo {author} {\bibfnamefont {L.~C.}\ \bibnamefont
  {Blum}}, \ and\ \bibinfo {author} {\bibfnamefont {J.-L.}\ \bibnamefont
  {Reymond}},\ }\bibfield  {title} {\enquote {\bibinfo {title} {Enumeration of
  166 billion organic small molecules in the chemical universe database
  gdb-17},}\ }\href@noop {} {\bibfield  {journal} {\bibinfo  {journal} {Journal
  of chemical information and modeling}\ }\textbf {\bibinfo {volume} {52}},\
  \bibinfo {pages} {2864--2875} (\bibinfo {year} {2012})}\BibitemShut {NoStop}%
\bibitem [{\citenamefont {Ertl}, \citenamefont {Rohde},\ and\ \citenamefont
  {Selzer}(2000)}]{ertl2000fast}%
  \BibitemOpen
  \bibfield  {author} {\bibinfo {author} {\bibfnamefont {P.}~\bibnamefont
  {Ertl}}, \bibinfo {author} {\bibfnamefont {B.}~\bibnamefont {Rohde}}, \ and\
  \bibinfo {author} {\bibfnamefont {P.}~\bibnamefont {Selzer}},\ }\bibfield
  {title} {\enquote {\bibinfo {title} {Fast calculation of molecular polar
  surface area as a sum of fragment-based contributions and its application to
  the prediction of drug transport properties},}\ }\href@noop {} {\bibfield
  {journal} {\bibinfo  {journal} {Journal of medicinal chemistry}\ }\textbf
  {\bibinfo {volume} {43}},\ \bibinfo {pages} {3714--3717} (\bibinfo {year}
  {2000})}\BibitemShut {NoStop}%
\bibitem [{\citenamefont {Kwon}(2001)}]{kwon2001handbook}%
  \BibitemOpen
  \bibfield  {author} {\bibinfo {author} {\bibfnamefont {Y.}~\bibnamefont
  {Kwon}},\ }\href@noop {} {\emph {\bibinfo {title} {Handbook of essential
  pharmacokinetics, pharmacodynamics and drug metabolism for industrial
  scientists}}}\ (\bibinfo  {publisher} {Springer Science \& Business Media},\
  \bibinfo {year} {2001})\BibitemShut {NoStop}%
\bibitem [{\citenamefont {Lipinski}\ \emph {et~al.}(1997)\citenamefont
  {Lipinski}, \citenamefont {Lombardo}, \citenamefont {Dominy},\ and\
  \citenamefont {Feeney}}]{lipinski1997experimental}%
  \BibitemOpen
  \bibfield  {author} {\bibinfo {author} {\bibfnamefont {C.~A.}\ \bibnamefont
  {Lipinski}}, \bibinfo {author} {\bibfnamefont {F.}~\bibnamefont {Lombardo}},
  \bibinfo {author} {\bibfnamefont {B.~W.}\ \bibnamefont {Dominy}}, \ and\
  \bibinfo {author} {\bibfnamefont {P.~J.}\ \bibnamefont {Feeney}},\ }\bibfield
   {title} {\enquote {\bibinfo {title} {Experimental and computational
  approaches to estimate solubility and permeability in drug discovery and
  development settings},}\ }\href@noop {} {\bibfield  {journal} {\bibinfo
  {journal} {Advanced drug delivery reviews}\ }\textbf {\bibinfo {volume}
  {23}},\ \bibinfo {pages} {3--25} (\bibinfo {year} {1997})}\BibitemShut
  {NoStop}%
\bibitem [{\citenamefont {Koltun}(1965)}]{koltun1965space}%
  \BibitemOpen
  \bibfield  {author} {\bibinfo {author} {\bibfnamefont {W.~L.}\ \bibnamefont
  {Koltun}},\ }\href@noop {} {\enquote {\bibinfo {title} {Space filling atomic
  units and connectors for molecular models},}\ } (\bibinfo {year} {1965}),\
  \bibinfo {note} {uS Patent 3,170,246}\BibitemShut {NoStop}%
\bibitem [{\citenamefont {Wildman}\ and\ \citenamefont
  {Crippen}(1999)}]{wildman1999prediction}%
  \BibitemOpen
  \bibfield  {author} {\bibinfo {author} {\bibfnamefont {S.~A.}\ \bibnamefont
  {Wildman}}\ and\ \bibinfo {author} {\bibfnamefont {G.~M.}\ \bibnamefont
  {Crippen}},\ }\bibfield  {title} {\enquote {\bibinfo {title} {Prediction of
  physicochemical parameters by atomic contributions},}\ }\href@noop {}
  {\bibfield  {journal} {\bibinfo  {journal} {Journal of chemical information
  and computer sciences}\ }\textbf {\bibinfo {volume} {39}},\ \bibinfo {pages}
  {868--873} (\bibinfo {year} {1999})}\BibitemShut {NoStop}%
\bibitem [{\citenamefont {Landrum}(2016)}]{landrum2016rdkit}%
  \BibitemOpen
  \bibfield  {author} {\bibinfo {author} {\bibfnamefont {G.}~\bibnamefont
  {Landrum}},\ }\href@noop {} {\enquote {\bibinfo {title} {Rdkit: Open-source
  cheminformatics software},}\ } (\bibinfo {year} {2016})\BibitemShut {NoStop}%
\bibitem [{\citenamefont {Zou}\ and\ \citenamefont
  {Lerman}(2018)}]{zou2018encoding}%
  \BibitemOpen
  \bibfield  {author} {\bibinfo {author} {\bibfnamefont {D.}~\bibnamefont
  {Zou}}\ and\ \bibinfo {author} {\bibfnamefont {G.}~\bibnamefont {Lerman}},\
  }\href@noop {} {\enquote {\bibinfo {title} {Encoding robust representation
  for graph generation},}\ } (\bibinfo {year} {2018}),\ \Eprint
  {http://arxiv.org/abs/1809.10851} {arXiv:1809.10851 [cs.LG]} \BibitemShut
  {NoStop}%
\end{thebibliography}%

\end{document}